\documentclass[a4paper, preprint, 5p, authoryear]{elsarticle}  

\pdfoutput=1
\usepackage{chngcntr}

\usepackage{amsfonts, amssymb, amsmath, mathtools, cuted, array, tabularx}
\usepackage[]{graphicx} 
\graphicspath{{figs/}}
\usepackage{subfig}
\usepackage{dblfloatfix}

\usepackage[open, openlevel=4, atend, numbered]{bookmark}
\usepackage{xcolor}
\usepackage{stackengine}
\usepackage{floatrow}
\usepackage[inline]{enumitem}
\usepackage[export]{adjustbox}
\usepackage{booktabs}
\usepackage{diagbox}

\usepackage{array}
\newcolumntype{L}[1]{>{\raggedright\let\newline\\\arraybackslash\hspace{0pt}}m{#1}}
\newcolumntype{C}[1]{>{\centering\let\newline\\\arraybackslash\hspace{0pt}}m{#1}}
\newcolumntype{R}[1]{>{\raggedleft\let\newline\\\arraybackslash\hspace{0pt}}m{#1}}

\newcounter{rowcntr}[table]
\renewcommand{\therowcntr}{\arabic{rowcntr}}
\newcolumntype{N}{>{\refstepcounter{rowcntr}}l<{(\therowcntr)}}
\AtBeginEnvironment{tabular}{\setcounter{rowcntr}{0}}

\usepackage{tikz}
\usetikzlibrary{backgrounds}
\usetikzlibrary{decorations.pathreplacing}
\usetikzlibrary{shapes}
\usetikzlibrary{arrows.meta}
\usetikzlibrary{calc}

\tikzset{
  mynode/.style={align=center, execute at begin node=\setlength{\baselineskip}{0.7em}}
}

\definecolor{myblue}{RGB}{44,123,182}

\newcommand{\dv}[2]{\frac{{\rm d}#1}{{\rm d}#2}}

\newcommand{\tr}{{\rm tr}}
\newcommand{\note}[1]{{#1}}

\newcommand{\smalltitle}[2]{\subsubsection{#1}\label{#2}}

\newcommand{\modelfigsize}{1}
\newcommand{\nullfigsize}{0.325}

\newcommand{\nullegrelfigsize}{0.7538}

\usepackage[titletoc, title]{appendix}
\usepackage{etoolbox}
\preto\subequations{\ifhmode\unskip\fi}
\patchcmd{\appendices}{\quad}{. }{}{}

\usepackage{textgreek} 
\usepackage{hyperref}
\hypersetup{unicode,colorlinks=true}
\usepackage[capitalise, noabbrev]{cleveref}
\crefname{equation}{}{}
\crefrangeformat{equation}{(#3#1#4)--(#5#2#6)}

\crefname{appsec}{Appendix}{Appendices}

\crefname{regitem}{region}{regions}
\Crefname{regitem}{Region}{Regions}
\crefformat{regitem}{region~(#2#1#3)}
\Crefformat{regitem}{Region~(#2#1#3)}
\Crefrangeformat{regitem}{(#3#1#4)--(#5#2#6)}
\crefrangeformat{regitem}{(#3#1#4)--(#5#2#6)}

\crefalias{enumi}{regitem}
\crefalias{rowcntr}{regitem}

\usepackage[normalem]{ulem}
\usepackage{lmodern}

\newcommand{\bref}[1]{(\ref{#1})}
\newcommand{\mycrefmultiregitem}[2]{Regions~\bref{#1} and~\bref{#2}}

\usepackage{microtype} 
\biboptions{round, semicolon} %


\journal{Journal of Theoretical Biology}

\date{\today} 

\begin{document}

	\begin{frontmatter}

		\title{Identifying and characterising the impact of excitability in a mathematical model of tumour-immune interactions}

		\author[add1]{Ana Osojnik\corref{cor1}}
		\ead{osojnik@maths.ox.ac.uk}
		\author[add1]{Eamonn A. Gaffney}
		\author[add2]{Michael Davies}
		\author[add2]{James W. T. Yates}
		\author[add1]{Helen M. Byrne}

		\address[add1]{Wolfson Centre for Mathematical Biology, Mathematical Institute, University of Oxford, Andrew Wiles Building, Woodstock Road, Oxford, OX2 6GG, UK}
		\address[add2]{DMPK, Early Oncology, Oncology R\&D, AstraZeneca, 
		Chesterford Research Park, Little Chesterford, Cambridge, CB10 1XL, UK}

		\cortext[cor1]{Corresponding Author: }

		\begin{abstract}

\noindent We study a five-compartment mathematical model originally proposed by \citet{Kuznetsov1994} to investigate the effect of nonlinear interactions between tumour and immune cells in the tumour microenvironment, whereby immune cells may induce tumour cell death, and tumour cells may inactivate immune cells. Exploiting a separation of timescales in the model, we use the method of matched asymptotics to derive a new two-dimensional, long-timescale, approximation of the full model, which differs from the quasi-steady-state approximation introduced by \citet{Kuznetsov1994}, but is validated against numerical solutions of the full model. Through a phase-plane analysis, we show that our reduced model is excitable, a feature not traditionally associated with tumour-immune dynamics. Through a systematic parameter sensitivity analysis, we demonstrate that excitability generates complex bifurcating dynamics in the model. These are consistent with a variety of clinically observed phenomena, and suggest that excitability may underpin tumour-immune interactions. The model exhibits the three stages of immunoediting -- elimination, equilibrium, and escape, via stable steady states with different tumour cell concentrations. Such heterogeneity in tumour cell numbers can stem from variability in initial conditions and/or model parameters that control the properties of the immune system and its response to the tumour. We identify different biophysical parameter targets that could be manipulated with immunotherapy in order to control tumour size, and we find that preferred strategies may differ between patients depending on the strength of their immune systems, as determined by patient-specific values of associated model parameters.
		\end{abstract}


	\end{frontmatter}


\section{Introduction}\label{sec:intro}

Cancer ranks as one of the leading causes of death worldwide, with millions of new cancer cases diagnosed every year, and incidence and mortality rates rapidly growing \citep{Bray2018}. 
Despite being toxic and lacking specificity for tumour cells, chemotherapy and surgery, together with radiotherapy, remain the standard of care for cancer patients. The ability of cancer to form metastases makes it challenging to control by these treatments alone, thereby raising the demand for complementary approaches to cancer therapy.

In recent years immunotherapy has become an increasingly important area of research. Understanding how treatments, which harness the immune system to treat cancer, may work provides the motivation for this study. The central idea underpinning cancer immunotherapy is cancer `immunosurveillance'%
, whereby tumour detection is enabled by tumour antigens. These antigens coat the tumour cell surface and, due to genetic alterations, are sufficiently different from other self-antigens so as to be immunogenic. The ability to detect tumour antigens allows the immune system to target cancer cells for destruction.

Initiation, execution and regulation of a specific adaptive immune response is performed by a group of immune cells called T cells. All T cells express T cell receptors (TCRs) that are specific for a particular antigen. When TCRs on a T cell match with cognate tumour antigens presented on the surface of phagocytotic dendritic cells that collect and process antigen material, the T cell is activated, and becomes an effector T cell. Different subtypes of effector T cells perform complementary functions to eliminate foreign cells. They proliferate, and secrete growth factors and cytokines, through which also other immune cells engage with the immune response to the tumour. Effector cytotoxic T cells (CTLs) can travel to the tumour, where they scan tumour cells until they find target cells, whose presented antigens match their TCRs. Upon recognition, CTLs programme target cells to die by apoptosis, either by delivering a mixture of perforin and granzyme to target cells or via engagement of the cell death surface receptor Fas \citep{Russell2002, Murphy2016}.

The above description summarises the key processes involved in successful immunosurveillance. In practice, however, multiple mechanisms enable tumour cells to escape immune elimination. The surviving, immunologically resistant tumour cells are maintained in a state of immune-mediated equilibrium, with some cells undergoing further division and editing. During this process, the selective pressure exerted by the immune system can result in the emergence of new tumour cell variants that can resist, avoid, or suppress immune attack \citep{Swann2007}. They may develop into clinically detectable tumours, and establish an immunosuppressive environment, enabling tumour progression. \citet{Dunn2004} termed this process immunoediting, proposing three possible outcomes of tumour-immune interactions, called `the three Es of immunoediting': tumour elimination, equilibrium, and escape.

In order to re-establish self-sustaining immunity to cancer, the  objective of some immunotherapies is to target specific mechanisms of tumour escape from immunosurveillance. 
A variety of strategies have been proposed and the most effective ones are now being combined with standard treatments in the clinic \citep{Emens2017}. These immunotherapies include: immune checkpoint inhibitors, which block immunosuppressive pathways, activated in a healthy organism to prevent over-inflammatory responses and minimize tissue damage \citep{Pardoll2012}; therapeutic dendritic cell vaccine, which boosts the population of dendritic cells that present tumour-associated antigens to effector T cells \citep{Hammerstrom2011}; and adoptive T cell transfer therapy, which induces tumour specificity in the population of effector T cells via genetic engineering or antigen-specific expansion \citep{Baruch2017}. 
Despite many successes, even the most promising immune checkpoint inhibitors have seen average response rates of less than 50\% \citep{Lipson2015, Grywalska2018}%
. To increase the fraction of responsive patients, better understanding of the complex interactions between the tumour and the immune system is needed.

With increasing recognition of mathematical modelling as a tool to gain mechanistic insight, a variety of models have been devised to explain clinically observed features of tumour-immune interactions. These models vary from spatial to non-spatial models, deterministic to stochastic, continuous to discrete, single scale to multiscale. The most commonly used framework, also employed in this paper, involves the use of ordinary differential equations (ODEs) that capture spatially-averaged, time-varying dynamics at the population level. 
Other models are formulated in terms of stochastic differential equations \citep{Lefever1979a, Bose2009, Xu2013, Liu2018}, partial differential equations \citep{Matzavinos2004, Webb2007, Lai2017, Friedman2018}, integro-differential equations \citep{Bellomo2007}, agent-based models \citep{Owen2011, Baar2016, Macfarlane2018} or hybrid models \citep{Mallet2006, Lopez2014, Gong2017}. The simplicity of ODEs renders them more analytically tractable and easier to validate than other approaches, while still capturing the essential features of tumour-immune interactions.
Some ODE models describe tumour-immune interactions by analogy with predator-prey models, with effector T cells as predators and tumour cells as prey \citep{Kuznetsov1994, Sotolongo-Costa2003, DeVladar2004, DOnofrio2005, Frascoli2014, Dritschel2018}. For an exhaustive review of ODE models of tumour-immune interactions we refer the reader to \citet{Eftimie2011} and \citet{Wilkie2013a}.

In this paper we study the model by \citet{Kuznetsov1994}, which was proposed to explain the phenomena of tumour dormancy/equilibrium, sneaking through %
\citep[when small tumours fail to induce immune responses and grow to escape, while medium tumours are contained by the immune system;][]{Gatenby1981}, and immunostimulation (when stimulation of immune response leads to progressive tumour growth rather than elimination). The five-compartment model describes interactions between effector and tumour cells through the formation of an intermediate effector-tumour complex. The model incorporates  immune killing of tumour cells, tumour-induced suppression and anergy of effector cells%
, and limits on immune recruitment and proliferation when the tumour grows large. \citet{Kuznetsov1994} perform model reduction by assuming that the timescale on which tumour and effector cells form complexes is much shorter than the timescale of other processes, such as effector cell proliferation and recruitment to the tumour site. A quasi-steady-state approximation (QSSA) is made, under which the model reduces to two ODEs that describe the dynamics of the effector and tumour cells; the complexes vary parametrically with these variables. 
\citeauthor{Kuznetsov1994}'s analysis of the QSSA model's bifurcations 
reveals how variation in the influx rate and death rate of effector cells affects system dynamics; the model exhibits tumour escape, equilibrium or elimination, or coexistence (bistability) of small-tumour or tumour-free steady states with escape. Occurrence of sneaking through and immunostimulatory effects in this model is found to depend on the rate of effector cell inactivation inflicted by tumour cells.

The QSSA model of \citet{Kuznetsov1994} has been used and modified in numerous studies. 
By relaxing assumptions about effector cell recruitment and proliferation, \citet{Gaach2003} proposed a simpler model that captures the three Es of immunoediting in different parameter regimes. 
 \citet{Roesch2014} adapted the QSSA model to study the impact of chemotherapy on large B cell lymphoma. 
\citet{Dritschel2018} introduced a helper T cell compartment to the QSSA model by \citet{Kuznetsov1994} to distinguish immune cell promotion and tumour cell killing, and implicitly incorporated tumour-modulated immunosupression by altering the functions describing immune recruitment and proliferation. In addition to the three Es of immunoediting, their model exhibits periodic growth and suppression of the immune
and tumour populations, unlike the QSSA model by \citet{Kuznetsov1994}. Oscillatory dynamics have been observed in extensions of the QSSA model that include time delays to account for the delay between antigen recognition and immune response \citep{Gaach2003, Rihan2014}. Other authors adapted the QSSA model by incorporating 
interactions with healthy cells \citep{DePillis2001} or natural killer (NK) cells \citep{DePillis2005}. While \citeauthor{Kuznetsov1994}'s QSSA model and its modifications have been studied extensively, the validity of the QSSA model or the dynamics of their original model have not been.

{
Parameters in such ODE models may vary for a variety of reasons. Parameters estimated from clinical data will have uncertainties associated with them, and parameters may vary between patients, depending on factors such as the aggressiveness and immunogenicity of their tumours, or the strength of their immune responses. Administration of chemo- and/or immunotherapy may also alter one or more model parameters. 
The effects of parameter perturbations on model behaviour are therefore analysed to uncover the full range of model dynamics. 
For example, \citet{Kirschner1998} investigated the impact of administering the cytokine interleukin-2 (IL-2) and adoptive cell therapy (ACT), separately and in combination. In their model, which is similar to the QSSA model with addition of an IL-2 compartment, parameters controlling the (constant) supply rates of IL-2 and immune cells were varied to model the effects of IL-2 and ACT. In this way ACT as a monotherapy or combined with IL-2 was found to be more beneficial for the patient than IL-2 alone as a monotherapy. \citet{Dritschel2018} thoroughly explored the parameter space of their model incorporating helper T cells to demonstrate that targetting the helper T cell population with ACT may be more effective than targetting the cytotoxic T cell population.
}

{In this paper we study \citeauthor{Kuznetsov1994}'s original model of tumour-immune interactions. We explore the validity of the QSSA, and investigate the complex tumour-immune dynamics that emerge from the original model. Using asymptotic, phase-plane and bifurcation analysis, in combination with numerical techniques, we identify parameter regimes in which the model admits similar dynamics to the QSSA model described above, and other regimes in which it admits limit cycles and excitability; the latter has to our knowledge not been traditionally identified as a feature of existing models of tumour-immune interactions. We place the resulting mechanistic insights within the context of immuno-oncology, and the theoretical prospect of implementing tumour control.
}

The remainder of this article is organised as follows. In \cref{sec:model} we describe the original model of \citet{Kuznetsov1994}. In \cref{sec:asymp} we exploit the separation of timescales,  which arises due to fast conjugate dynamics, to perform an asymptotic reduction of the original model. The reduced model we obtain differs from the QSSA model derived in \citet{Kuznetsov1994}, but agrees with numerical simulations of the original model where the asymptotic approximation is valid. 
In \cref{sec:phase_plane} we use phase-plane methods to show that our reduced model is excitable, while in \cref{sec:bifurcation_analysis} we characterise the model's bifurcation structure as the supply and death rates of effector cells vary, and show how sensitive this is to variability in parameters associated with the immune reponse to the tumour. We discuss our findings in \cref{sec:discussion}, and explain how they can be used to understand tumour-immune interactions and to identify immunotherapeutic targets, represented by model parameters, that may successfully eliminate a tumour.


\section{Mathematical model}\label{sec:model}
In this section, we summarise the original model proposed by \citet{Kuznetsov1994}, describe the assumptions made about parameter values, and derive a dimensionless version of the model. This will facilitate its simplification and analysis of its long-term behaviour in later sections of the paper. We end by stating the parameter values that will be used for model analysis.

\subsection[Kuznetsov et al.'s original five-compartment model]{\citeauthor{Kuznetsov1994}'s original five-compartment model}

\begin{figure}[t!]
	\includegraphics[width=\modelfigsize\columnwidth]{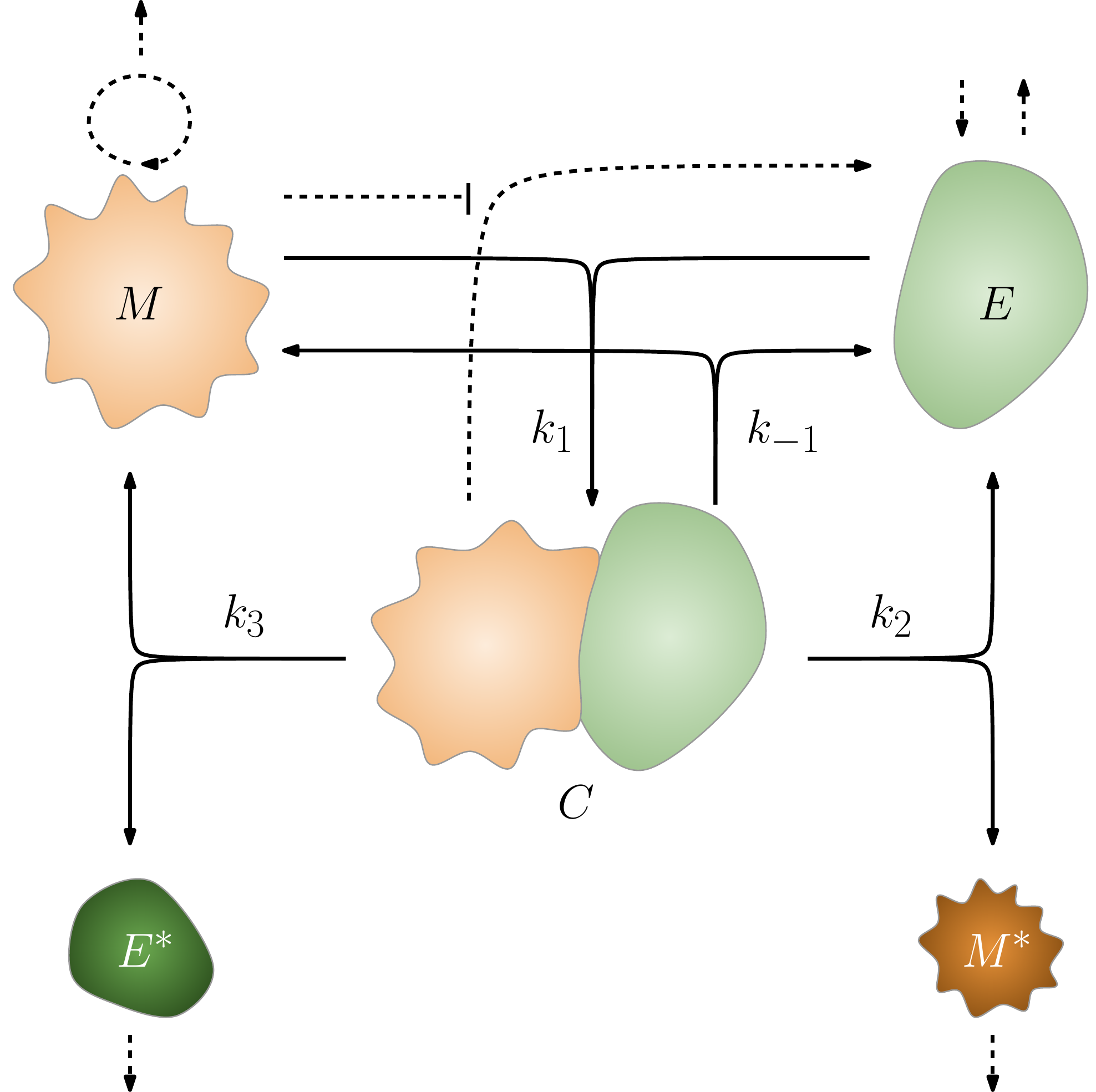}
	\caption{Schematic showing how effector (green) and tumour (orange) cells interact in the model of \protect\citet{Kuznetsov1994}. Kinetic interactions are denoted with solid arrows, while birth, death and migration processes are denoted with dashed ones. Arrows with a straight line in place of an arrow head represent inhibition of a process.}
	\label{fig:kinetic_scheme}
\end{figure}

\citet{Kuznetsov1994} proposed a system of five, time-dependent, nonlinear ordinary differential equations (ODEs) to describe temporal interactions between effector and tumour cells in a well-mixed (spatially-uniform) region, which represents the interface of the tumour and its microenvironment, see \cref{fig:kinetic_scheme}. Here the generic term effector (immune) cells is used to refer to all cell types capable of destroying tumour cells when activated, noting that these cells are anticipated to be predominantly CTLs. 
In their model, both tumour and effector cells may exist in a functional state, in which they may interact with other cells, or in a dysfunctional state, in which they are unable to interact with other cells. Accordingly, the tumour cell population is split into normal tumour cells, $M(t)$, and lethally hit (marked for death) tumour cells, $M^*(t)$, while the effector cell population is divided into activated effector cells, $E(t)$ and inactivated effector cells, $E^*(t)$. The model also includes an intermediate state, $C(t)$, associated with the formation of conjugates, when an activated effector cell and a normal tumour cell come in contact.

The evolution of the normal tumour cell population is governed by their proliferation, and programmed cell death due to interactions with effector cells. Proliferation is modelled via a logistic growth term $aM(1-bM)$, where the positive constant $a$ (units: $\text{day}^{-1}$) is the basal growth rate, and the reciprocal of the positive parameter $b$ (units: $\text{cells}^{-1}$) is the carrying capacity of the tumour site. 
Tumour cell programmed death, induced by effector cells, is modelled as a transition from the functional state, $M$, to the dysfunctional state, $M^*$, that is assumed to occur via effector-tumour conjugates, $C$; the inactivation of effector cells occurs in a similar manner. Conjugates are assumed to form at rate $k_1 EM$, which is proportional to the functional effector and tumour cell numbers, with constant of proportionality $k_1$ (units: $\text{cell}^{-1} \text{day}^{-1}$). There are three possible outcomes for the resulting conjugate, which may dissociate into (see solid arrows in \cref{fig:kinetic_scheme}): 
\begin{enumerate*}[label=(\roman*)]
	\item a tumour cell, $M$, and an activated effector cell, $E$, at rate $k_{-1} C$, proportional to conjugate numbers, with constant of proportionality $k_{-1}$ (units: $\text{day}^{-1}$); 
	\item a lethally hit (marked for programmed death) tumour cell, $M^*$,  and an activated effector cell, $E$, at rate $k_2 C$, which is proportional to conjugate numbers, with constant of proportionality $k_2$ (units: $\text{day}^{-1}$); 
	\item an inactivated (suppressed) effector cell, $E^*$, and a tumour cell, $M$, at rate $k_3 C$, proportional to conjugate numbers, with constant of proportionality $k_3$ (units: $\text{day}^{-1}$). 
\end{enumerate*}
Dysfunctional cells are assumed to arise only by transitioning, via conjugates, from their respective functional compartments, and their ultimate fate is death or migration out of the region. Programmed death of lethally hit tumour cells, $M^*$, is assumed to occur at rate $d_3$ (units: $\text{day}^{-1}$).

The evolution of the activated effector cells is assumed to be dominated by their migration from the lymph nodes to the tumour site, natural cell death, and inactivation by tumour cells; 
the latter is assumed to occur as described above. Migration to the tumour is modelled via a supply term with two parts: a positive constant rate, $s$ (units: $\text{cells } \text{day}^{-1}$), that represents a continuous baseline supply of activated effector cells to the tumour, and a tumour-stimulated rate, $
\frac{f C}{g + M} 
$, 
that accounts for an increase in the rate of effector recruitment, proliferation, and infiltration due to their interactions with tumour cells within conjugates; $f$ (units: cells $\text{day}^{-1}$) and $g$ (units: cells) are positive constants, such that $f/g$ is the maximum effector supply rate per complex, while $g$ is the tumour cell concentration, at which the influx rate per complex is half-maximal. 
The factor $(g + M)^{-1}$ ensures that the tumour-stimulated rate of influx saturates as the number of tumour cells increases, and accounts for limitations in the rate of stimulation of the immune system, or in the rate of transport to and through the tumour.
Activated effector cells are assumed to die at rate $d_1$ (units: $\text{day}^{-1}$). 
Similarly as for $M^*(t)$, the dysfunctional effector cells are assumed to die/migrate out at rate $d_2$ (units: $\text{day}^{-1}$).

By combining the processes outlined above, we recover the following ODEs for the different cell species that were proposed by \citet{Kuznetsov1994}:
\begin{subequations}
	\begin{align}
			&\dv{M}{t} = \underbrace{aM(1-bM_{\rm tot}
			)}_{\text{net proliferation}} \,- \underbrace{k_1 EM}_{\substack{\text{conjugate} \\\text{formation}}}
		 	+ \,\underbrace{(k_{-1} + k_3) C}_{\substack{\text{conjugate}\\\text{dissociation}\\ \text{w/o damage to}\\ \text{tumour cell}}}\,,\label{eq:orig_T}
		\\
		\begin{split}
			&\dv{E}{t} = \underbrace{s}_{\substack{\text{constant}\\ \text{supply}}} + \underbrace{\frac{f C}{g + M}}_{\substack{\text{tumour-}\\\text{stimulated} \\ \text{supply}}} - \,\underbrace{d_1 E}_{\text{death}}
			- \underbrace{k_1 EM}_{\substack{\text{conjugate}\\ \text{formation}}} 
			\\&\phantom{\dv{E}{t} =}
			+ \underbrace{(k_{-1} + k_2)C}_{\,\,\substack{\text{conjugate} \\ \text{dissociation}\\ \text{w/o damage to}\\ \text{effector cell}}\,\,}\, , \label{eq:orig_E}
		\end{split}
		\\
		&\dv{C}{t} = \underbrace{k_1 EM}_{\substack{\text{conjugate}\\ \text{formation}}} - \underbrace{(k_{-1} + k_2 + k_3) C}_{\text{conjugate dissociation}} \, ,\label{eq:orig_C}\\
		&\dv{M^*}{t} = \underbrace{k_2 C}_{\substack{\text{tumour cell}\\\text{marking}\\\text{for death}}} - \underbrace{d_3 M^*}_{\substack{\text{programmed}\\\text{death}}} , \label{eq:orig_T*}\\
		&\dv{E^*}{t} = \underbrace{k_3 C}_{\substack{\text{effector cell} \\ \text{inactivation}}} - \underbrace{d_2 E^*}_{\substack{\text{death or} \\ \text{migration}}} . \label{eq:orig_E*}
	\end{align}
We close equations \crefrange{eq:orig_T}{eq:orig_E*} by prescribing the following initial conditions:
	\begin{gather}
		E(0) = E_0,\quad M(0)=M_0,\quad C(0)=0, 
		\\
		E^*(0)=0, \quad \text{and} \quad M^*(0)=0, \label{eq:ini_dim}
	\end{gather}
	\label{eq:sys_dim}%
\end{subequations}
where the positive constants $E_0$ and $M_0$ denote the initial concentrations of free tumour cells and effector cells respectively. 
Noting the impact on the subsequent model solutions is essentially negligible, no conjugates, $C$, and dysfunctional cells, $M^*$ and $E^*$, are present initially. 
We note from equations \eqref{eq:sys_dim} that the dynamics of $M^*$ and $E^*$ are slave to those of $M$, $E$ and $C$. Since the system behaviour can be determined from equations \cref{eq:orig_E,eq:orig_T,eq:orig_C}, we henceforth focus on these.

\subsection{Parameter assumptions and the dimensionless model}\label{sec:assump}

\begin{table*}[b!]
	\centering
	\footnotesize
	\begin{tabular}{llll}
		\toprule
		\textbf{Parameter} & \textbf{Description} & \textbf{Value} \\
		\midrule
		$a$ & net tumour growth rate & $0.18\,\, \text{day}^{-1}$ \\
		$b$ & inverse carrying capacity & $2.0 \times 10^{-9}\,\,\text{cell}^{-1}$  \\
		$d_1$ & per cell death rate of activated effector cells & $0.0412\,\,\text{day}^{-1} $ \\
		$f$ & `maximum'' tumour-stimulated effector supply rate per complex &  $1.245 \times 10^{5}\,\,\text{cells}\,\,\text{day}^{-1}$ \\
		$g$ &  Michaelis-Menten-type constant & $2.019 \times 10^{7}\,\,\text{cells}$ \\
		$k_2$ & rate of effector inactivation & $0.1101 \,\, \text{day}^{-1}$\\
		$k_3$  & rate of tumour kill &  $3.422 \times 10^{-4} \,\, \text{day}^{-1}$\\
		$s$ & constant effector supply rate to the tumour site & $1.3\times 10^{4}\,\,\text{cells}\,\,\text{day}^{-1}$ \\
		$\hat{E}, \hat{M}, \hat{C}, \frac{K}{k_1}$ & typical tumour cell numbers & $10^6\,\,\text{cells}$ \\
		\bottomrule
	\end{tabular}
	\caption{Summary of the dimensional parameters that appear in the dimensional model \eqref{eq:sys_dim}, as estimated in \protect\cite{Kuznetsov1994}. The logistic growth parameters, $a$ and $b$, were estimated from tumour growth data for a non-chimeric mouse, where tumour mass was assumed to grow in the absence of immune response. Under premises that the system is in steady state in tumour absence, with the steady state number of effector cells in the region being $E^*=3.2\times 10^5$, and that the lifetime of cytotoxic T cells is 30 days or more, i.e. $d_1\approx 1/30 \text{ days}^{-1}$, the constant influx rate $s$ is estimated to be $s\approx E^* d_1$. Other parameters are estimated using data for a chimeric mouse with an immune system. Values for scalings of variables, $\hat{E}$ and $\hat{M}$, are chosen from experiments to be order-of-magnitude numbers of tumour cells, this is $10^{6}$ cells.}
	\label{tab:param_dim}
\end{table*}
\begin{table*}[t!]
	\centering
	\footnotesize
	\begin{tabular}{llll}
		\toprule
		\textbf{Parameter} & \textbf{Description} & \textbf{Value} \\
		\midrule
		$\alpha$ &  net tumour growth rate & $1.6348$ \\
		$\beta$ & inverse carrying capacity & $0.002$  \\
		$\delta$ & per cell death rate of activated effector cells & $0.374$ \\
		$\rho$ & ``maximum'' tumour-stimulated effector supply rate per complex &  $1.131$ \\
		$\eta$ & Michaelis-Menten-type constant  & $20.19$ \\
		$\mu$  & rate of effector inactivation &  $0.003$\\
		$\sigma$ &  constant effector supply rate to the tumour site & $0.118$ \\
		$\kappa_1, \kappa_2$ & ratios of variable scales & $1$\\
		\bottomrule
	\end{tabular}
	\caption{Summary of the dimensionless parameters that appear in the nondimensional model \eqref{eq:sys_full}. Estimates of their values are computed from dimensional parameters in Table \ref{tab:param_dim} according to parameter groupings \eqref{eq:param_nondim}.}
	\label{tab:param_nondim}
\end{table*}

We nondimensionalise equations \crefrange{eq:orig_T}{eq:orig_C} and \cref{eq:ini_dim} by rescaling model variables as follows:
\begin{subequations}
\begin{align}
	e = E/\hat{E}, \quad m = M/\hat{M}, \quad c = C/\hat{C}, \quad \tau = k_2 t,%
\end{align}%
where $\hat{E}$ and $\hat{M}$ are positive constants (see Table \ref{tab:param_dim}).  The scaling for $C$ is chosen as $\hat{C}=\frac{k_1\hat{E}\hat{M}}{K}$ with $K=k_{-1} + k_2 + k_3$, in order to balance terms in equation \eqref{eq:orig_C} for $C$.
We define the dimensionless parameter groupings as%
\begin{gather}
		\sigma = \frac{s}{k_2 \hat{E}}, \quad \rho = \frac{f k_1}{k_2 K}, \quad \eta=\frac{g}{\hat{M}}, \quad \delta = \frac{d_1}{k_2}, 
		\\ 
		\mu = \frac{k_3 k_1 \hat{M}}{k_2 K},\quad \kappa_1 =\frac{k_1 \hat{M}}{K},
		\quad
		\alpha=\frac{a}{k_2}, \quad \beta = b \hat{M}, 
		\\  
		\kappa_2 =\frac{k_1 \hat{E}}{K} =\kappa_1 \frac{\hat{E}}{\hat{M}}.
\end{gather}
Additionally, we identify a small dimensionless parameter grouping%
\begin{align}
	0<\varepsilon = \frac{k_2}{k_{-1} + k_2 + k_3}=\frac{k_2}{K} \ll 1.%
\end{align}%
\label{eq:param_nondim}%
\end{subequations}
\nolinebreak
\indent The choice of $\varepsilon$ is based on the observation that while effector cells constantly scan and interact with tumour cells in order to locate their target cell, only a small proportion of these interactions result in the marking of tumour cells for death, or inactivation of effector cells. For this to happen, the tumour antigen must be cognate to the TCR, but this is not the case at every effector-tumour interaction due to heterogeneity in repertoires of tumour antigens \citep{Chen2013,Rajasagi2014,Boyer1989} and TCRs at the tumour site \citep{Dovedi2017}. The average kinetics of tumour cell programmed death and effector cell inactivation is therefore slow in comparison to conjugate formation and dissociation without damage. 
In terms of parameter values, we thus assume that the per cell rates $k_{-1}$ and $k_{1} \hat{E}$, are large in comparison to $k_2$ and $k_3$, i.e.
$
k_{1} \hat{E} = \mathcal{O}(k_{-1})$, $ k_{2} \ll k_{-1}$, and  $k_3 = \mathcal{O}(k_2) \ll k_{-1}.
$ 
Given our interest in the long-term behaviour, we scale the model with one of the longer timescales, choosing $k_2^{-1}$ as in \citet{Kuznetsov1994}. 
Following their reasoning for applying the QSSA, we further assume that kinetics of processes such as cell death, proliferation and migration to the tumour microenvironement are also much slower than kinetics of conjugate formation and dissociation without damage, and we assume they are approximately comparable to those of tumour cell programmed death and effector cell inactivation; we therefore have $\tfrac{s}{\hat{E}},\, \tfrac{f k_1}{K}, \,d_1, \,a = \mathcal{O}(k_2) \ll k_{-1}$. 
In summary, there is a separation of timescales, with a number of events associated with the complexes occurring rapidly. 
We will see in \cref{sec:param} that our parameter assumptions correspond with parameter values from \citet{Kuznetsov1994}, given in \cref{tab:param_dim}. We note that these assumptions represent a distinguished limit, and estimates of parameters $k_1$ and $k_{-1}$ are needed to provide additional support for it.

Under the assumptions outlined above, equations \eqref{eq:sys_dim} can be rewritten in dimesionless form using rescalings in \eqref{eq:param_nondim}:
\begin{subequations}
	\label{eq:sys_full}%
	\begin{align}
	\dv{e}{\tau} &=\sigma + \frac{\rho c}{\eta + m} - \delta e - \mu c - \frac{\kappa_1}{\varepsilon}(em-c),\label{eq:sys_full_x}\\
	\dv{m}{\tau} &= \alpha m (1-\beta m) - \kappa_2 c - \frac{\kappa_2}{\varepsilon}(em-c),\label{eq:sys_full_y}\\
	\dv{c}{\tau} &= \frac{1}{\varepsilon}(em-c),\label{eq:sys_full_z}
	\end{align}
	with initial conditions
	\begin{align}
		\begin{gathered}
			e(0)=E_0/\hat{E}=:e_0, \quad m(0)=M_0/\hat{M}=:m_0,
			\\ 
			\text{and}\quad c(0)=0.
		\end{gathered}%
		\label{eq:ini_nondim}
	\end{align}%
\end{subequations}
In the remainder of this work, we use dimensionless forms of variables and parameters.

\subsection{Parameter values}\label{sec:param}

We use the dimensional parameter values given in \cref{tab:param_dim}, which were estimated in \citet{Kuznetsov1994} from tumour growth curves for B cell lymphoma in the spleen of mice \citep{Siu1986}, and using information about effector cell kinetics in tumour absence. 
The corresponding dimensionless parameter values are computed in \cref{tab:param_nondim}, by assuming that the ratio $K/k_1=10^{6}$ cells, or equivalently $K/k_1 \hat{E} \approx 1$. This agrees with our assumptions about parameters stated in \cref{sec:assump}, and also implies that $\kappa_1=\kappa_2=1$.

%
\section{Matched asymptotic model approximation}\label{sec:asymp}
The dependence  of equations \cref{eq:sys_full} on the small parameter $0 < \varepsilon \ll 1$ means that the system operates on at least two timescales. Exploiting this, we apply the method of matched asymptotic expansions to equations \cref{eq:sys_full} in order to derive approximate equations that are simpler to analyse.

We first rewrite the model as a slow-fast system of ordinary differential equations, with a clear difference in the timescales governing the dependent model variables. In other words, we seek to eliminate terms in \cref{eq:sys_full_x} and \cref{eq:sys_full_y} that scale with $1/\varepsilon$. We define
\begin{subequations}
	\begin{align}
		&
		v(\tau) := c(\tau),
		\\ 
		\text{and} \qquad
		&
		\boldsymbol{w}(\tau) := \begin{bmatrix}
		w_1(\tau) \\ w_2(\tau)
		\end{bmatrix} = \begin{bmatrix}
		e(\tau) + \kappa_1 c(\tau)\\
		m(\tau) + \kappa_2 c(\tau)
		\end{bmatrix},	
	\end{align}%
	\label{eq:map_variables}%
\end{subequations}
where $w_1$ and $w_2$ are, in descriptive terms, the total number of effector and tumour cells respectively, and $v$ is the number of conjugates at the tumour site. With significant algebra, differential equations that describe the evolution of $\boldsymbol{u}(\tau):=[v(\tau), \boldsymbol{w}(\tau)^T]^T$ take the desired form of a slow-fast system of ODEs given by
\begin{subequations}
	\begin{align}
	\varepsilon\dv{v}{\tau} = f(v,\boldsymbol{w}),\label{eq:v}\\
	\dv{\boldsymbol{w}}{\tau} = \boldsymbol{g} (v,\boldsymbol{w}),\label{eq:w}
	\end{align}
with initial conditions
	\begin{align}
		v(0) = 0, \qquad
		\boldsymbol{w}(0)=[w_{1,0}, w_{2,0}]^T=[e_0, m_0]^T,
	\end{align}
and
	\begin{align}
	f(v, \boldsymbol{w}) &:= em - c = (w_1 - \kappa_1 v)(w_2 - \kappa_2 v) - v,\label{eq:f}\\
	\boldsymbol{g}(v, \boldsymbol{w}) &:= \begin{bmatrix}
	\sigma + \frac{\rho v}{\eta + w_2 - \kappa_2 v} - \delta (w_1 - \kappa_1 v) - \mu v\\
	\alpha(w_2 - \kappa_2 v)(1- \beta(w_2-\kappa_2v)) - \kappa_2 v
	\end{bmatrix}.\label{eq:g}
	\end{align}%
	\label{eq:sys_sing}%
\end{subequations}%
\nolinebreak
\indent The singular nature of system \cref{eq:sys_sing}, where some derivatives are multiplied by the small parameter $\varepsilon$, means that we cannot construct a uniformly valid solution by taking the limit $\varepsilon\rightarrow 0$. We must consider separately approximations on the short and the long timescales, and match them to obtain a uniformly valid approximation of the full model solution $\boldsymbol{u}(\tau)$. We denote the inner and outer solutions by $\boldsymbol{u}_{\rm I}(\tau_{f})$ and $\boldsymbol{u}_{\rm O}(\tau)$ respectively, where $\tau_f$ denotes the rescaled time variable on the short timescale, proportional to the slow time variable $\tau$ via $\tau_f:=\tau/\varepsilon$. The composite matched solution $\boldsymbol{u}_{\rm C}(\tau)$ \citep{Hinch1991} is then given by
\begin{subequations}
\begin{align}
	\boldsymbol{u}_{\rm C}(\tau) &= \boldsymbol{u}_{\rm I}(\tau/\varepsilon) + \boldsymbol{u}_{\rm O}(\tau) - \boldsymbol{u}_{\rm overlap},	\label{eq:full_sol_w}
\end{align}
where, to satisfy the initial conditions \cref{eq:ini_nondim}, we  impose Prandtl's matching condition 
\begin{align}
	\underset{\tau_f \rightarrow \infty}{\lim} \boldsymbol{u}_{\rm I}(\tau_f)= \underset{\tau \rightarrow 0}{\lim} \,\, \boldsymbol{u}_{\rm O}(\tau)=\boldsymbol{u}_{\rm overlap}.\label{eq:matching_cond}
\end{align}
\label{eq:matching}%
\end{subequations}
We determine the inner and outer solutions in \cref{sec:fast_dynamics,,sec:slow_dynamics}.

\subsection{Asymptotic approximation of fast dynamics}\label{sec:fast_dynamics}
To analyse behaviour on the short timescale, we rescale time in equations \cref{eq:sys_sing} via $\tau_{f}=\tau/\varepsilon$, and obtain the system
\begin{subequations}
	\begin{align}
		\dv{v}{\tau_f} &= f(v, \boldsymbol{w}),\\
		\dv{\boldsymbol{w}}{\tau_f} &= \varepsilon\boldsymbol{g}(v, \boldsymbol{w}).
	\end{align}\label{eq:sys_sing_short}%
\end{subequations}
Let us suppose that
	\begin{align}
		v_{\rm I}(\tau_f; \varepsilon) = \sum_{i=0}^{\infty} \varepsilon^{i} v^{(i)}_{\rm I}(\tau_f), \quad 
		\text{and} \quad
		\boldsymbol{w}_{\rm I}(\tau_f; \varepsilon) = \sum_{i=0}^{\infty} \varepsilon^{i} \boldsymbol{w}^{(i)}_{\rm I}(\tau_f), \label{eq:asymptotic_expansions}
	\end{align}
are asymptotic expansions in $\varepsilon$, where subscript ${\rm I}$ denotes the inner (short-timescale) model solution.
By substituting from \eqref{eq:asymptotic_expansions} in \eqref{eq:sys_sing_short} and equating terms of $\mathcal{O}(\varepsilon^{i})$, we obtain a sequence of differential equations.
At leading order in $\varepsilon$ we obtain
\begin{subequations}
	\begin{align}
		\dv{{v}_{\rm I}}{\tau_f} &= f\left(v_{\rm I}, \boldsymbol{w}_{\rm I}\right),\\
		\dv{{\boldsymbol{w}}_{\rm I}}{\tau_f} &= 0,
	\end{align}
	\label{eq:sys_fast}%
	with initial conditions
	\begin{align}
		v_{\rm I}(0) = 0,\qquad 
		\boldsymbol{w}_{\rm I}(0)=\boldsymbol{w}_0=[w_{1,0}, w_{2, 0}]^T, \label{eq:sys_sing_fast_ini}
	\end{align}
\end{subequations}
where $\boldsymbol{u}_{\rm I}(\tau_f;\varepsilon)\sim\boldsymbol{u}_{\rm I}^{(0)}(\tau_{f})$. For simplicity henceforth $\boldsymbol{u}_{\rm I}$ will always represent the leading order approximation $\boldsymbol{u}_{\rm I}^{(0)}$. 
Equations \eqref{eq:sys_fast} imply that 
\begin{align}
	\boldsymbol{w}_{\rm I}(\tau_f) &:= \boldsymbol{w}_{\rm I}(0)
	=\boldsymbol{w}_{0}= [e_0, m_0]^T=\text{constant}
	\label{eq:sys_sing_fast}%
\end{align}
while dynamics of conjugates, $v_{\rm I}$, are governed by
\begin{subequations}
	\begin{align}
		\dv{v_{\rm I}}{\tau_f} &= A v_{\rm I}^2 - Bv_{\rm I} + C, %
	\end{align}
with%
	\begin{align}
		A=\kappa_1 \kappa_2, \,\,\, 
		B=1 + \kappa_2 w_{1, {\rm I}} + \kappa_1 w_{2, {\rm I}},\,\,\,
		C=w_{1, {\rm I}} w_{2, {\rm I}},
	\end{align}\label{eq:z_fast_simp}%
\end{subequations}
where $w_{1, {\rm I}}$ and $w_{2, {\rm I}}$ are  constants corresponding to initial conditions \eqref{eq:sys_sing_fast_ini}.
For positive initial conditions $\boldsymbol{w}_0>\boldsymbol{0}$ and $v_0=0$, we deduce that $v_{\rm I}$ tends to a positive steady state  $v_{\rm I}^*=v^\dagger(\boldsymbol{w}_0)$ satisfying $f(v^*_{\rm I},\boldsymbol{w}_{0})=0$ as $\tau_f \rightarrow \infty$, where we define
\begin{align}
	\begin{split}
		v^\dagger(\boldsymbol{w}) &= \frac{1}{2 \kappa_1 \kappa_2}\left(1 + \kappa_1 w_{2} + \kappa_2 w_1  
		\right.\\&\phantom{=}\left.
		- \sqrt{  (1 + \kappa_1 w_2 + \kappa_2 w_1 )^2 -4 \kappa_1 \kappa_2 w_1 w_2}\right) . 
	\end{split}
	\label{eq:z_star}
\end{align}
We conclude that, on the short timescale at leading order in $\varepsilon$, the number of conjugates, $v_{\rm I}$, rapidly relaxes from $0$ to $v^*_{\rm I}=v^\dagger(\boldsymbol{w}_0)$, while the total numbers of tumour and effector cells undergo negligible changes and are so conserved at $\boldsymbol{w}^*_{\rm I}=\boldsymbol{w}_0$. 
The numbers of free effector and tumour cells, $e$ and $m$, are depleted due to conjugate binding to their steady state values $e^*_{\rm I}=e^\dagger(\boldsymbol{w}_0)$ and $m^*_{\rm I}=m^\dagger(\boldsymbol{w}_0)$ respectively, where
\begin{subequations}
	\begin{align}
		\begin{split}
			e^\dagger(\boldsymbol{w}) &= w_1 - \kappa_1 v^\dagger(\boldsymbol{w}) 
			= \frac{1}{2\kappa_2}\left(\kappa_2 w_1 -1 - \kappa_1 w_2
			\right.\\&\phantom{=}\left.
			+ \sqrt{4 \kappa_2 w_1 + ( \kappa_2 w_1 -1 - \kappa_1 w_2 )^2}\right),
		\end{split}
		\\
		\begin{split}
			m^\dagger(\boldsymbol{w}) &= w_2 - \kappa_2 v^\dagger(\boldsymbol{w}) = \frac{1}{2\kappa_1}\left(\kappa_1 w_2 -1 - \kappa_2 w_1 
			\right.\\&\phantom{=}\left.
			+ \sqrt{4 \kappa_1 w_2 + ( \kappa_1 w_2 -1 - \kappa_2 w_1 )^2}\right).
		\end{split}
	\end{align}
	\label{eq:ss_fast_xy}%
\end{subequations}
Note that when the steady state $[e^*_{\rm I}, m^*_{\rm I}, c^*_{\rm I}]^T$ is reached, $\dv{c_{\rm I}}{\tau_f} = \dv{v_{\rm I}}{\tau_f}=f(v^*_{\rm I}, \boldsymbol{w}^*_{\rm I})=0$, or $v^*_{\rm I}=c^*_{\rm I}=e^*_{\rm I} m^*_{\rm I}$, and expressions \eqref{eq:z_star} and \eqref{eq:ss_fast_xy} satisfy 
	$v^\dagger(\boldsymbol{w})=e^\dagger(\boldsymbol{w})m^\dagger(\boldsymbol{w})$.

\subsection{Asymptotic approximation of slow dynamics}\label{sec:slow_dynamics}
We now focus on the long timescale, for which the dynamics of $v$ are slave to the dynamics of $\boldsymbol{w}$, i.e. $v$ changes parametrically with respect to $\boldsymbol{w}$.
On this timescale the total effector and tumour cell numbers, $w_1$ and $w_2$, vary significantly, since $\mathcal{O}(\varepsilon)$ terms in \eqref{eq:sys_sing_short} become non-negligible.
Formally, we consider the long-timescale system \eqref{eq:sys_sing}, and seek solutions for $v_{\rm O}$ and $\boldsymbol{w}_{\rm O}$ of the form
	\begin{align}
		v_{\rm O}(\tau; \varepsilon) = \sum_{i=0}^{\infty} \varepsilon^{i} v^{(i)}_{\rm O}(\tau), \quad
		\text{and} \quad
		\boldsymbol{w}_{\rm O}(\tau; \varepsilon) = \sum_{i=0}^{\infty} \varepsilon^{i} \boldsymbol{w}^{(i)}_{\rm O}(\tau).
	\end{align}
At leading order in $\varepsilon$, we obtain the following equations that approximate dynamics on the slow manifold,
\begin{subequations}
	\begin{gather}
		\dv{\boldsymbol{w}_{\rm O}}{\tau} = \boldsymbol{g}(v_{\rm O}, \boldsymbol{w}_{\rm O}), \\
		f(v_{\rm O}, \boldsymbol{w}_{\rm O})=0.\label{eq:manifold_condition}
	\end{gather}
According to \eqref{eq:manifold_condition}, $v_{\rm O}$ must be a positive solution to the quadratic \eqref{eq:f}, which is the case exactly when 
	$v_{\rm O}=v^\dagger(\boldsymbol{w}_{\rm O})$
	 \label{eq:slow_man}
\end{subequations}
as defined in \eqref{eq:z_star}, except that now $\boldsymbol{w}_{\rm O}$ is a time-dependent variable on the long timescale. Initial conditions for the outer (slow manifold) solution are given by ${\lim}_{\tau\rightarrow 0}=[v^\dagger(\boldsymbol{w}_0), \boldsymbol{w}_0^T]^T$, equivalent to the inner solution steady state $\boldsymbol{u}_{\rm I}^*=[v_{\rm I}^*, {\boldsymbol{w}_{\rm I}^*}^T]^T$ given in \cref{sec:fast_dynamics}, 
by which the matching condition \eqref{eq:matching_cond} is automatically satisfied. 
On the long timescale, the model reduces to
\begin{align}
	\dv{\boldsymbol{w}}{\tau}=\boldsymbol{g}(v^\dagger(\boldsymbol{w}), \boldsymbol{w}),
\end{align}
or equivalently
\begin{subequations}
	\begin{align}
	\dv{w_1}{\tau}&=\sigma + \frac{\rho e^\dagger(\boldsymbol{w}) m^\dagger(\boldsymbol{w})}{\eta + m^\dagger(\boldsymbol{w})} - \delta e^\dagger(\boldsymbol{w}) - \mu e^\dagger(\boldsymbol{w}) m^\dagger(\boldsymbol{w}),\label{eq:sys_slow2_w1}\\
	\dv{w_2}{\tau} &= \alpha m^\dagger(\boldsymbol{w})(1-\beta m^\dagger(\boldsymbol{w})) - \kappa_2 e^\dagger(\boldsymbol{w}) m^\dagger(\boldsymbol{w})\label{eq:sys_slow2_w2},
	\end{align}
	\label{eq:sys_slow2}%
\end{subequations}
where we drop the subscript O notation for simplicity.
Equations \eqref{eq:sys_slow2} are highly nonlinear, deeming their analysis difficult. To simplify the steady state and linear stability analysis in later sections, we rewrite these (slow manifold) equations in terms of the dimensionless variables $e$ and $m$, and obtain the following system of ODEs:
\begin{align}
	\begin{bmatrix}
	\dot{e} \\  \dot{m}
	\end{bmatrix} &= 
	\begin{bmatrix}
	1+\kappa_1 m & \kappa_1 e \\ \kappa_2 m & 1+\kappa_2 e
	\end{bmatrix}^{-1}
	\underbrace{\begin{bmatrix}
		\sigma + \frac{\rho em}{\eta + m} - \delta e- \mu em \\ \alpha m(1-\beta m) - \kappa_2 em
		\end{bmatrix}}_{=:\boldsymbol{q}(e,m)}.
	\label{eq:sys_slow_xy}
\end{align}
This contrasts with the \citeauthor{Kuznetsov1994}'s QSSA model given
by equations \cref{eq:sys_full} with $c = em$.

\subsection{Comparison of numerical solutions}\label{sec:num_sim}

\begin{figure*}[t!]
	\subfloat[]{%
	\includegraphics[width=\columnwidth]{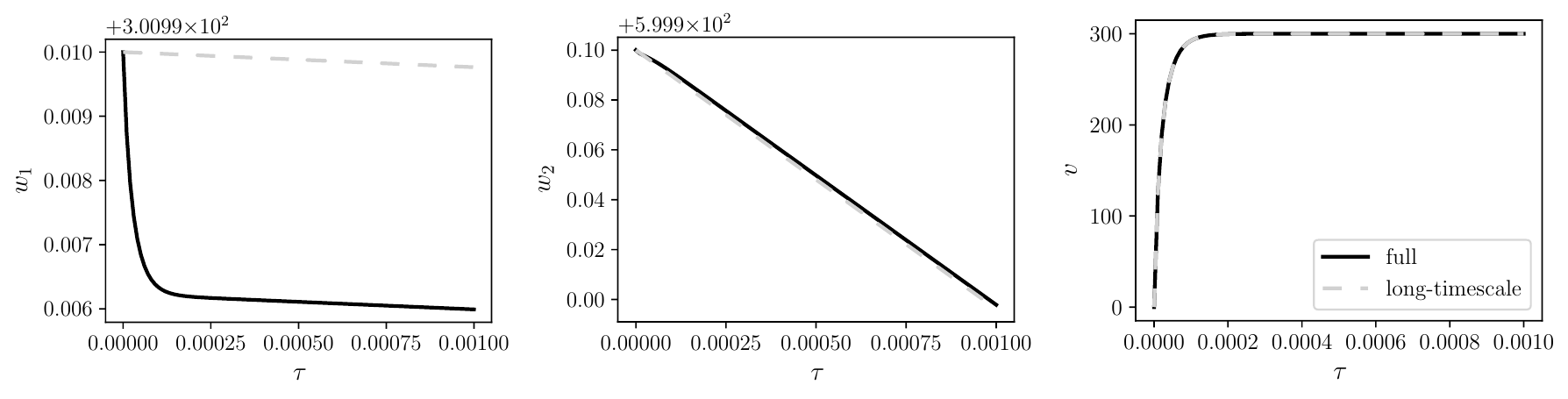}%
	}\\
	\subfloat[]{%
	\includegraphics[width=\columnwidth]{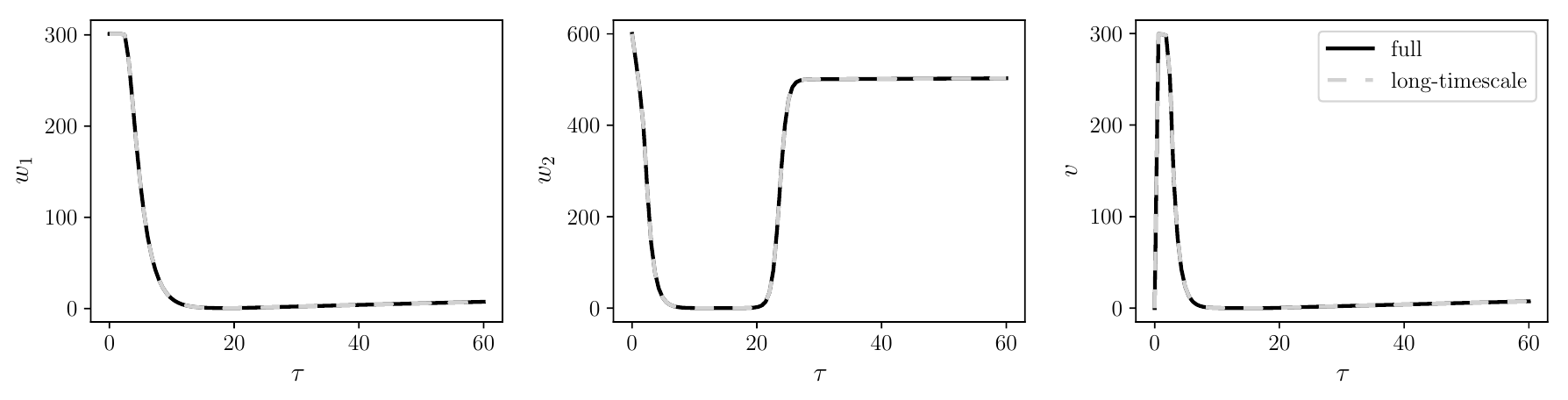}%
	\label{fig:num_long_qssa}
}
	\caption{Series of plots showing good agreement between the composite and full solutions of equations \eqref{eq:sys_sing} on (a) the short timescale ($0\leq \tau \leq \varepsilon/10$), where the inner solution dominates, and on (b) the long timescale ($\varepsilon/10 \leq \tau \leq 60$), where the outer solution dominates. The left, middle and right panel plots respectively show the solutions for model variables $w_1(\tau)$, $w_2(\tau)$ and $v(\tau)$ over time. The solutions are computed numerically as described in \cref{sec:num_sim}, for initial conditions $w_1(0)=301$, $w_2(0)=600$, $v(0)=0$, and default parameter values with $\varepsilon=10^{-2}$.
	}
	\label{fig:num}
\end{figure*}

To confirm the accuracy of the asymptotic reduction, we compare the solution of the full system \eqref{eq:sys_full} with the composite solution \cref{eq:matching}. We solve the full system by integrating equations \eqref{eq:sys_full} using the Python function \texttt{scipy.in\-te\-gra\-te.ode\-int}, a wrapper for the \texttt{lso\-da} solver from Fortran's library \texttt{ode\-pack}. This ODE solver automatically switches between stiff and nonstiff methods, noting the system is stiff when the dynamics are fast. To construct the composite solution, we use the same solver to generate numerical solutions for the fast and slow systems, whereby integration of the governing equations for both systems is performed with respect to the slow timescale $\tau$. We then combine numerical solutions using the matching condition \eqref{eq:matching}.

The results presented in \cref{fig:num} show that the full and composite solutions are in good agreement for small and large times, i.e. during the fast and slow dynamics. As anticipated, $\mathcal{O}(\varepsilon)$ discrepancies in the composite solution are observed on the short timescale. We conclude that the matched asymptotic approximation of the full solution is accurate within the expected error range, and note that the solutions exhibit a rapid relaxation onto the slow manifold dynamics. Based on these findings, we assert that the reduced model associated with the long timescale characterises the long-term dynamics of the full model, by which we motivate its analysis in Sections \ref{sec:phase_plane} and \ref{sec:bifurcation_analysis}.

Details on how the full and asymptotic long-timescale solutions, i.e. solutions of equations \cref{eq:sys_full} and \cref{eq:sys_slow2} respectively, compare with solutions of the QSSA model from \citet{Kuznetsov1994} are provided in \cref{app:qssa}.

\section{Phase-plane analysis of slow dynamics}\label{sec:phase_plane}

We now characterise the long-term behaviour of the composite solution (derived in \cref{sec:asymp}), by analysing the reduced outer model defined by \eqref{eq:sys_slow2} using approaches similar to those for the study of the \textit{Fitzhugh-Nagumo} (FN) \textit{equations}, which model excitability in neuronal dynamics \citep{Fitzhugh1961}. A system is excitable if a small perturbation from a stable steady state results in a large excursion before the system relaxes back to the same or a different steady state. Alternatively, the system may relax onto a limit cycle with large excursions. In this section, we show that our reduced system \eqref{eq:sys_slow2} exhibits excitable  tumour-immune dynamics.

\subsection{Nullclines and excitability}\label{sec:nullclines}

\begin{figure*}[p]
	\centering
	\subfloat[default parameters]{%
		\begin{minipage}[t]{\nullfigsize\textwidth}
			\includegraphics[width=\textwidth, valign=t]{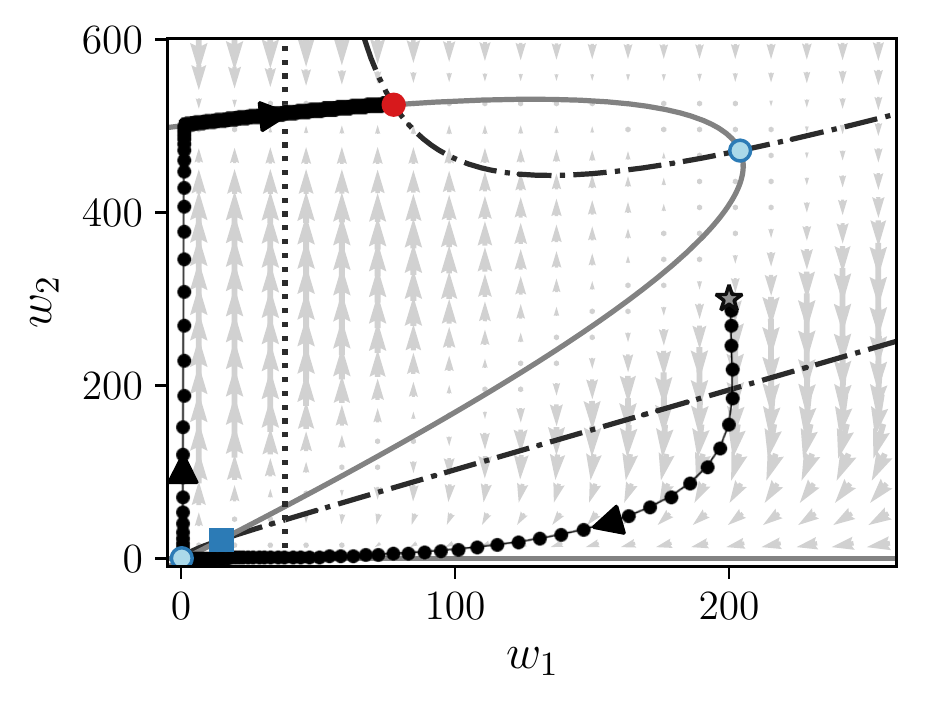}
			\includegraphics[width=\textwidth, valign=t]{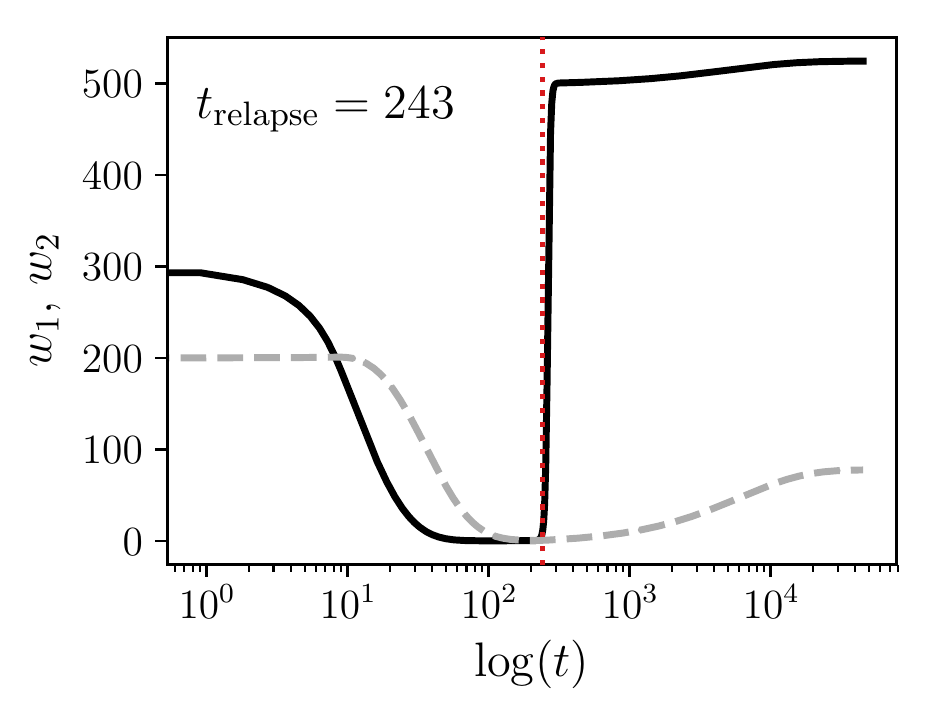}%
		\end{minipage}\label{fig:null_default}}%
	\subfloat[{$\alpha=0.5$}]{%
		\begin{minipage}[t]{\nullfigsize\textwidth}
			\includegraphics[width=\textwidth, valign=t]{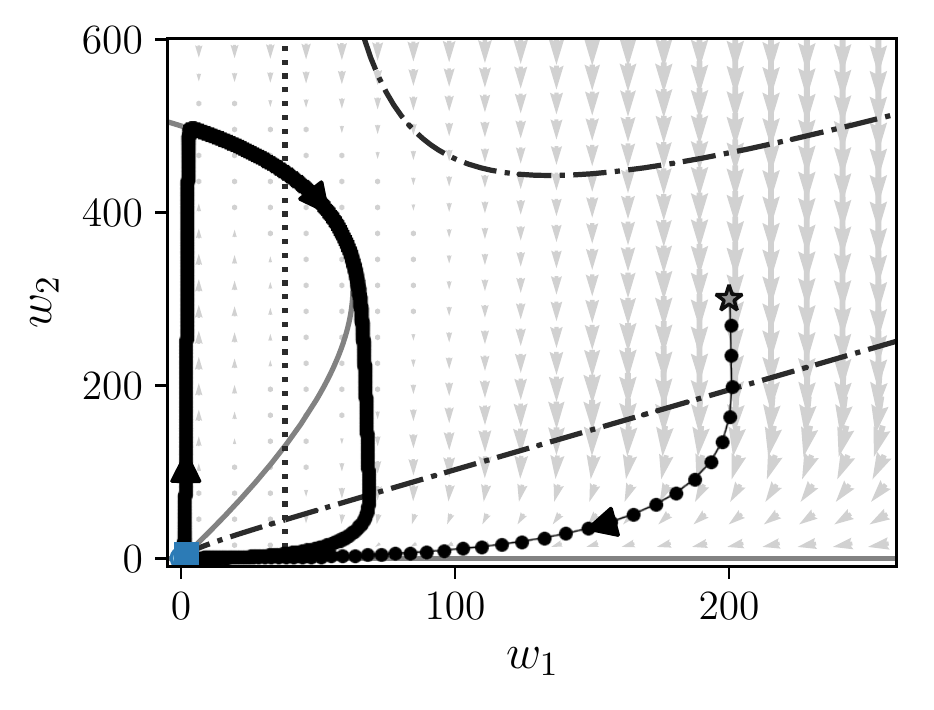}
			\includegraphics[width=\textwidth, valign=t]{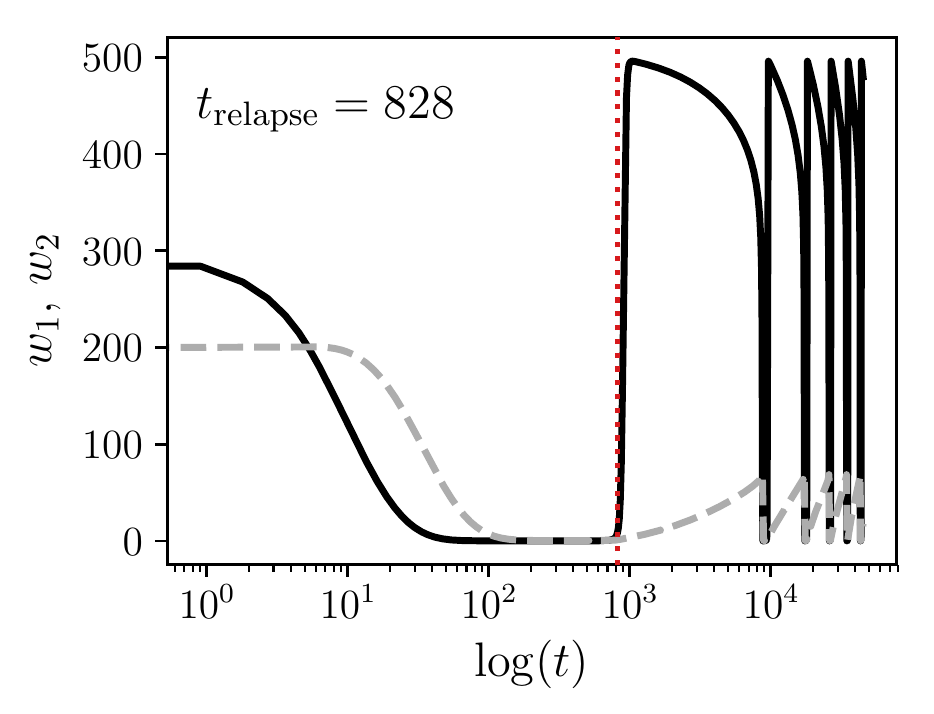}%
		\end{minipage}\label{fig:lim_cycle1}}%
	\subfloat[{$\alpha=2.0$}]{%
		\begin{minipage}[t]{\nullfigsize\textwidth}
			\includegraphics[width=\textwidth, valign=t]{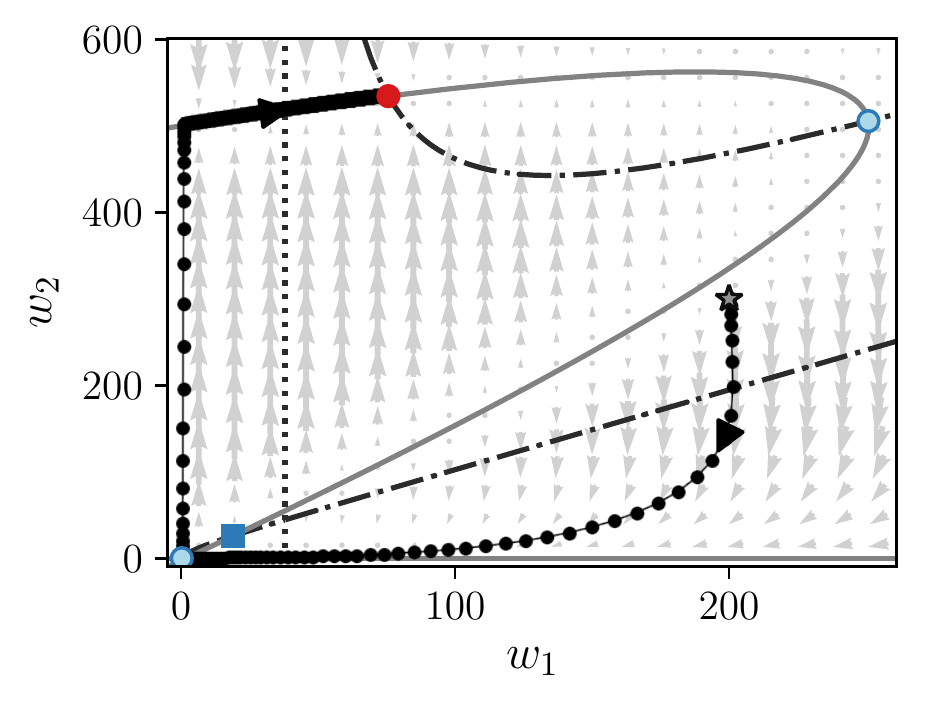}
			\includegraphics[width=\textwidth, valign=t]{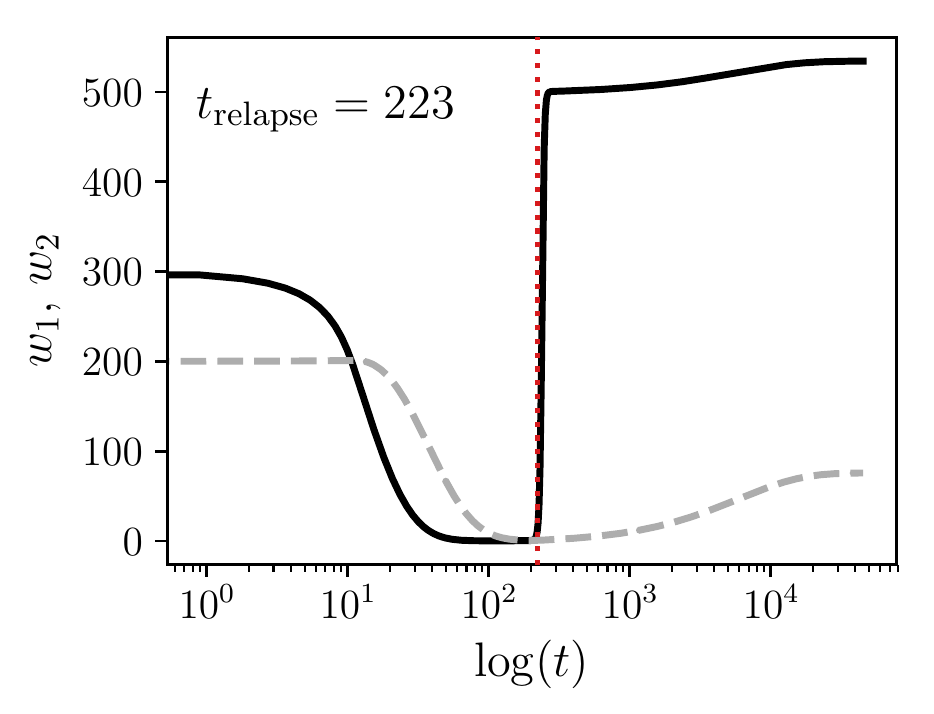}
		\end{minipage}\label{fig:null_large_alpha}}\\
	\subfloat[{$\delta=0.11$}]{%
		\begin{minipage}[t]{\nullfigsize\textwidth}
			\includegraphics[width=\textwidth, valign=t]{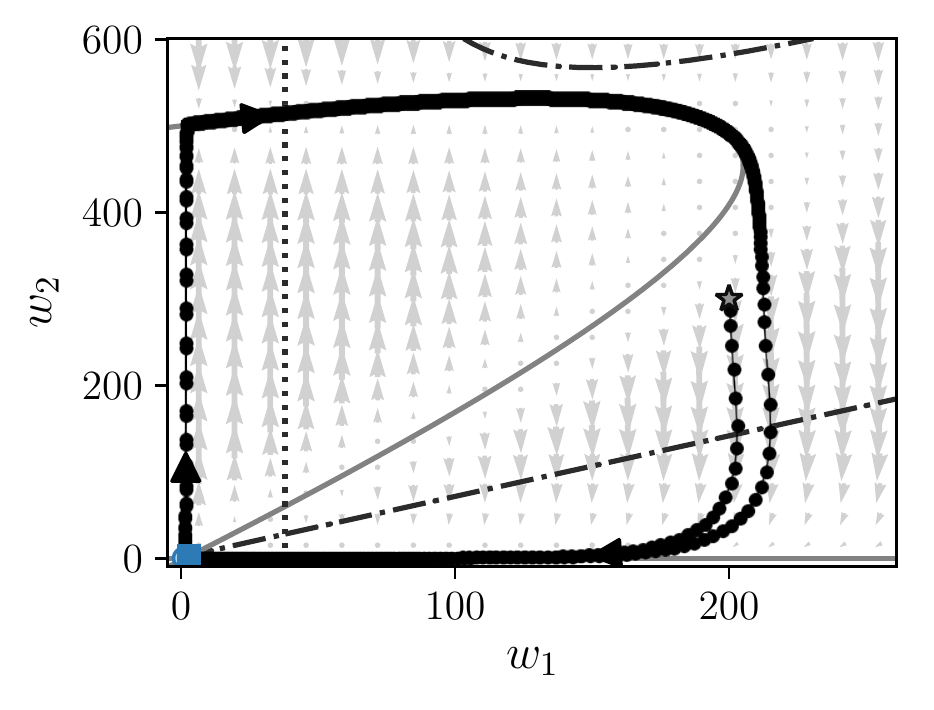}
			\includegraphics[width=\textwidth, valign=t]{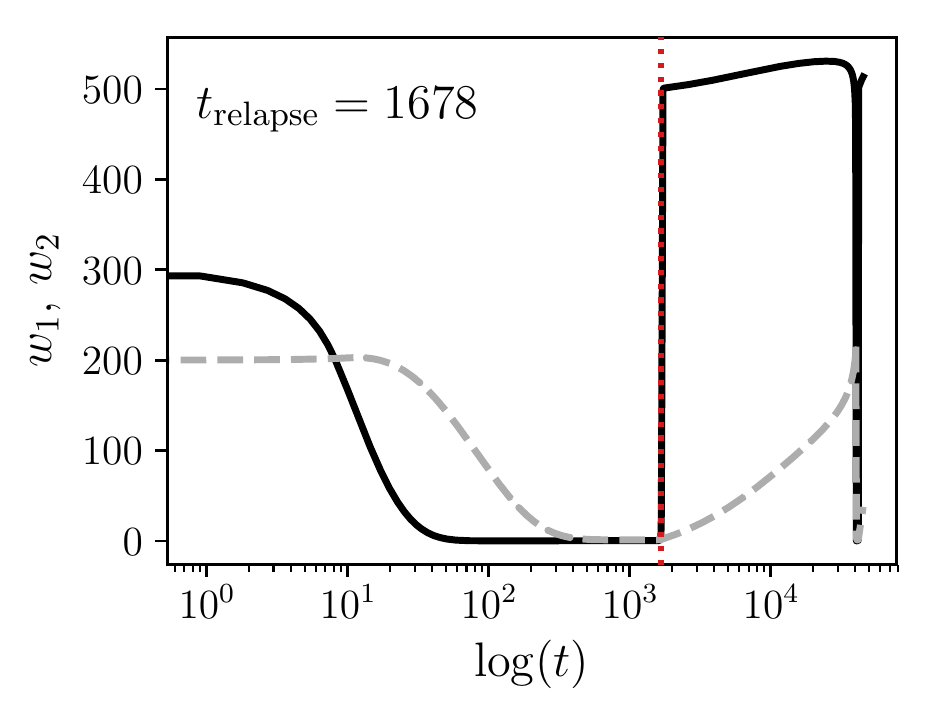}%
		\end{minipage}\label{fig:lim_cycle2}}%
	\subfloat[{$\delta=1.1$}]{%
		\begin{minipage}[t]{\nullfigsize\textwidth}
			\includegraphics[width=\textwidth, valign=t]{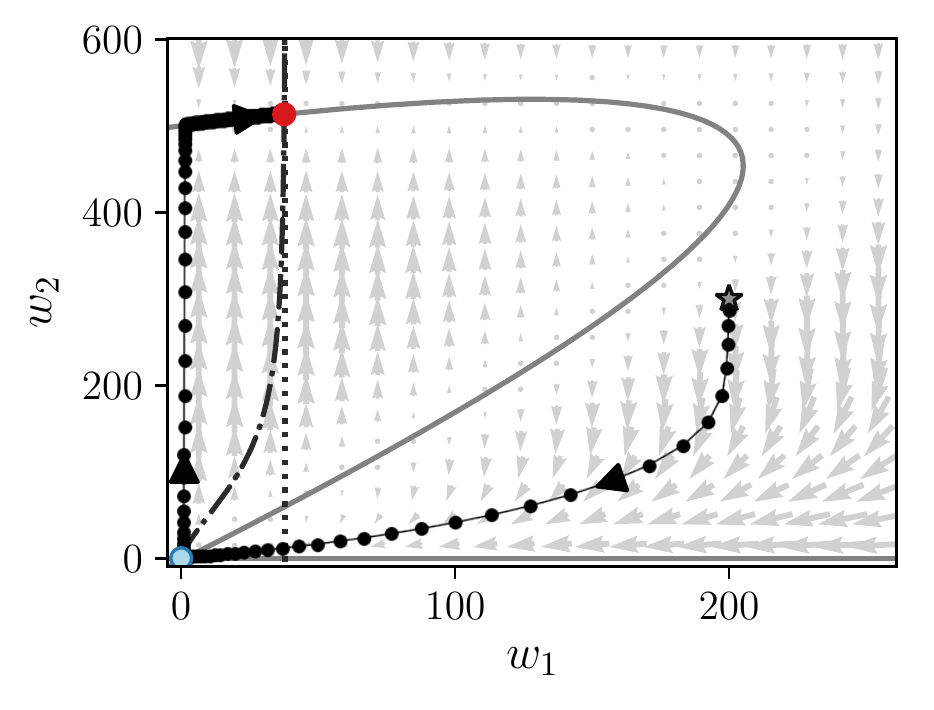}
			\includegraphics[width=\textwidth, valign=t]{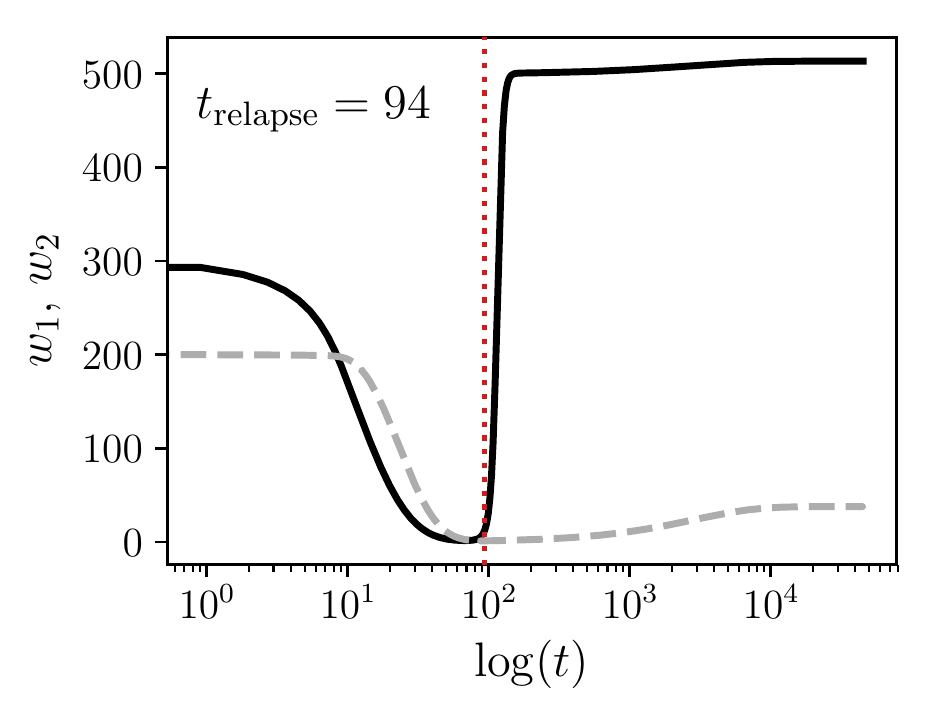}%
		\end{minipage}\label{fig:null_large_delta}}%
	\subfloat{%
		\begin{minipage}[t]{\nullfigsize\textwidth}
			\centering
			\hspace*{0.13\textwidth}
			\vspace*{-0.5\floatsep}
			\includegraphics[width=\nullegrelfigsize\textwidth, valign=t]{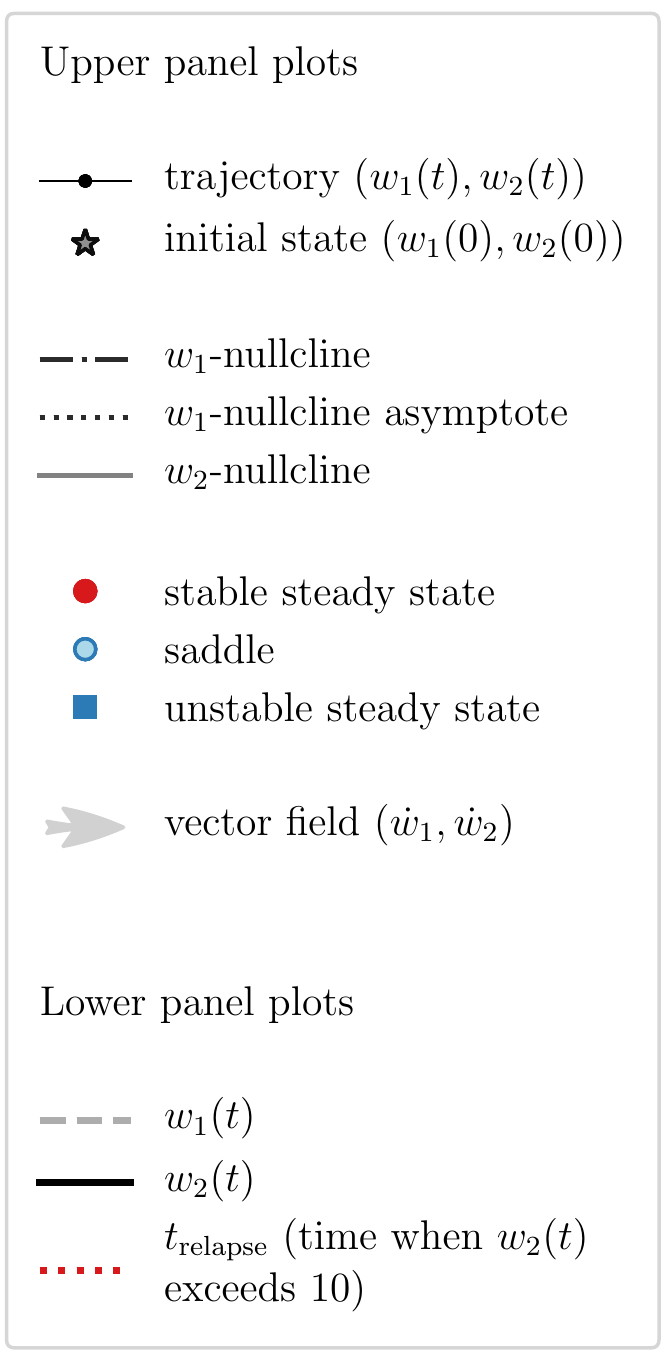}%
		\end{minipage}}
	\caption{Behaviour of the long-timescale model \eqref{eq:sys_slow2}, shown via $(w_1, w_2)$ phase portraits and dynamic plots of example solutions, is excitable, and remains excitable as individual parameters are varied. Plot (a) shows model behaviour for default parameter values (see \cref{tab:param_nondim}). In plots (b) and (c), tumour growth rate parameter $\alpha$ is varied, whereas effector death rate parameter $\delta$ is varied in plots (d) and (e), with other parameters fixed at their default values in all plots. $(w_1, w_2)$ phase portraits in upper panels are equipped with nullclines, steady states, flow field $(\dot{w_1}, \dot{w_2})$ and an example solution trajectory. The latter, starting at $w_1(0)=200$, $w_2(0)=300$, is obtained numerically using the same method as in \cref{sec:num_sim}. The time step between plotted trajectory points is uniform; points far apart indicate fast dynamics, whereas points close together show slow dynamics. In lower panels the solution trajectory is plotted against time for $0< t \leq  36331$. The red dotted vertical line denotes the time, $t_{\rm relapse}$, at which tumour cell density, $w_2$, exceeds 10 after the first shrinkage; $t_{\rm relapse}$ increases when $\delta$ or $\alpha$ are decreased. }
	\label{fig:phase_portraits_vary}
\end{figure*}

On the $w_1$- and $w_2$-nullclines of the slow system \eqref{eq:sys_slow2} we have $\dot{w_1}=0$ and $\dot{w_2}=0$ respectively. We find that $w_1$-nullclines are a subset of solutions to a quadratic in $w_2$, which has polynomials in $w_1$ as its coefficients. The $w_2$-nullclines are the line $w_2=0$ and a subset of solutions to another quadratic in $w_2$, with coefficients independent of $w_1$. For the sake of brevity, the equations of nullclines are given in \cref{app:nullclines}. 
For default parameter values in \cref{fig:null_default}, the nontrivial  $w_2$-nullcline has a shape of a rotated and skewed U that, together with the $w_2=0$ line, resembles a cubic; this is intersected in different ways by the $w_1$-nullcline. The described nullcline geometry is similar to that of the FN model, where a cubic nullcline of one model variable is intersected by a straight line nullcline of the other variable \citep{Fitzhugh1961}.

We investigate how robust nullclines and behaviour of the system \eqref{eq:sys_slow2} are to small changes in individual parameter values. The U shape of the non-trivial $w_2$-nullcline is always preserved under changes in tumour kinetic constants $\alpha$ (see \crefrange{fig:null_default}{fig:null_large_alpha}) or $\beta$. Under variation of parameters associated with the immune system and its response to the tumour, the shape of the $w_1$-nullcline ranges from approximately linear, as in the FN model, to curved (see \cref{fig:null_default,fig:lim_cycle2,fig:null_large_delta}); it intersects the $w_2$-nullcline a different number of times and in a number of ways, giving variable numbers of steady states with a range of stability behaviours. The shape of the $w_2$-nullcline is therefore robust to parameter changes, whereas that of the $w_1$-nullcline is not.

Across all the plots in \cref{fig:phase_portraits_vary}, solution trajectories exhibit excitable behaviour with large excursions in the phase plane and switching kinetics. A typical trajectory exhibits rapid vertical dynamics, during which the tumour quickly grows or shrinks, until reaching the $w_2$-nullcline. It then moves on a slow timescale approximately horizontally on this nullcline, with the number of effector cells either slowly increasing or decreasing. 
When the horizontal direction can no longer be sustained, the trajectory is pushed onto a different branch of the $w_2$-nullcline via fast vertical movement. The rapid switches between nullcline branches continue until a steady state or a stable limit cycle is reached. 
We notice that the variability of the $w_1$-nullcline shape under single parameter variations does not greatly impact the large excursions of trajectories in the phase plane, but it does impact the long-term behaviour as time progresses.

\subsection{Separation of timescales and its implications}

\begin{figure*}[t!]
	\begin{minipage}[t]{0.325\columnwidth}
		\subfloat[{$\zeta=0.1$}]{%
		\includegraphics[width=\columnwidth, valign=t]{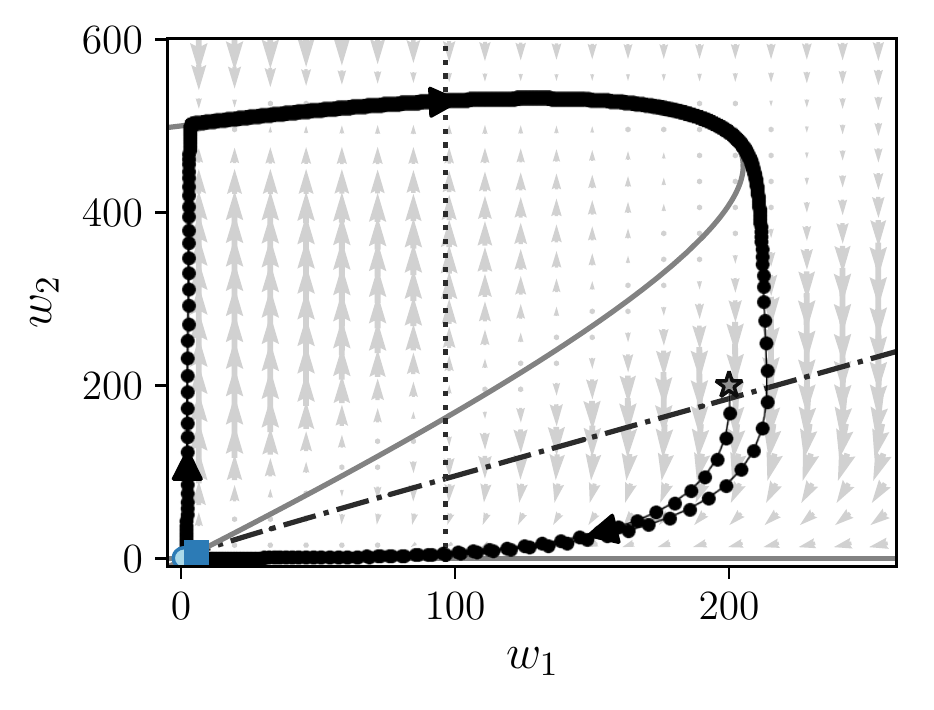}
		\label{fig:zeta_small}}
	\end{minipage}
	\begin{minipage}[t]{0.325\columnwidth}
		\subfloat[{$\zeta=1$}]{%
		\includegraphics[width=\columnwidth, valign=t]{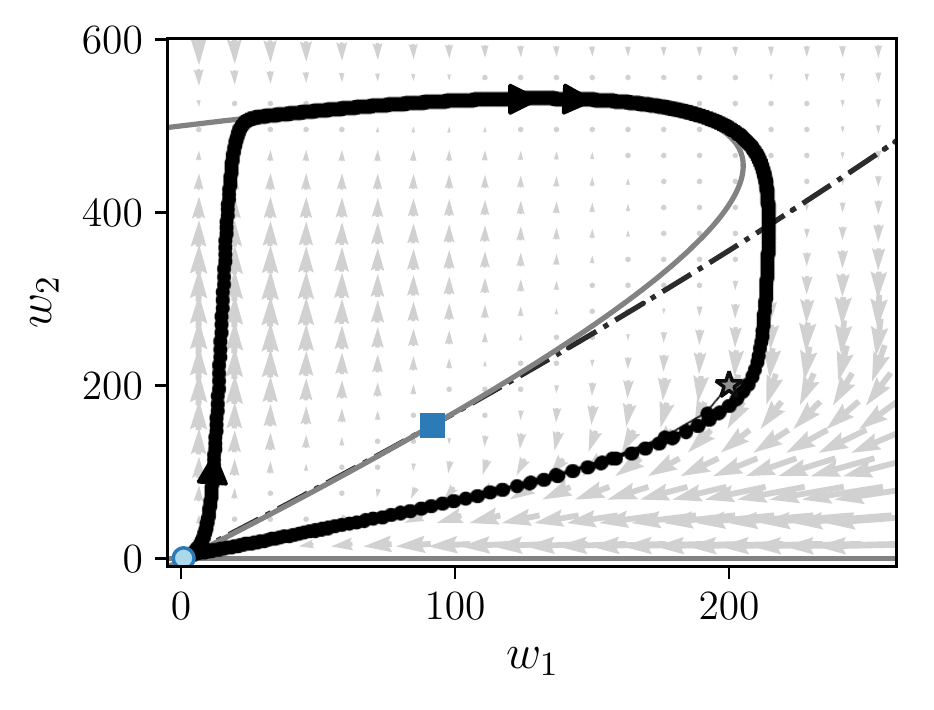}
		}
	\end{minipage}
	\begin{minipage}[t]{0.325\columnwidth}
		\subfloat[{{$\zeta=10$}}]{%
		\includegraphics[width=\columnwidth, valign=t]{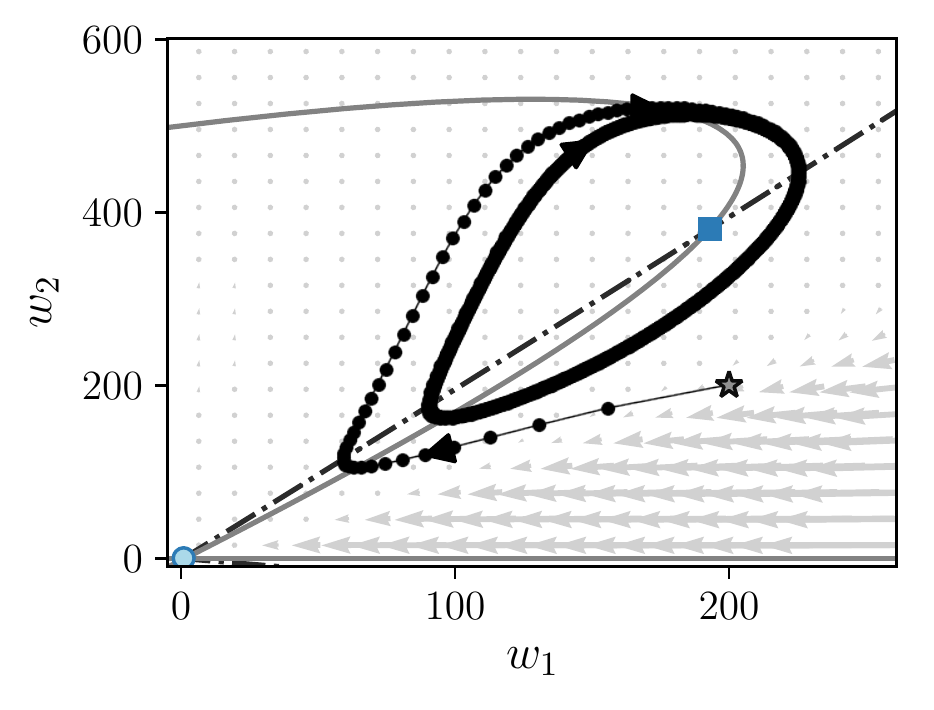}
		\label{fig:zeta_large}}
	\end{minipage}
	\caption{Dynamics of the long-timescale model \cref{eq:sys_slow3}, shown via $(w_1, w_2)$ phase portraits, evolve from excitable to non-excitable as $\zeta$ increases from $0.1$ to $10$; $\zeta$ is a dimensionless parameter that scales parameters associated with effector cells (see equation \cref{eq:assump_mu_beta}). Large excursions in the phase plane in plots (a) and (b) transform into regular oscillations in plot (c). The plots are generated using $\hat{\delta}=\hat{\sigma}=3$, while other parameters are kept at their default values. Trajectories start at $w_1(0)=w_2(0)=200$. See \cref{fig:phase_portraits_vary} for legend.}
	\label{fig:zeta}
\end{figure*}

In \cref{fig:phase_portraits_vary}, we observe that the transient fast kinetics arise due to the dominance of the vertical direction $(0, \dot{w_2})$ in the flow field, except in the vicinity of the $w_2$-nullcline where $\dot{w_1}$ is significant, as in the FN model. This indicates a separation of timescales in the model, whereby the $w_2$ variable has a shorter timescale than the timescale of the $w_1$ variable. As parameters are varied, the length of time trajectories spend moving on the $w_2$-nullcline also varies. The lower the pro-tumour/anti-immune parameters, such as the rate of effector cell death/migration out of the tumour region, $\delta$, or tumour growth rate, $\alpha$, the more time passes until tumour relapse. While this indicates changes in the difference of timescales as we vary the parameters, and a degree of sensitivity of excitable solution trajectories, small changes in individual parameters appear insufficient for the system to lose its excitable nature. 
By systematically manipulating the parameters controlling the kinetics of total effector cells, we will investigate the origin and robustness of timescale separation in model \eqref{eq:sys_slow2}, and the associated fast-slow dynamics, one of our aims being to identify parameter regimes in which excitable dynamics are less pronounced.

Inspection of the default parameter values in Table \ref{tab:param_nondim} reveals that we can represent their sizes by writing each of them as an $\mathcal{O}(1)$ parameter scaled with another appropriately sized parameter. We introduce two positive small parameters, $\zeta=\mathcal{O}(10^{-1})$ and $\xi=\mathcal{O}(10^{-3})$, which we treat independently in order to preserve the richness of the model when their values are close to zero. We relate them to parameters from slow manifold equations \eqref{eq:sys_slow2} via
\begin{gather}
	\sigma = \zeta \hat{\sigma}, \quad \delta=\zeta \hat{\delta}, \quad  \eta = \hat{\eta}/\zeta, \quad 
	\mu = \xi \hat{\mu}, \quad  \beta = \xi \hat{\beta},
	\label{eq:assump_mu_beta}
\end{gather}
noting that values of $\alpha$ and $\rho$ are both $\mathcal{O}(1)$. 
We then rewrite equations \eqref{eq:sys_slow2} as 
\begin{subequations}
	\begin{align}
		&\dot{w_1}=\zeta\left(\hat{\sigma} + \frac{\rho e m}{\hat{\eta} + \zeta m} - \hat{\delta} e - \frac{\xi}{\zeta}\hat{\mu} e m\right)=: \zeta g_1(w_1,w_2),\label{eq:sys_slow3_w1}\\
		&\dot{w_2} = \alpha m - \kappa_2 e m -\xi\alpha \hat{\beta} m^2=:g_2(w_1,w_2) ,\label{eq:sys_slow3_w2}
	\end{align}
	\label{eq:sys_slow3}%
\end{subequations}
where we drop the notation $e=e^\dagger(\boldsymbol{w})$ and $m=m^\dagger(\boldsymbol{w})$ from \eqref{eq:sys_slow2} for simplicity.

Equations \eqref{eq:sys_slow3} are singular, and in the limiting case $0<\xi<\zeta \ll 1$ we can apply asymptotic methods, as in \cref{sec:asymp}, to accurately further approximate the fast and slow dynamics. For reasons of brevity, we only sketch this analysis here. By taking the limit $\zeta, \xi \rightarrow 0$, we approximate the fast manifold equations for $\boldsymbol{w}_{\rm F}$ to leading order in $\zeta$ and $\xi$ via
\begin{subequations}
	\begin{align}
	\dot{w}_{1, {\rm F}} &= 0,\\
	\dot{w}_{2, {\rm F}} &= \alpha y^\dagger(\boldsymbol{w}_{\rm F}) - \kappa_2 x^\dagger(\boldsymbol{w}_{\rm F}) y^\dagger(\boldsymbol{w}_{\rm F}).\label{eq:zeta_fast}
	\end{align}%
	\label{eq:sys_slow3_fast}%
\end{subequations}
From equations \cref{eq:sys_slow3_fast}, we notice that the number of total effector cells, $w_{1, {\rm F}}$, is approximately constant on the fast manifold, while dynamics of total tumour cells, $w_{2, {\rm F}}$, are governed by exponential-like terms, which emerge from the difference between exponential growth of tumour cells and their loss due to immune-induced death of tumour cells in equation \cref{eq:sys_full_y} for $\dot{m}$. We deduce that on the fast timescale the total number of tumour cells diverges to infinity if $w_{1, {\rm F}}$ falls below the threshold value $w_{1}^\times:=\frac{\alpha}{\kappa_2}(1+\frac{\alpha\kappa_1 w_2}{1+\alpha})$, whereas the system is attracted to the tumour-free steady state if $w_{1, {\rm F}}>w_1^\times$, i.e. enough effector cells are recruited relative to the tumour size. 

Once a trajectory that follows the fast manifold reaches the neighbourhood of the $w_2$-nullcline, it starts moving on it, and we say it moves on the slow manifold of outer equations \eqref{eq:sys_slow2}. By rescaling time via $\tilde{\tau} = \tau/\zeta$ in equations \eqref{eq:sys_slow3}, we obtain the following slow manifold equations for $\boldsymbol{w}_{\rm S}$ with respect to $\tilde{\tau}$:
\begin{subequations}
	\begin{align}
	&\dot{w}_{1, {\rm S}}=g_1(w_{1, {\rm S}},w_{2, {\rm S}}),\\
	&g_2(w_{1, {\rm S}},w_{2, {\rm S}})=0, \label{eq:sys_slow3_slow_null}
	\end{align}
\end{subequations}
where equation \eqref{eq:sys_slow3_slow_null} constrains the trajectory to the $w_2$-nullcline, i.e. the dynamics of $w_{2, {\rm S}}$ are slave to those of $w_{1, {\rm S}}$. 
In contrast to the asymptotic solutions derived in \cref{sec:asymp}, the trajectories here do not remain on the slow manifold, but jump between the fast and slow manifolds. When moving on the slow manifold, i.e. on a branch of the $w_2$-nullclines, the number of total effector cells changes slowly, which eventually causes a change in the attracting steady state of the fast manifold equations. The time point, at which the threshold $w_1^\times$ is hit, is approximately when solution trajectories switch via fast dynamics to a different branch of the $w_2$-nullclines; see \cref{fig:phase_portraits_vary}. We conclude that the separation of timescales and the logistic growth assumption allow tumour cells to exploit small changes in the effector cell numbers to rapidly expand to carrying capacity in the parameter regime considered.

We show in \cref{fig:zeta} how the system dynamics may change from the excitable regime to a non-excitable one as we increase the value of $\zeta$. For smaller values of $\zeta$, transitions between the fast and slow manifolds are sharp, and the trajectory moves on the above derived manifolds (see \cref{fig:zeta_small}). For larger values of $\zeta$, the excitable dynamics are less pronounced; they resemble classic oscillations, with smaller excursions and periods of oscillation, and they do not follow the nullclines closely (see \cref{fig:zeta_large}). In this case, 
the timescale of effector dynamics approaches those of the tumour, rather than the tumour dominating and the immune system slowly catching up, as in the excitable regime.

In summary, our results show that the model's excitable behaviour can be attributed to a combination of the robust, cubic-like, shape of the $w_2$-nullcline, the separation of timescales, and the logistic growth assumption. Excitability is observed for $\zeta$ as large as $\mathcal{O}(10^{-1})$, and we have shown that a considerable inflation of the parameters associated with the function of the immune system and its response is needed to lose the excitable behaviour (see \cref{fig:phase_portraits_vary,fig:zeta}). We conclude that in the neighbourhood of the default parameter regime the system \eqref{eq:sys_slow2} is inherently prone to excitable behaviour, analogously to the FN model. 

\section{Bifurcation analysis of slow dynamics} \label{sec:bifurcation_analysis}

Having shown that the model can exhibit excitable dynamics, we now investigate how parameter changes impact the long-term behaviour. 
Our aim is to show that system \cref{eq:sys_slow2}, or equivalently \cref{eq:sys_slow_xy}, captures immunoediting behaviour and also a variety of other complex tumour-immune dynamics. 
We first identify and characterise the system's steady states via asymptotic and linear stability analysis. We then confirm and extend these results numerically to describe the system's global bifurcation structure for different values of model parameters influencing effector cell supply and decay, $\sigma$ and $\delta$. In the end we investigate the sensitivity of the bifurcation structure to changes in model parameters controlling tumour-effector cell interactions, $\mu$, $\rho$ and $\eta$, in order to gain insight into tumour responses to different immunotherapies, which may perturb one or more of these model parameters depending on their mechanism of action.

\subsection{Characterisation of steady state solutions}\label{sec:ss}

Identifying the steady state solutions of the slow system is most tractable when working with equations \eqref{eq:sys_slow_xy} for variables $e$ and $m$. Steady state solutions solve $\boldsymbol{q}(e,m)=0$ 
and are identical to steady state solutions of the full system \eqref{eq:sys_full}. We find that there is  \textit{one tumour-free steady state} $(e_1^*, 0)$, with 
\begin{align}
	e_1^*=\frac{\sigma}{\delta},
\end{align}
and \textit{at most three nonzero-tumour steady states} $(e_{2,3,4}^*, m_{2,3,4}^*)$ (physically realistic when both $e^*$ and $m^*$ are real and nonnegative) given by
\begin{align}
	e^*_{2, 3, 4} = \frac{\alpha}{\kappa_2}(1-\beta m) \qquad \text{and} \qquad m^*_{2, 3, 4} = m,
\end{align}
where $m$ is a root of the cubic
\begin{subequations}
\begin{align}
	p(m):=a_3 m^3 + a_2 m^2 + a_1 m + a_0\label{eq:ss_cubic}
\end{align}
with coefficients%
\begin{align}
	\begin{aligned}
		&a_0 = \eta\left(\tfrac{\sigma \kappa_2}{\alpha} - \delta\right),\\
		&a_2 =-\mu + (\mu \eta + \delta - \rho)\beta,
	\end{aligned}
	\quad
	\!
	\begin{aligned}
		&a_1 = \tfrac{\sigma \kappa_2}{\alpha} + \rho - \mu \eta - \delta + \delta \eta \beta,\\
		&a_3 = \mu \beta.%
	\end{aligned}\label{eq:ss}%
\end{align}%
\end{subequations}%
\nolinebreak
\subsubsection{Asymptotic approximation of steady states}\label{sec:asymp_ss}
\begin{figure*}[t!]
	%
	%
	\includegraphics{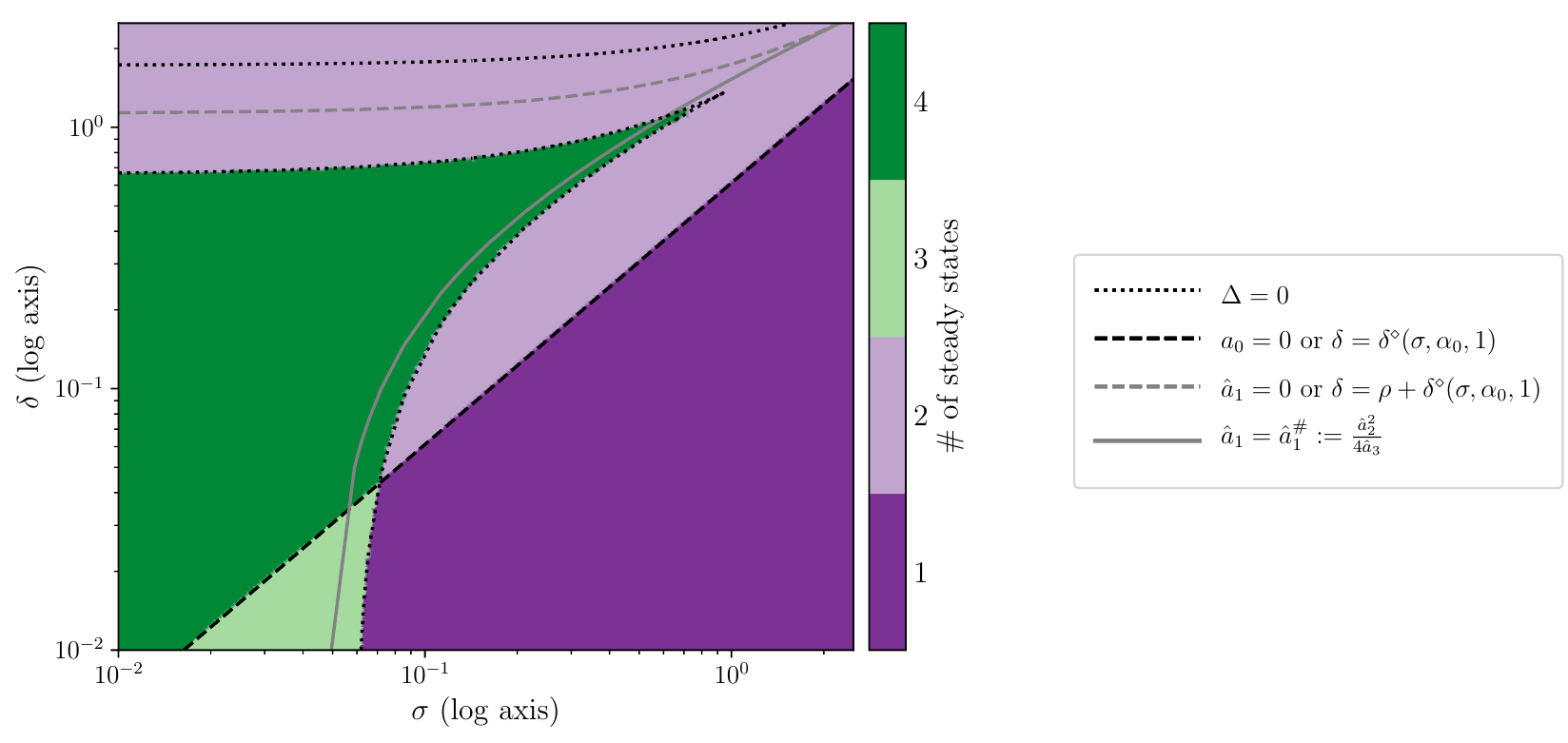}
	\caption{The number of physically realistic steady states in system \eqref{eq:sys_slow_xy} changes as model parameters vary. Accordingly, $(\sigma, \delta)$ parameter space is decomposed into distinct regions which are colour-coded in line with the number of physically realistic steady states that exist. The plot is generated by numerically computing the number of physically realistic steady states of system \cref{eq:sys_slow_xy} at different values of $\sigma$ and $\delta$ that lie on a triangular mesh of the $(\sigma, \delta)$ plane, with other parameters fixed at their default values (see \cref{tab:param_nondim}). We also plot analytically obtained region boundaries (see \cref{sec:ss_phys} and \cref{app:ss}). The black curves, $\Delta=0$ and $\hat{a}_0=0$, trace region borders well, while the grey curves, $\hat{a}_1=0$ and $\hat{a}_1=\hat{a}_1^\#$, capture them only qualitatively. 
	}
	\label{fig:ss_borders}
\end{figure*}

Recalling that $\mu$ and $\beta$ are small for the default parameter values (see \cref{tab:param_nondim}), we introduce the artificial small parameter $\xi=\mathcal{O}(10^{-3})$ as before, and write $\mu=\xi\hat{\mu}$ and $\beta=\xi\hat{\beta}$. This allows us to 
asymptotically approximate the roots $m$ of the cubic \cref{eq:ss_cubic}, 
and obtain estimates for the nonzero-tumour steady states $(e_{2,3,4}^*, m_{2,3,4}^*)$. We detail this analysis in \cref{app:ss}, and summarise the key results here.  

When $0<\xi\ll 1$, 
there are at most three types of steady state solutions: 
\begin{enumerate}[label=(\roman*)]
	\item the \textit{tumour-free steady state} $(e_1^*, 0)=(\frac{\sigma}{\delta}, 0)$,\label{enum:ss1}
	\item at most one \textit{intermediate-sized-tumour steady state} $(e_2^*, m_2^*)$, where 
		\begin{align}
			e_2^*\sim\frac{\alpha}{\kappa_2} \quad \text{and} \quad m_2^* \sim-\frac{a_0}{\hat{a}_1}
			\label{eq:ss2}
		\end{align}
		are both $\mathcal{O}(1)$, and\label{enum:ss2}
	\item at most two \textit{large-tumour steady states} $(e_{3, 4}^*, m_{3, 4}^*)$, where 
		\begin{subequations}
			\begin{gather}
				e_{3, 4}^*\sim\frac{\alpha}{\kappa_2}(1-\hat{\beta}\hat{m}_{3, 4}) \quad \text{and} \quad 
				m_{3, 4}^*\sim\frac{1}{\xi}\hat{m}_{3, 4}, \\
				\text{with}\qquad		
				\hat{m}_{3, 4}=\frac{-\hat{a}_2 \pm \sqrt{\hat{a}_2^2 - 4\hat{a}_3 \hat{a}_1}}{2\hat{a}_3},
			\end{gather}\label{eq:ss34}%
		\end{subequations}
		are $\mathcal{O}(1)$ and $\mathcal{O}(\xi^{-1})$ respectively.\label{enum:ss34}
\end{enumerate}
The coefficients used above are all $\mathcal{O}(1)$, and given as follows
\begin{align}
	\begin{aligned}
	\hat{a}_1&=\frac{\sigma\kappa_2}{\alpha} + \rho - \delta,
	\\
	\hat{a}_2&=-\hat{\mu} + \hat{\beta}(\delta - \rho),
	\\
	\hat{a}_3&=\hat{\mu}\hat{\beta}.
	\end{aligned}
\end{align}

In regions of parameter space where these steady states are physically realistic and stable,  we may identify them with the elimination, equilibrium/dor\-mancy and escape phases that are associated with the three Es of immunoediting. To find these regions of the parameter space, we study in \cref{sec:ss_phys} where the different steady states are physically realistic, whereas we examine their stability analytically in \cref{sec:lin_stab}, and numerically in \cref{sec:bifurcation_structure}.

\subsubsection{Physicality of steady states}\label{sec:ss_phys}

Assuming that all model parameters are positive, we observe that the tumour-free steady state $(\frac{\sigma}{\delta}, 0)$ exists for all positive values of $\sigma$ and $\delta$. From \cref{eq:ss2} we deduce that the intermediate-sized-tumour steady state is nonnegative if
\begin{gather}
\delta^{\diamond}(\sigma, \alpha, \kappa_2)\leq\delta<\rho + \delta^{\diamond}(\sigma, \alpha, \kappa_2), \label{eq:cond_ss2_real}\\
\text{where} \qquad \delta^{\diamond}(\sigma, \alpha, \kappa_2)  :=\frac{\sigma\kappa_2}{\delta},
\end{gather}
or equivalently $\frac{a_0}{\eta} \leq 0 < \hat{a}_1$. We use \cref{eq:ss34} to derive similar conditions for the large-tumour steady states; details are provided in \cref{app:ss}. In short, provided $\hat{m}_{3,4} \leq \beta^{-1}$, there is a unique large-tumour steady state solution $(e^*_3, m^*_3)$ when $\hat{a}_1<0$, two such solutions when $0<\hat{a}_1<\hat{a}_2^2/4\hat{a}_3=:\hat{a}^\#$ and $\hat{a}_2<0$, and none otherwise.

In \cref{fig:ss_borders} we plot the borders of the asymptotic physically realistic regions in the $(\sigma,\delta)$ plane, and compare them to a colour plot of the $(\sigma, \delta)$ plane, which shows how the number of physically realistic steady states varies in this parameter space.
The line $a_0=0$ accurately captures the lower border of the region, in which the intermediate-sized-tumour steady state is physically realistic, while the curve $\hat{a}_1=0$ only qualitatively captures the upper border of this region. 
The border $\hat{a}_1=\hat{a}_1^\#$, where the asymptotic large-tumour steady states $(e_3^*, m_3^*)$ and $(e_4^*, m_4^*)$ collide and become complex conjugates, shows similar inaccuracies (see \cref{fig:ss_borders}).
These discrepancies occur because our asymptotic approximations of the nonzero-tumour steady states, given in \cref{sec:asymp_ss}, lose accuracy in the neighbourhood of parameter regimes where the roots of the cubic \cref{eq:ss_cubic} collide, e.g. as the parameters approach the curve $\hat{a}_1=0$. 

To avoid these difficulties, we study the multiplicity of roots of the original cubic \eqref{eq:ss_cubic}. We identify parameter sets for which multiplicity of the cubic roots changes by considering the discriminant,
\begin{align}
	\Delta := a_2^2 a_1^2 - 4 a_3 a_1^3 - 4 a_2^3 a_0 - 27 a_3^2 a_0^2 + 18 a_3 a_2 a_1 a_0,
\end{align}
of the cubic \eqref{eq:ss_cubic}. 
Using Mathematica we numerically compute curves in $(\sigma, \delta)$ parameter space on which $\Delta=0$. These curves separate regions which have different numbers of physically realistic steady state solutions 
(see \cref{fig:ss_borders}). At the lower boundary of the dark green region, roots $m_3^*$ and $m^*_4$ collide and become complex, while this occurs for roots $m_2^*$ and $m^*_4$ at the upper boundary of the dark green region, as mentioned previously. On the black dotted curve in the light purple region, the complex roots $m_2^*$ and $m^*_4$ collide again and become real, but since they are real and negative (unrealistic), the steady state multiplicity does not change there. Accurately computing curves on which steady states coalesce is important, since they indicate parameter regimes where saddle-node-type bifurcations occur, i.e. when there is a collision of two physically realistic steady states, one stable, and the other a saddle; we demonstrate this in \cref{sec:bifurcation_structure,sec:bifurcation_structure_param_sens}.

\subsubsection{Linear stability of steady states} \label{sec:lin_stab}

To demonstrate the system exhibits immunoediting behaviour via the three types of steady states identified in \cref{sec:asymp_ss}, our goal in this section is to identify parameter regions in which each steady state is both stable and physically realistic.

We determine steady state stability via linear stability analysis, complemented with numerics. As in \cref{sec:asymp_ss}, we assume throughout that parameters $\mu$ and $\beta$ scale with the small parameter $0<\xi\ll 1$. We introduce a second small parameter $0< \nu \ll \xi$, and seek solutions of the form
\begin{subequations}
	\begin{align}
			e(\tau) 
			&
			= e^* + \nu e_1(\tau) + \mathcal{O}(\nu^2),
			\\
		\label{eq:lin_x}
			m(\tau) 
			&
			= m^* + \nu m_1(\tau) + \mathcal{O}(\nu^2),
	\end{align}%
	\label{eq:lin_exp}%
\end{subequations}
where $(e^*, m^*)$ are steady states.
By substituting \eqref{eq:lin_exp} into equations \eqref{eq:sys_slow_xy}, 
equating terms of $\mathcal{O}(\nu)$ to zero and neglecting terms of $\mathcal{O}(\xi)$ or higher orders of $\nu$, we obtain a linearised system $\dot{\boldsymbol{x}}=\mathcal{J}(e^*,m^*)\boldsymbol{x}$ for $\boldsymbol{x}=(e_1,m_1)$, with 
\begin{align}
\begin{split}
	&
	\mathcal{J}(e^*, m^*)=
	\\&\phantom{=}
	{\underbrace{\begin{bmatrix}
			1+\kappa_1 m^* & \kappa_1 e^* \\ \kappa_2 m^* & 1+ \kappa_2 e^*
			\end{bmatrix}}_{=:\mathcal{B}(e^*,m^*)}}^{-1} \underbrace{\begin{bmatrix}
		\frac{\rho m^*}{\eta + m^*} - \delta & \frac{\rho \eta e^*}{(\eta+ m^*)^2}\\
		-\kappa_2 m^* & \alpha - \kappa_2 e^*
		\end{bmatrix}}_{=:\mathcal{A}(e^*,m^*)} \label{eq:jac_intermediate}
\end{split}
\end{align}
as the Jacobian, and $(e^*, m^*)$ as the steady state asymptotic approximation to leading order in $\xi$, such as given in \cref{sec:asymp_ss}. 
The eigenvalues $\lambda_1$ and $\lambda_2$ of $\mathcal{J}(e^*, m^*)$ predict exponentially decaying behaviour of the linearised system, and therefore linear stability of the steady state $(e^*,m^*)$ whenever the condition ${\max}_{i=1,2}\,{\rm Re}(\lambda_i)<0$ is satisfied. 

For the tumour-free steady state $(e_1^*, m_1^*)=(\frac{\sigma}{\delta},0)$
\begin{align}
\lambda_1 = -\delta \qquad \text{and}\qquad \lambda_2 = \alpha - \frac{\sigma\kappa_2}{\delta},
\end{align}
so this steady state is linearly stable when the parameters satisfy
\begin{align}
	\delta < \frac{\sigma\kappa_2}{\alpha}=\delta^\diamond(\sigma, \alpha, \kappa_2),
	\label{eq:stab_cond_zero}
\end{align}
This means that the ratio of the basal effector influx rate, $\sigma$, and effector death rate, $\delta$, must be larger than the net tumour growth rate, $\alpha$, for the tumour-fr{}ee steady state to be stable, i.e. for tumour eradication.

Considering \cref{eq:cond_ss2_real} and \cref{eq:stab_cond_zero}, we observe that the region in which the intermediate-sized-tumour steady state $(e^*_2, m^*_2)$ is physically realistic is separated from the region in which the tumour-free steady state is stable by the line $\delta=\delta^\diamond(\sigma, \alpha, \kappa_2)$. 
This implies that the tumour-free and the intermediate-sized-tumour steady states may undergo a transcritical bifurcation when $\delta=\delta^\diamond(\sigma, \alpha, \kappa_2)$, provided the intermediate-sized-tumour steady state is stable once it has emerged as physically realistic.

When characterising the local stability of the intermediate-sized-tumour steady state, we consider the trace and the determinant of the Jacobian evaluated at this steady state. We note that
\begin{align}
	\det(\mathcal{J}(e^*, m^*)) = \frac{\det(\mathcal{A}(e^*, m^*))}{\det(\mathcal{B}(e^*, m^*))}, 
\end{align}
and
\begin{align}
	{\rm tr}(\mathcal{J}(e^*, m^*)) = \frac{\tr(\tilde{\mathcal{B}}(e^*, m^*) \mathcal{A}(e^*, m^*))}{\det(\mathcal{B}(e^*, m^*))},
\end{align}
where 
\hfill 
$\tilde{\mathcal{B}}(e^*, m^*) = \det(\mathcal{B}(e^*, m^*)) (\mathcal{B}(e^*, m^*))^{-1}$, 
\hfill
 and 
\\
$\det(\mathcal{B}(e^*, \allowbreak m^*)) = 1 + \kappa_2 e^* + \kappa_1 m^* >0$ for any nonnegative steady state. Therefore, the signs of $\det(\mathcal{J}(e^*, m^*))$ and ${\rm tr}(\mathcal{J}(e^*, m^*))$ are determined by the signs of $\det(\mathcal{A}(e^*, m^*))$ and ${\rm tr}(\tilde{\mathcal{B}}(e^*, m^*) \mathcal{A}(e^*, m^*))$ respectively. When $(e^*, m^*)=(e^*_2, m^*_2)$ from \cref{eq:ss2}, we have
\begin{align}
	\det(\mathcal{A}(e^*_2, m^*_2))
	 =\frac{\alpha}{\eta \rho}\left(\frac{\sigma\kappa_2}{\alpha}-\delta + \rho\right)^2 m^*_2
 	\label{eq:det2}
\end{align}
which is positive when $m^*_2$ is physically realistic. Stability of $(e^*_2, m^*_2)$ thus reduces to satisfying the trace condition
\begin{align}
	\begin{split}
	&\tr(\tilde{\mathcal{B}}(e^*_2, m^*_2)\mathcal{A}(e^*_2, m^*_2)) = 
		\\&\phantom{=}
	-\frac{\alpha \eta \left(\frac{\sigma\kappa_2}{\alpha} - \delta\right)}{\hat{a}_1}\left(\kappa_1 -\frac{1}{\eta\rho}\hat{a}_1^2\right)
	-(\alpha +1)\frac{ \sigma\kappa_2  }{\alpha } 
	<0.
	\end{split}
	\label{eq:trace_int1}%
\end{align}
We conclude that in regions, where $\delta > \delta^\diamond(\sigma, \alpha,\kappa_2)$ and inequality \eqref{eq:trace_int1} are simultaneously satisfied, the intermediate-sized-tumour steady state is stable. 

When $\delta=\delta^\diamond(\sigma, \alpha, \kappa_2)$, 
\begin{align}
	\tr(\tilde{\mathcal{B}}(e^*_2, m^*_2)\mathcal{A}(e^*_2, m^*_2))=-(\alpha + 1)\frac{\sigma\kappa_2}{\alpha}<0,
\end{align}
so our analysis predicts that the tumour-free and the inter\-me\-diate-sized-tumour steady states undergo a transcritical bifurcation on this curve. Moreover, when $\tr(\tilde{\mathcal{B}}(e^*_2, m^*_2)\allowbreak\mathcal{A}(e^*_2, m^*_2))=0$, the Jacobian eigenvalues are purely imaginary, and the intermediate-sized-tumour steady state is predicted to undergo a Hopf bifurcation. We numerically validate these predictions in \cref{sec:bifurcation_structure}, and note that further work is required to demonstrate existence of the Hopf bifurcation.

\begin{figure*}[b!]
	\centering
	\vspace{-\floatsep}
	\subfloat[number of free effector cells]{%
		\includegraphics[width=0.42\columnwidth]{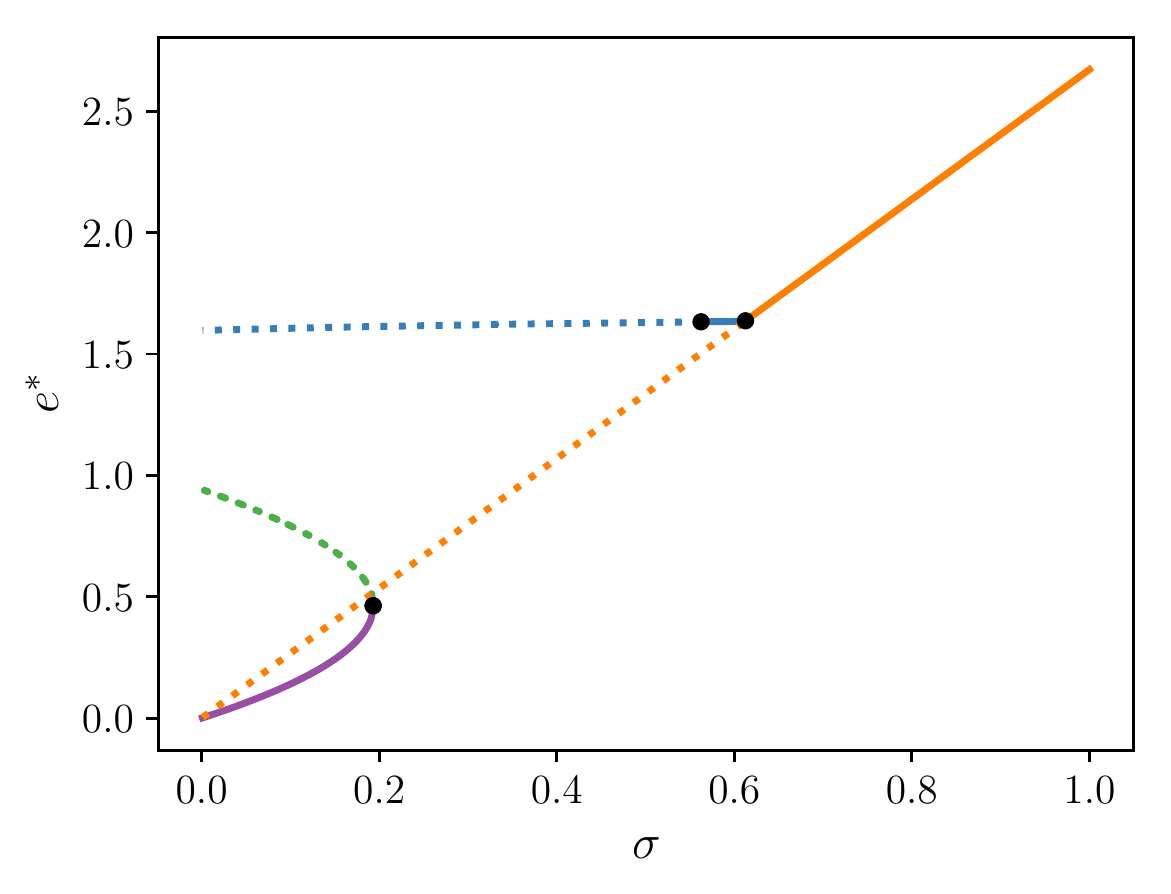}%
	}
	\subfloat[number of free tumour cells]{%
		\includegraphics[width=0.42\columnwidth]{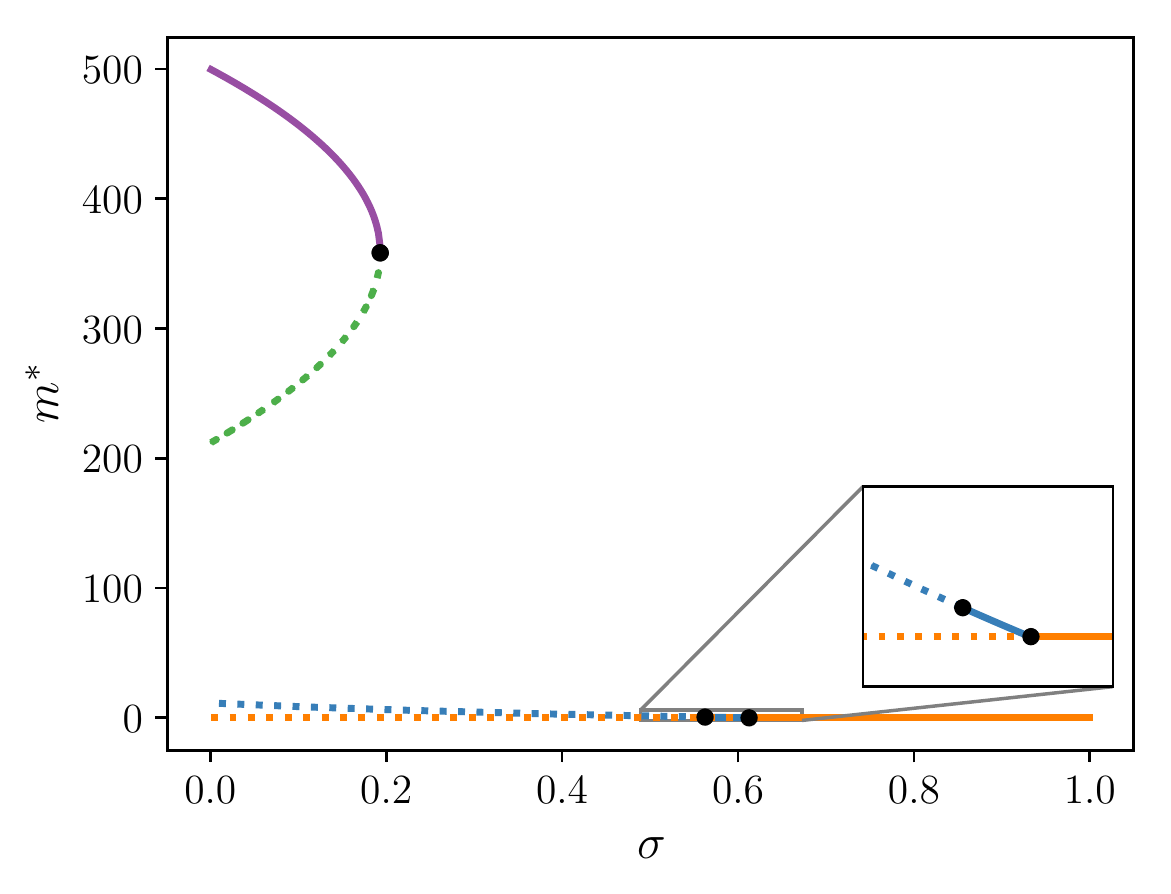}%
		\label{fig:bifurcation_diagram_default_y}%
	}
	\caption{The long-timescale system \eqref{eq:sys_slow_xy} (or equivalently \eqref{eq:sys_slow2}) exhibits the three Es of immunoediting through stable steady states of varying tumour cell densities. We can see this in the bifurcation diagrams, which are obtained numerically by sweeping through the $\sigma$ parameter space, and computing the system steady states and their stability at discrete values of $\sigma$ between 0 and 1. Other parameters are fixed at their default values (see \cref{tab:param_nondim}). Plots (a) and (b) respectively show the steady state numbers of free effector and tumour cells. Purple and green curves denote large-tumour steady states, blue curves denote intermediate-sized-tumour steady states, and orange curves denote tumour-free steady states. Local stability of steady state is distinguished by linestyle, whereby solid curves denote stable solutions, while dashed curves denote unstable ones. Black points mark bifurcations of steady state solutions, i.e. where solution branches change stability or collide.}
	\label{fig:bifurcation_diagram_default}
\end{figure*}
\begin{figure*}[b!]
	\includegraphics[]{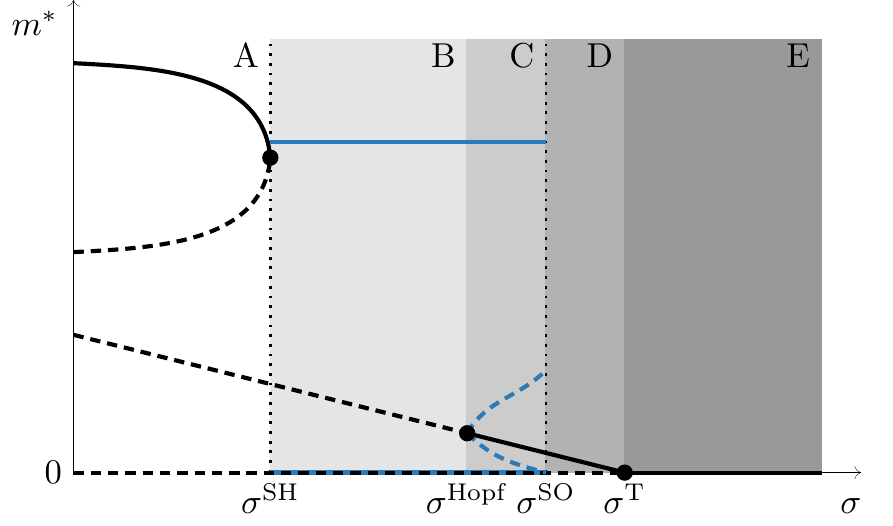}
	\caption{
		Schematic (not to scale) of the bifurcation diagram for scaled tumour cell numbers in \cref{fig:bifurcation_diagram_default_y}, indicating how periodic solutions arise in the long-timescale system \eqref{eq:sys_slow2} with respect to its steady states, as the parameter $\sigma$ is varied. Black curves indicate tumour cell numbers of steady state solutions, whereas blue curves indicate maximum and minimum tumour cell numbers of periodic solutions. Local stability of solutions is distinguished by linestyle, whereby solid curves denote stable solutions, while dashed curves denote unstable ones. As $\sigma$ is increased, the system undergoes the following bifurcations: SH = saddle-node homoclinic bifurcation, Hopf = subcritical Hopf bifurcation, SO = saddle-node of orbits, T = transcritical bifurcation. Bifurcations of steady state solutions are denoted with black dots, while bifurcations of periodic solutions are denoted with dotted vertical lines. These bifurcations split the parameter space into regions of qualitatively different model behaviours with respect to its steady states and orbits; the different regions are shaded with grey of varying intensities.
	}
	\label{fig:bifurcation_diagram_schematic}
\end{figure*}

\renewcommand{\thesubfigure}{\Alph{subfigure}}
\begin{figure*}[t!]
		\centering
		\hspace*{-\floatsep}
		\subfloat[$0<\sigma<\sigma^{\rm SH}$]{
			\includegraphics[width=0.33\columnwidth, valign=b]{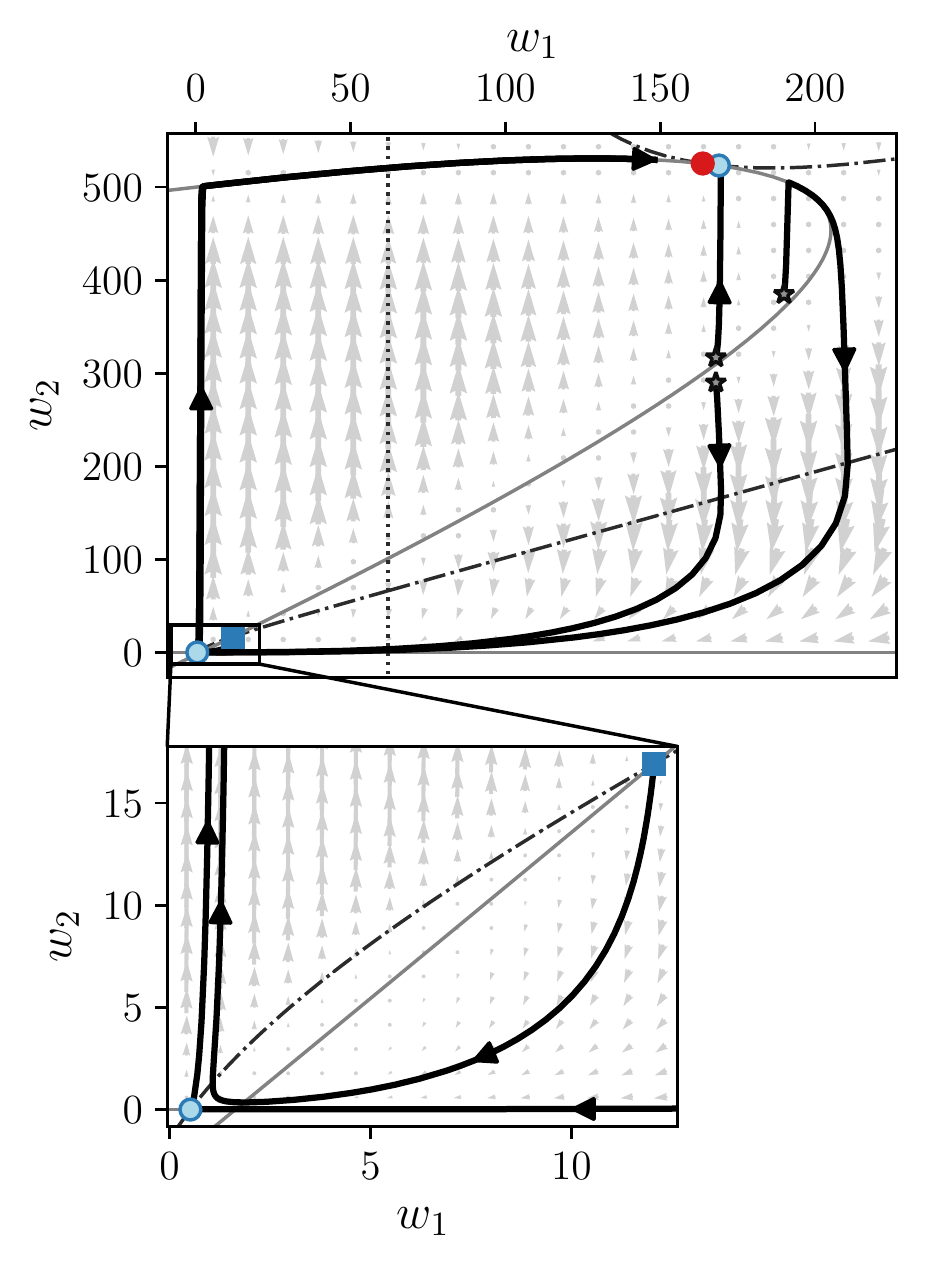}
			\label{fig:pp1}
		}
		\hspace*{-\floatsep}
		\subfloat[$\sigma^{\rm SH}<\sigma<\sigma^{\rm Hopf}$]{
			\includegraphics[width=0.33\columnwidth, valign=b]{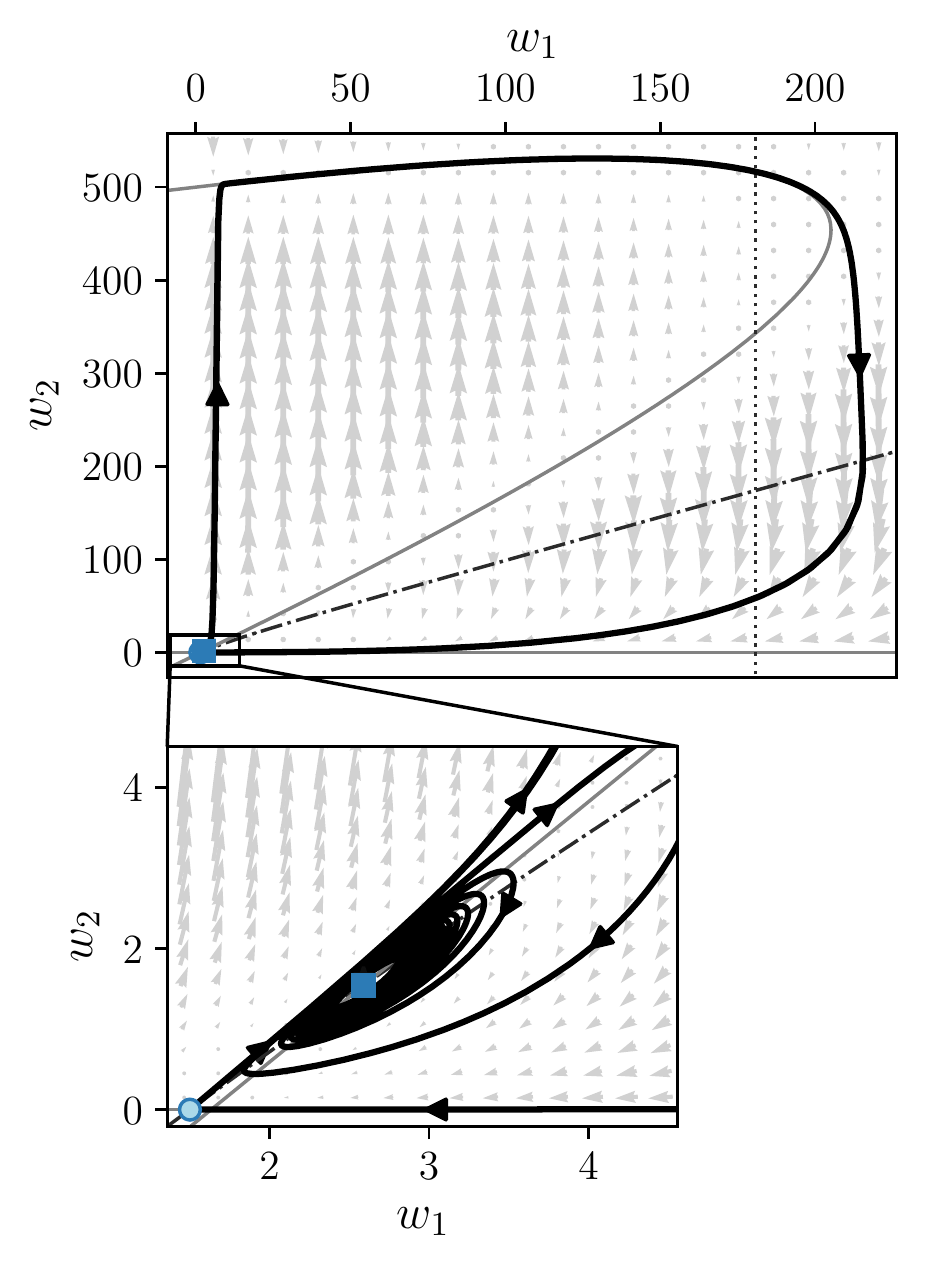}
			\label{fig:pp2}
		}
		\hspace*{-\floatsep}
		\subfloat[$\sigma^{\rm Hopf}<\sigma<\sigma^{\rm SO}$]{
			\includegraphics[width=0.33\columnwidth, valign=b]{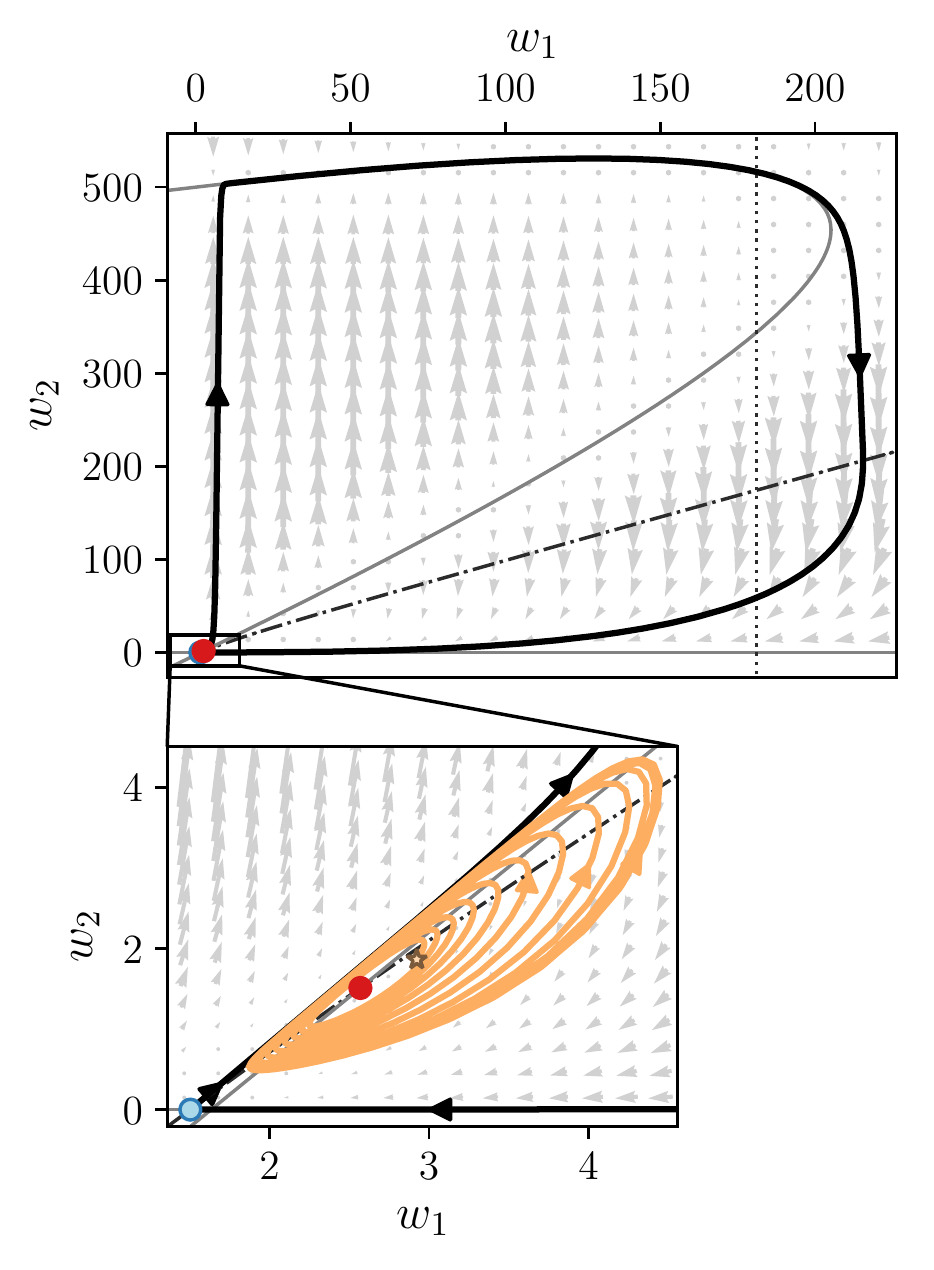}
			\label{fig:pp3}
		}\\
		\hspace*{-\floatsep}
		\subfloat[$\sigma^{\rm SO}<\sigma<\sigma^{\rm T}$]{
			\includegraphics[width=0.33\columnwidth, valign=c]{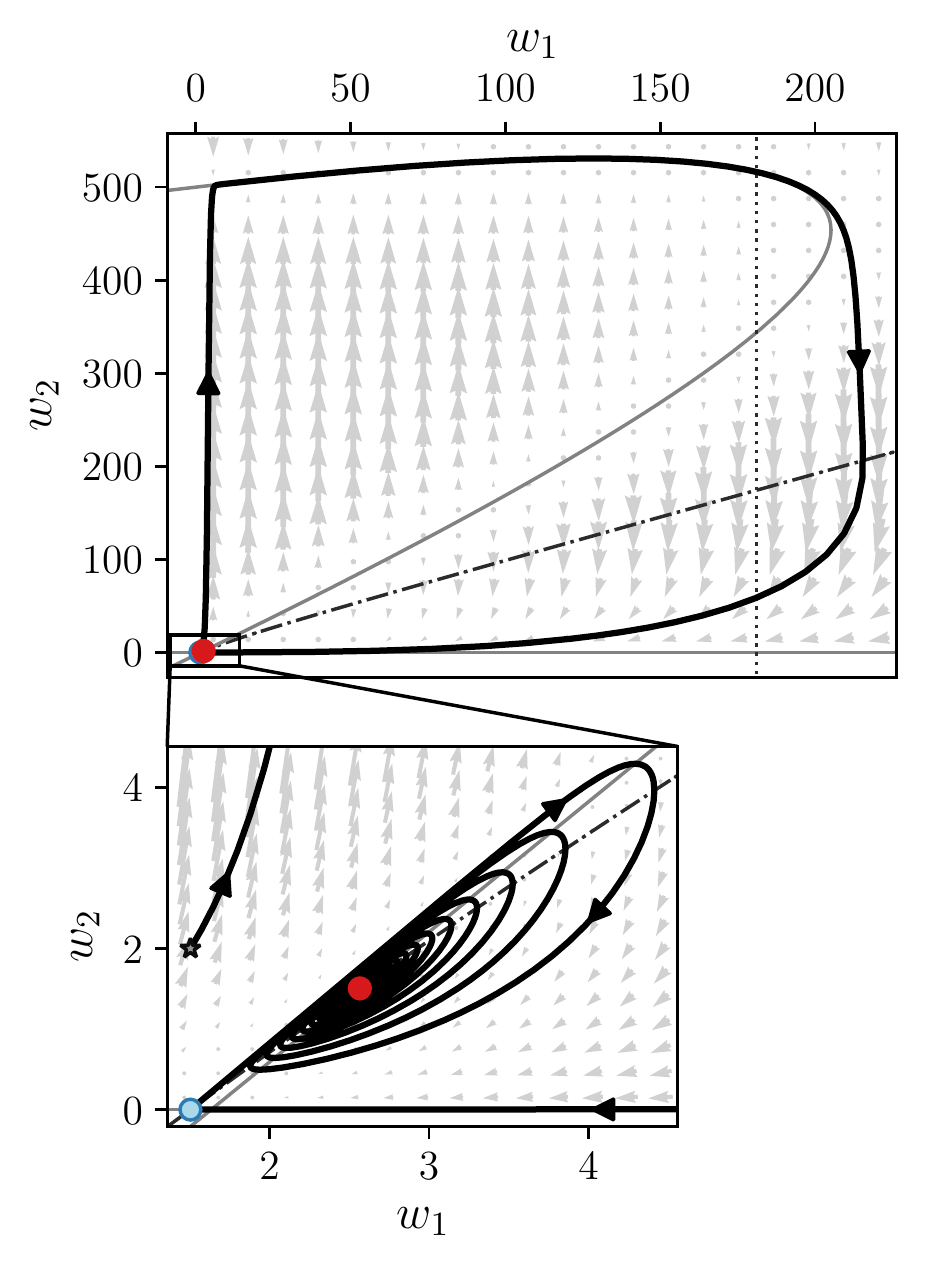}
			\label{fig:pp4}
		}
		\hspace*{-\floatsep}
		\subfloat[$\sigma>\sigma^{\rm T}$]{
			\includegraphics[width=0.33\columnwidth, valign=c]{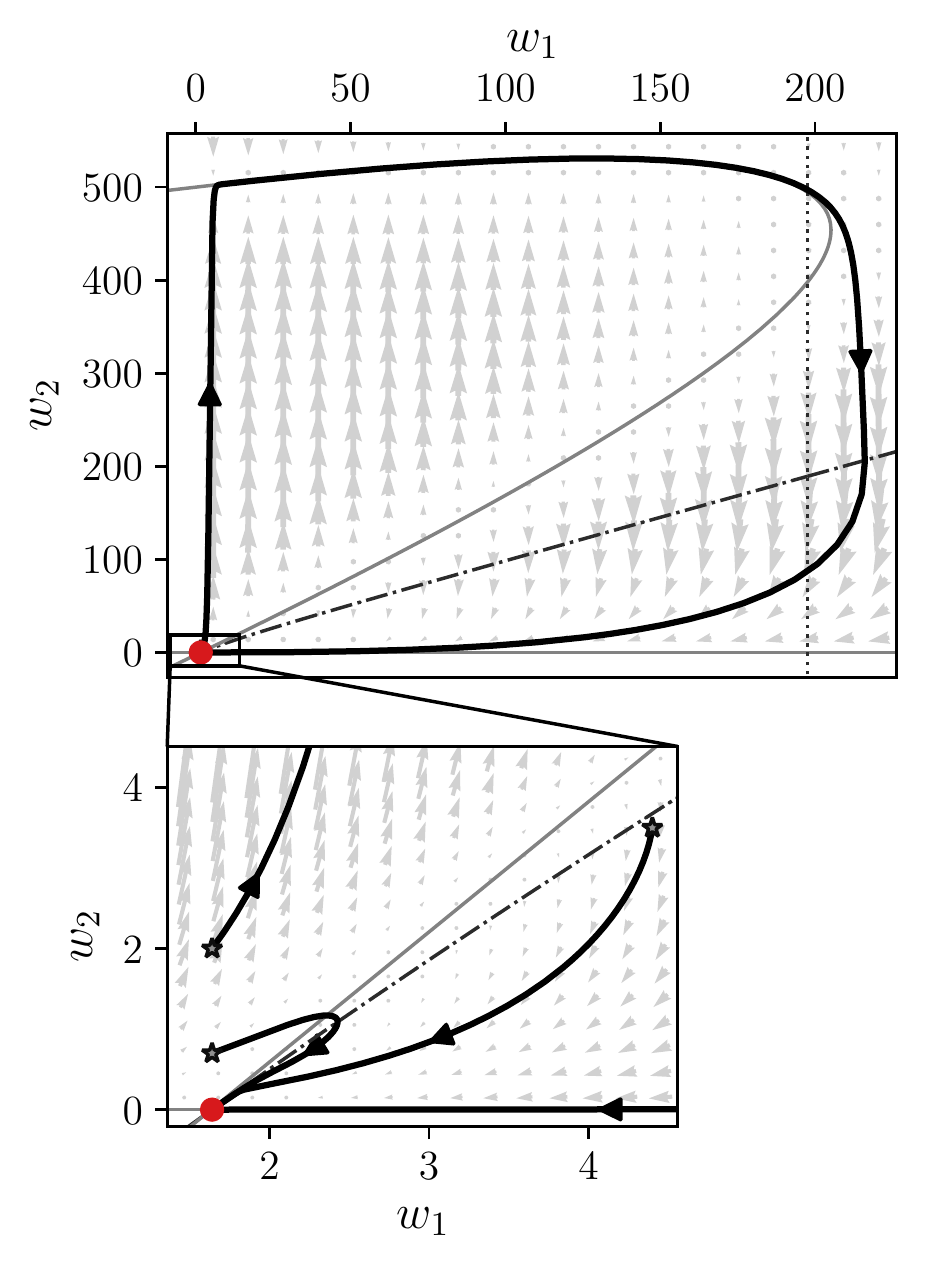}
			\label{fig:pp5}
		}
		\hspace*{-\floatsep}
			\subfloat{\hspace*{0.045\columnwidth}%
			\includegraphics[width=0.26\columnwidth, valign=c]{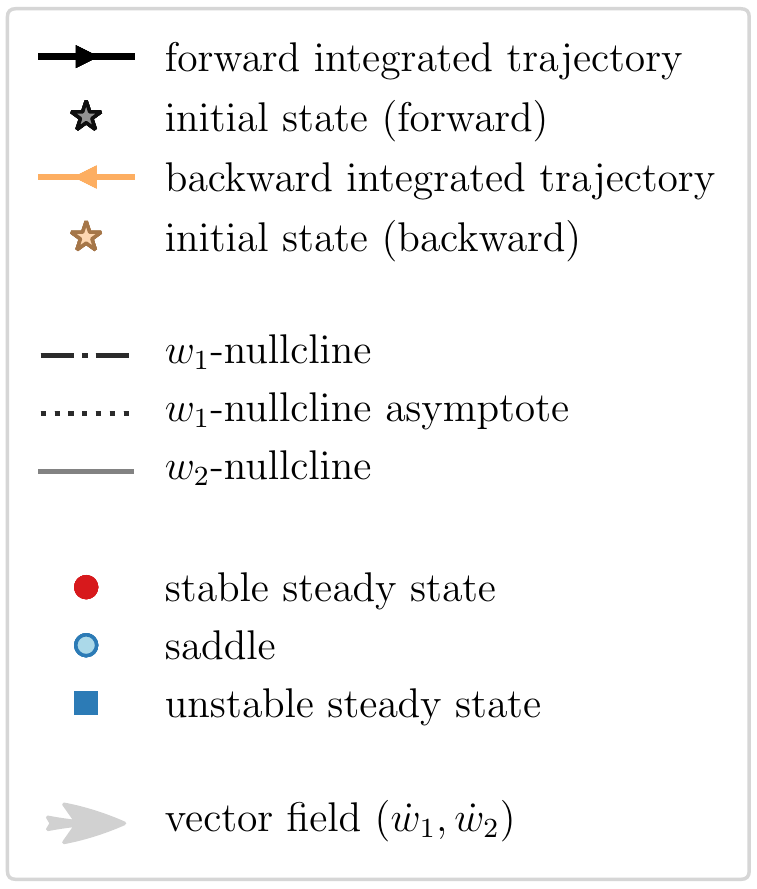}%
			}
	\caption{As $\sigma$ is varied in Regions \crefrange{enum:regA}{enum:regE} of the parameter space (see the schematic in \cref{fig:bifurcation_diagram_schematic}), phase portrait plots (A)--(E) of the long-timescale model \eqref{eq:sys_slow2} in $(w_1, w_2)$ plane reveal that the model exhibits rich bifurcating behaviours, which include emergence of orbits and large excursions in phase space. The values of $\sigma$ for which the phase portraits are generated lie within intervals indicated in respective plot captions, while other parameters are fixed at their default values (see \cref{tab:param_nondim}). The lower panel plots show model behaviour in the vicinity of the tumour-free steady state by magnifying this area of phase space. Solid black curves show forward integrated trajectories (stable), whereas solid orange curves denote trajectories integrated backward in time (unstable) in order to uncover the dynamics close to unstable orbits. For schematics of these phase portrait plots, see \cref{app:schematics_phase_portraits}.
	}
	\label{fig:phase_planes_regions}
\end{figure*}
\renewcommand{\thesubfigure}{\alph{subfigure}}

\begin{table*}[b!]
	\centering
	\footnotesize
	\begin{tabular}{Nlll}
		\toprule
		\multicolumn{1}{l}{\begin{tabular}{@{}l@{}} \textbf{Region} \\ \textbf{No.} \end{tabular}} & \begin{tabular}{@{}l@{}} \textbf{{}Denotation in $(\sigma, \delta)$ parameter}\\ \textbf{space of Fig. \ref{fig:2Dplot_default} and the left} \\\textbf{panels of Figs. \ref{fig:2Dplot_rho}, \ref{fig:2Dplot_eta} and \ref{fig:2Dplot_mu}} \end{tabular} & \begin{tabular}{@{}l@{}}\textbf{Description in Sec. \ref{sec:bifurcation_structure}} \\ \textbf{and phase portrait in}\\ \textbf{Fig. \ref{fig:phase_planes_regions}}\end{tabular} & \textbf{Qualitative behaviour} \\
		\midrule
		\label{enum:reg1}& yellow with diagonal lines & \bref{enum:regE} & elimination \\
		\label{enum:reg2}& yellow with cross-hatching & joint \bref{enum:regC} and \bref{enum:regD} & equilibrium (or oscillations -- bistable)$^{*}$ \\
		\label{enum:reg3}& orange & \bref{enum:regB} & oscillations  \\
		\label{enum:reg4}& yellow below upper curve $\Delta=0$ & \bref{enum:regA} & escape  \\
		\label{enum:reg5tab}& yellow above upper curve $\Delta=0$ & ${**}$ & escape   \\
		\label{enum:reg6tab}& blue with diagonal lines & ${**}$ & escape or elimination -- bistable\\
		\label{enum:reg7tab}& blue with cross-hatching & ${**}$ & escape or equilibrium -- bistable\\
		\bottomrule
	\end{tabular}
	\caption{Distinct regions of $(\sigma, \delta)$ parameter space, as denoted with different colours and hatching (see the second column) in \cref{fig:2Dplot_default} and in the left-panel plots of Figures \ref{fig:2Dplot_rho}, \ref{fig:2Dplot_eta} and \ref{fig:2Dplot_mu}, exhibit different qualitative behaviours that range from elimination to escape (see the fourth column). The third column indicates which regions from the bifurcation schematic in \cref{fig:bifurcation_diagram_schematic}, where only $\sigma$ varies, some regions of the $(\sigma, \delta)$ parameter space in \cref{fig:2Dplot_default} correspond to. $^{*}$\Cref{enum:reg2} exhibits two distinct qualitative behaviours, and should be split into two disjoint parameter subregions depending on whether the stable homoclinic orbit with large amplitude and period exists. In the parameter subregion where this orbit exists, trajectories may evolve to an oscillatory solution in the vicinity of the orbit. $^{**}$For a description of model behaviour in Regions \crefrange{enum:reg5}{enum:reg7} see \cref{sec:vary_sigma_delta}. Phase portraits for these regions are omitted for brevity.}
	\label{tab:region_descr}
\end{table*}

Since exact analytical expressions for the roots of the cubic \cref{eq:ss_cubic} are not practical to work with, we have only characterised the stability of nonzero-tumour steady states using their asymptotic approximations \cref{eq:ss2,eq:ss34}, as shown above for the intermediate-sized-tumour steady state. Similar analysis for the large-tumour steady states is difficult due to the complexity of the asymptotic expressions \cref{eq:ss34} (see \cref{app:ss}). Therefore, in \cref{sec:bifurcation_structure}, we numerically identify parameter regimes in which these steady states are stable.

\subsection{Bifurcation structure for default parameter values as the basal effector supply rate varies}\label{sec:bifurcation_structure}

Our aims here are to use numerical methods to validate analytical results from \cref{sec:ss}, and to detect global bifurcations not evident from the linear stability analysis. 
We focus on $\sigma$, the nondimensional constant supply rate of effector cells, as the key bifurcation parameter. A larger value of $\sigma$ means the immune system is more effective in recruiting effector cells into the tumour microenvironment, and the recruited cells better infiltrate the tumour. In practice, $\sigma$ is expected to vary between patients since those properties of the immune system on which this parameter depends on, such as the specificity of the repertoire of effector cells and the degree of their infiltration of the tumour, exhibit high inter-patient variability \citep{Fridman2012,Rosenthal2019}. The parameter $\sigma$ may also vary in response to immunotherapies, such as vaccination and adoptive T cell therapy, both of which provide an external boost to the immune system \citep{Farkona2016}. These two therapies enhance the population of tumour antigen specific effector T cells, by increasing exposure to tumour antigens away from the tumour microenvironment described by our system (e.g. in lymph nodes or \textit{ex vivo}).

The system's bifurcation structure is depicted with bifurcation diagrams showing how (scaled) steady state tumour and effector cell numbers vary with $\sigma$ (see Figure \ref{fig:bifurcation_diagram_default}). We generate these diagrams numerically, by considering discrete values of $\sigma$ that lie in the interval $[0,1]$, while fixing other parameters at their default values (see \cref{tab:param_nondim}). We calculate for each value of $\sigma$ the steady states of system \cref{eq:sys_slow_xy}, and their corresponding eigenvalues and eigenvectors to characterise local stability. Moreover, by perturbing a steady state along the direction of an eigenvector associated with a positive/negative eigenvalue, and using the perturbed point as an initial state for forward/backward integration of equations \cref{eq:sys_slow_xy}, we also generate unstable/stable manifolds of the computed steady states. 
We plot the steady states and manifolds in $(w_1, w_2)$ phase portraits at different values of $\sigma$ (see \cref{fig:phase_planes_regions} and \cref{fig:phase_portrait_schematics}), and observe that in some regions of parameter space the phase portraits exhibit limit cycles. By tracing where limit cycles emerge and where steady states change multiplicity or stability as $\sigma$ varies, we identify five regions with qualitatively different model behaviour, separated by four bifurcation points $\sigma^{\rm SH}$, $\sigma^{\rm Hopf}$, $\sigma^{\rm SO}$ and $\sigma^{\rm T}$ (see schematic in \cref{fig:bifurcation_diagram_schematic}). 
The behaviours in the five regions \crefrange{enum:regA}{enum:regE} are described below:
\begin{enumerate}[label=(\Alph*), ref=\Alph*]
	\item $0 < \sigma < \sigma^{\rm SH}$: The system is monostable; for all positive initial conditions trajectories evolve to the larger of the large-tumour steady states (\textit{tumour escape}). There are also three unstable steady states; the tumour-free and the smaller of the large-tumour steady states are saddles, linked via a heteroclinic connection (see \cref{fig:pp1,fig:regA_schematic}). \label{enum:regA}
	\item[(AB)] $\sigma = \sigma^{\rm SH}$: The large-tumour steady states collide at a \textit{saddle-node homoclinic bifurcation} (or infinite period bifurcation) \citep{Nekorkin2015}; the saddle and node annihilate each other, and the tumour-free steady state admits a large-amplitude, stable homoclinic orbit.
	\item $\sigma^{\rm SH} < \sigma < \sigma^{\rm Hopf}$: For all positive initial conditions trajectories evolve to the stable homoclinic orbit associated with the tumour-free steady state, via \textit{oscillations with large amplitudes} (see \cref{fig:pp2,fig:regB_schematic}). \label{enum:regB}
	\item[(BC)] $\sigma = \sigma^{\rm Hopf}$: There is a \textit{subcritical Hopf bifurcation}, at which the intermediate-sized-tumour steady state becomes stable and is surrounded by an unstable, small-amplitude limit cycle that emerges from it.
	\item $\sigma^{\rm Hopf} < \sigma < \sigma^{\rm SO}$: The system exhibits \textit{bistability} between the intermediate-sized-tumour steady state and the homoclinic orbit; trajectories starting within the region of the phase space mapped out by the unstable, small-amplitude limit cycle, oscillate towards the intermediate-sized-tumour steady state (\textit{tumour dormancy}); otherwise the system evolves towards the homoclinic orbit via \textit{large-amplitude oscillations} (see \cref{fig:pp3,fig:regC_schematic}).  \label{enum:regC}
	\item[(CD)] $\sigma = \sigma^{\rm SO}$: There is a \textit{saddle-node of limit cycles} \citep{Nekorkin2015}, at which the stable homoclinic and the unstable Hopf orbits collide, and then annihilate each other.
	\item $\sigma^{\rm SO} < \sigma < \sigma^{\rm T}$: The system is monostable; trajectories evolve towards the stable intermediate-sized-tumour steady state (\textit{tumour dormancy}) via damped oscillations, possibly preceded by a large excursion (see \cref{fig:pp4,fig:regD_schematic}). \label{enum:regD}
	\item[(DE)] $\sigma = \sigma^{\rm T}$: There is a \textit{transcritical bifurcation}, at which the unstable tumour-free and the stable intermediate-sized-tumour steady state exchange stability, and the latter steady state becomes physically unrealistic (negative).
	\item $\sigma > \sigma^{\rm T}$: The system is monostable; all trajectories evolve to the tumour-free steady state (\textit{tumour elimination}). (See \cref{fig:pp5,fig:regE_schematic}) \label{enum:regE}
\end{enumerate}
The local bifurcations established numerically agree with the analytical predictions from \cref{sec:ss}. 
Further work is needed mathematically to show existence of the identified global bifurcations, and the local subcritical Hopf bifurcation (e.g. weakly nonlinear analysis); such analyses are beyond the aim and scope of this paper.

We have shown how the system dynamics change as the basal supply rate of effector cells, $\sigma$, and thereby the strength of the immune system, is increased. When $\sigma$ is small (in \Cref{enum:regA}), effector cells at sufficient numbers infiltrate tumours and decrease their volume, but the response cannot be sustained; tumour escape eventually occurs for all tumours. 
In \Cref{enum:regB} effector cells infiltrate all tumours, and markedly decrease their size. As this happens, effector cell numbers also decline, enabling the immune-suppressed tumour to relapse and rapidly grow towards its carrying capacity until the immune system is stimulated again. In this way, the cycle repeats, with tumour and effector cells oscillating in an excitable manner as they traverse phase space in the vicinity of the large-amplitude homoclinic orbit, passing through regions with very small and large tumour mass. 
Increasing $\sigma$ further (\Cref{enum:regC}) leads to the emergence of a small region of phase space, in which the immune system controls the tumour size -- tumour equilibrium or dormancy. In \Cref{enum:regC} there is a fine balance between tumour and effector cell numbers, and all solutions starting within it oscillate towards the dormant steady state as the amplitude of oscillations decreases. In order to enter this dormant region, the tumour must be suppressed in a specific way, since otherwise, the system is attracted to the homoclinic loop of escape and elimination, and we say that the tumour is ``sneaking through''. 
As $\sigma$ increases, the basin of attraction of the small tumour steady state increases so that in \Cref{enum:regD}, the immune response is strong enough to drive all tumours to a dormant state, albeit after a large phase-plane excursion (see \cref{fig:pp4}). Ideally, we would seek to increase $\sigma$ (e.g. via therapy) so that the system enters \Cref{enum:regE}, where the model predicts all tumours will be eradicated by the immune system.

Taken together, our results suggest that, as model parameters vary, the system exhibits the three Es of immunoediting -- elimination, equilibrium and escape. 
Our analysis reveals that a therapy, which increases the basal effector supply rate (e.g. adoptive cell therapy or cancer vaccine), could push the system from tumour escape to elimination or equilibrium at the intermediate-sized-tumour steady state. Such changes in system behaviour occur via a series of local bifurcations (saddle-node, Hopf and transcritical).

The system also admits stable oscillatory solutions with large excursions in the phase plane, arising from its intrinsic excitability (see \cref{sec:phase_plane}). We have shown numerically that these solutions emerge and disappear via global bifurcations of a homoclinic orbit, a feature that is common in excitable systems, but cannot be detected via local analysis. The excitable homoclinic orbit is seen as the main driver of excitable dynamics in all regions, as large excursions are observed not only in regions where the orbit exists, but also in other regions (see \cref{fig:phase_planes_regions,fig:phase_portrait_schematics}), where the associated (disconnected) steady state manifolds retain the approximate shape of the orbit.

\subsection{Changes in the bifurcation structure as default parameters vary}\label{sec:bifurcation_structure_param_sens}

\begin{figure*}[t!]
	\centering
	%
	%
	\includegraphics{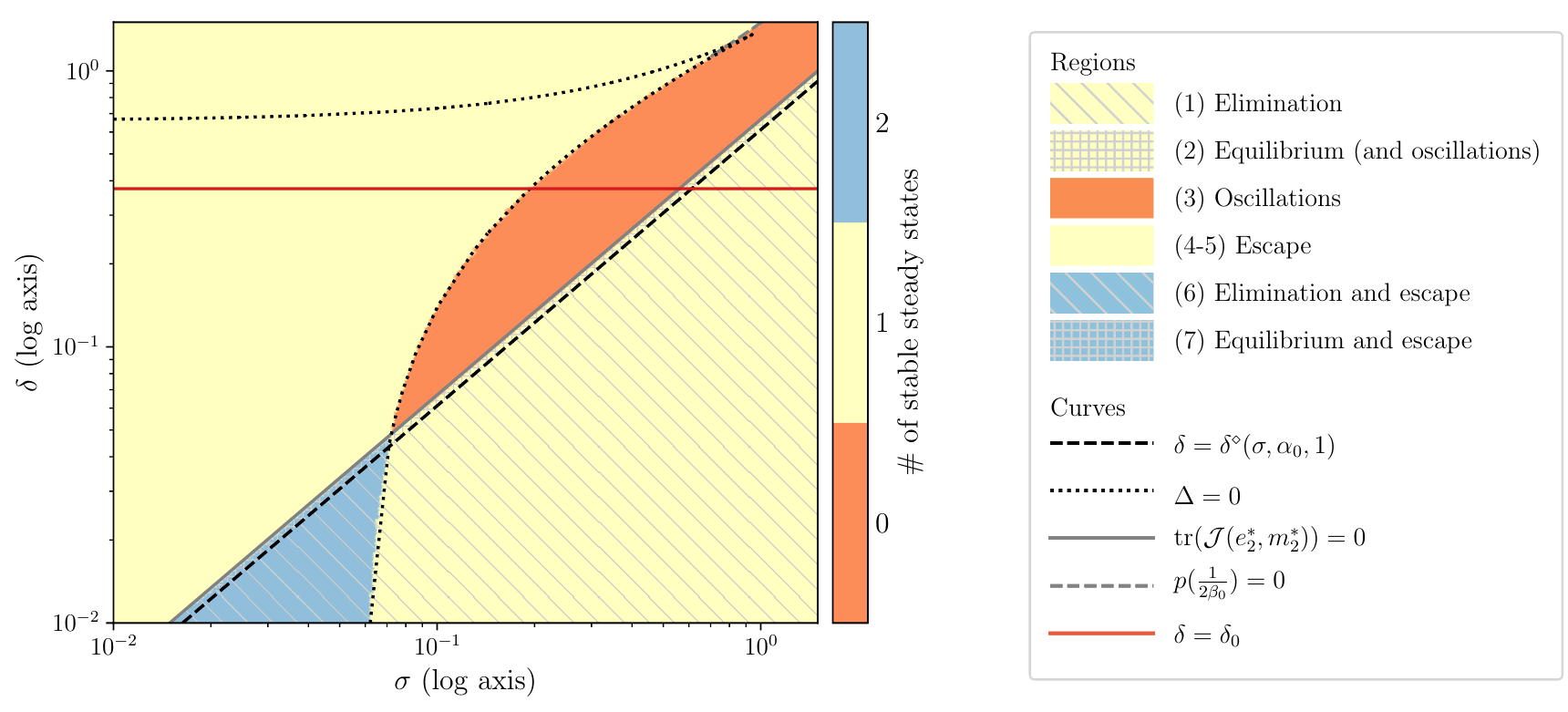}%
	\caption{The long-timescale model \eqref{eq:sys_slow2} exhibits qualitatively different behaviours as parameters $\sigma$ and $\delta$ are varied, and the model undergoes bifurcations. The plot illustrates this by showing how $(\sigma, \delta)$ parameter space is split into distinct regions depending on existence and stability of model's steady states. 
	Regions with different numbers of stable steady states are distinguished using a colour map. It is generated by numerically computing the number of stable steady states of system \cref{eq:sys_slow_xy} at discrete values of $\sigma$ and $\delta$ that lie on a triangular mesh of $(\sigma, \delta)$ plane, with other parameters fixed at their default values (see \cref{tab:param_nondim}). 
	Boundaries of distinct regions are traced by different curves on which bifurcations may occur. 
	Dotted black curves are numerically computed curves satisfying $\Delta=0$, and therefore indicate where the multiplicity of steady states changes (see \cref{sec:ss_phys}). Dashed black lines indicate where the tumour-free steady state changes its local stability and intermediate-sized-tumour steady state becomes physically unrealistic, i.e. when $\delta=\delta^\diamond(\sigma, \alpha, \kappa_2)$. Solid grey lines indicate where the local stability of the latter steady state changes, i.e. when $\tr{(\mathcal{J}(e^*_2, m^*_2))}=0$ (see \cref{sec:lin_stab}). 
	Dashed grey curves, for example by the transition between \mycrefmultiregitem{enum:reg3}{enum:reg5} for large $\delta$, indicate where a large-tumour steady state of $\mathcal{O}(\xi^{-1})$, with $0<\xi\ll 1$ (as in \cref{sec:asymp_ss}), changes stability; \note{in analysis not presented in this paper we derive that this is given by the curve $p((2\beta)^{-1})=0$, where $p$ is the cubic \eqref{eq:ss_cubic}}.
	Hatching with diagonal lines and grey cross-hatching (between solid grey and dashed black lines, see the magnified area in the plot) respectively denote regions in which the tumour-free and intermediate-sized-tumour steady states are locally stable. 
	The red horizontal line marks the default value of $\delta$, $\delta_0=0.374$ (see \cref{fig:bifurcation_diagram_default,fig:bifurcation_diagram_schematic} for bifurcation diagrams at this value of $\delta$). 
	}
	\label{fig:2Dplot_default}
\end{figure*}

We now consider how variation of other model parameters affects the bifurcation structure. 
We focus on the parameters $\mu$, $\rho$ and $\eta$, which together determine the extent of the immune response to the tumour. This is because they may be altered by immunotherapies that impact tumour-immune interactions directly, or reverse the effects of tumour-modulated immunosuppression; for example, immune checkpoint therapies that inhibit the PD-1/PD-L1 axis may, in a tumour-dependent manner, decrease T cell apoptosis, exhaustion and anergy, while also increase T cell proliferation \citep{He2015}. 
At the same time, we investigate the efficacy of such treatments across populations of patients whose immune systems 
(in the absence of a tumour) can vary in strength; a patient might have a compromised immune system (corresponding a small $\sigma$ or a large effector death rate $\delta$); they may have been vaccinated or undergone adoptive cell therapy (large $\sigma$).

\begin{figure*}[b!]
	\centering
	\subfloat[$\rho=\rho_0/113=0.01$]{%
		%
		%
		\includegraphics[width=\columnwidth]{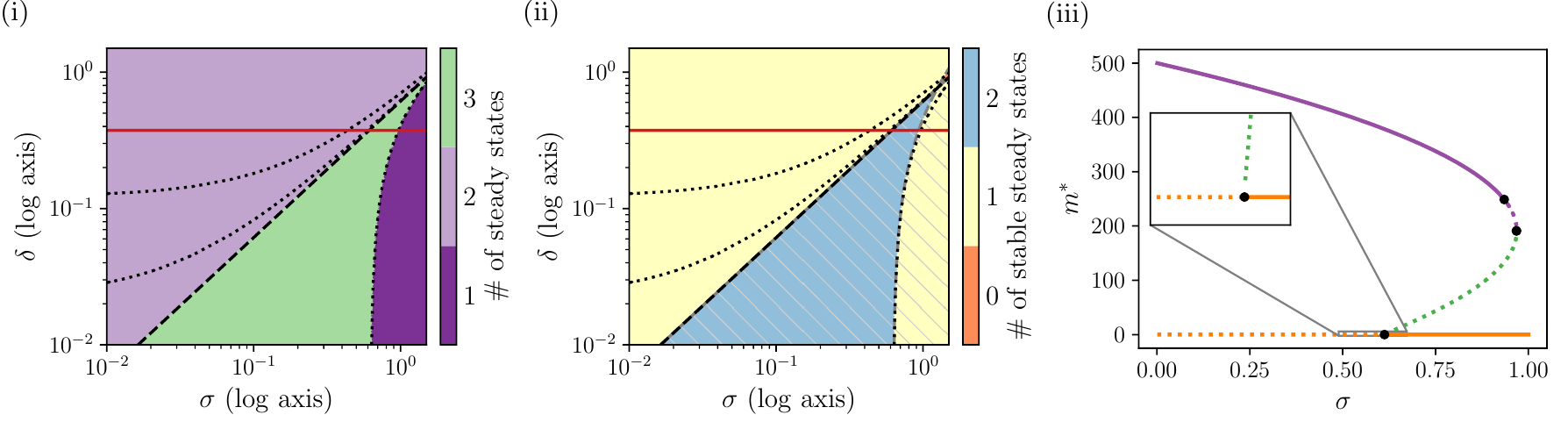}%
		\label{fig:rho_0-01}%
	}\\
	\subfloat[$\rho=\rho_0/22.6=0.5$]{%
		\includegraphics[width=\columnwidth]{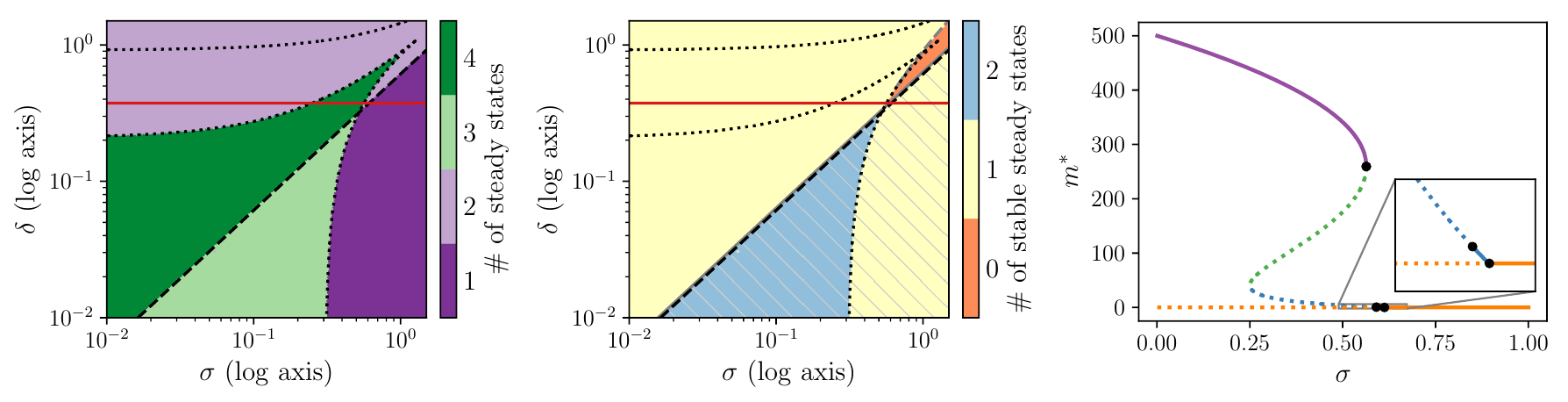}%
	}\\
	\subfloat[$\rho=1.33\rho_0=1.5$]{%
		\includegraphics[width=\columnwidth]{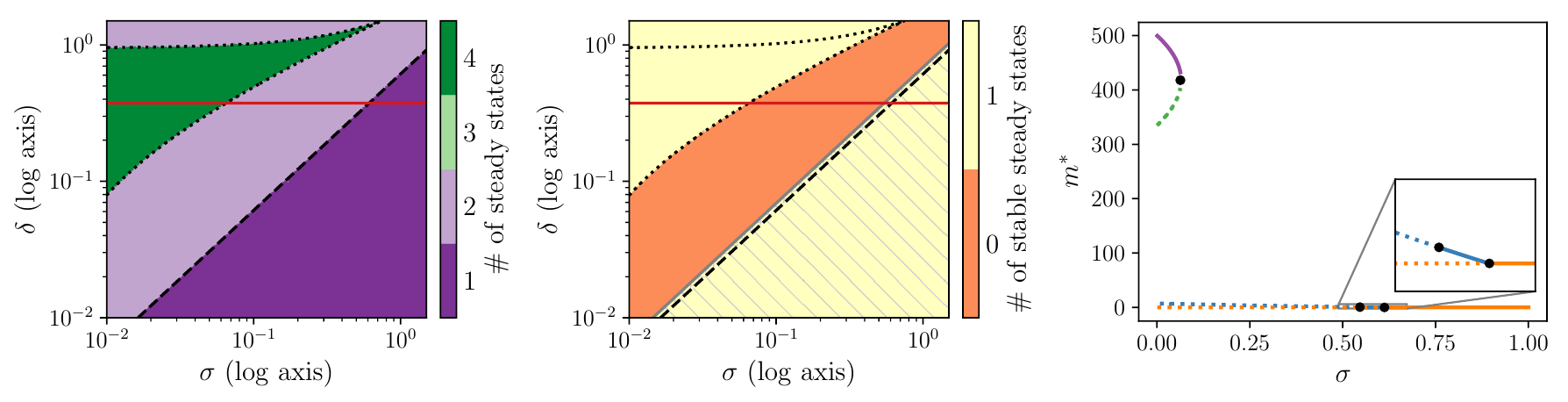}%
		\label{fig:rho_1-5}%
	}
	\caption{Series of plots (a)--(c) showing how the bifurcation structure of the long-timescale system \cref{eq:sys_slow2} changes as $\rho$ is varied from its default value $\rho_0=1.131$. As $\rho$ is increased, Regions \crefrange{enum:reg4}{enum:reg6tab} in which the model may exhibit tumour escape shrink, due to which \Cref{enum:reg1} with tumour elimination expands up to its maximum size below the black dashed line, $\delta=\delta^\diamond(\sigma, \alpha, \kappa_2)$. Above this line, while small regions in which tumour dormancy may occur (\mycrefmultiregitem{enum:reg2}{enum:reg7tab}) emerge, the dominating region is \Cref{enum:reg3} in which the model exhibits large-amplitude oscillations. For each value of $\rho$, we present three bifurcation diagrams: plots of $(\sigma, \delta)$ parameter space showing
	(i) regions with different numbers of physically realistic steady states (see \cref{fig:ss_borders} for legend and details), and (ii) regions with different numbers of stable steady states (see \cref{fig:2Dplot_default} for legend and details); 
	(iii) bifurcation diagrams of tumour steady state numbers, $m^*$, as $\sigma$ is varied and $\delta$ is fixed at its default value (see \cref{fig:bifurcation_diagram_default} for legend and details). }
	\label{fig:2Dplot_rho}
\end{figure*}

\begin{figure*}[t!]
	\centering
	\subfloat[$\eta=\eta_0/50=0.404$]{%
		%
		%
		\includegraphics[width=\columnwidth]{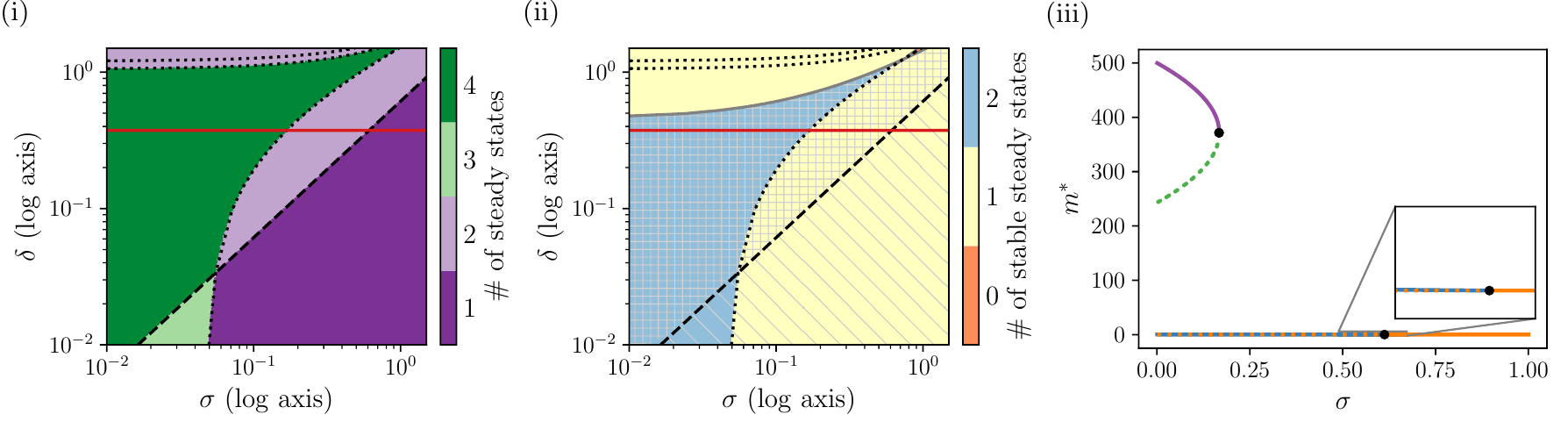}%
	}\\
	\subfloat[$\eta=2\eta_0=40.38$]{
		\includegraphics[width=\columnwidth]{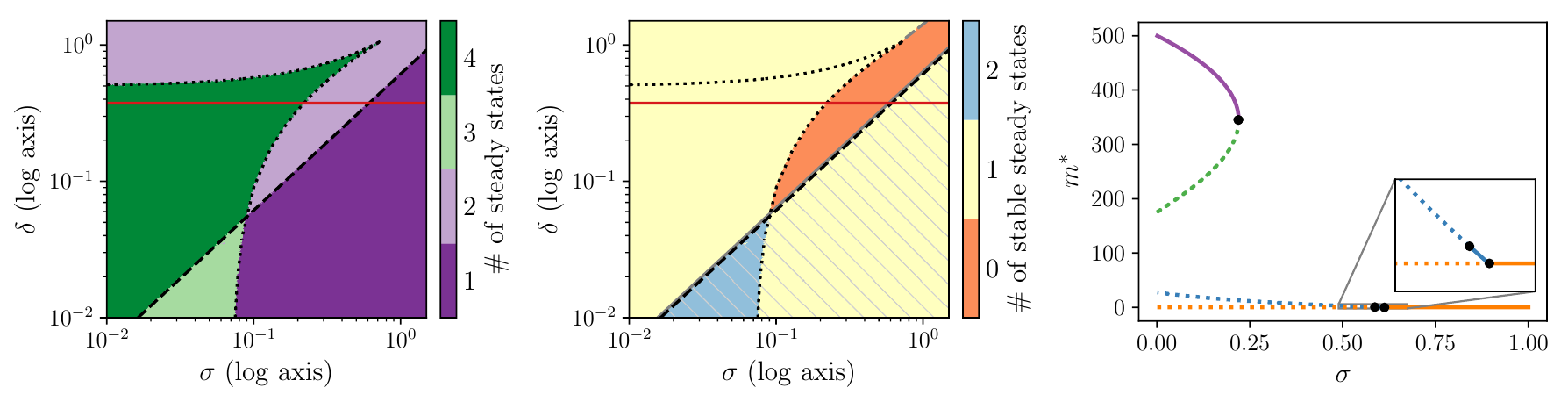}%
	}
	\caption{Series of plots (a) and (b) showing how the bifurcation structure of the long-timescale system \cref{eq:sys_slow2} changes as $\eta$ is varied from its default value $\eta_0=20.19$. 
	As $\eta$ is decreased, \mycrefmultiregitem{enum:reg2}{enum:reg7tab} in which the dormant steady state of system \cref{eq:sys_slow2} is locally stable expand. For each value of $\eta$, we present three bifurcation diagrams: plots of $(\sigma, \delta)$ parameter space showing
	(i) regions with different numbers of physically realistic steady states (see \cref{fig:ss_borders} for legend and details), and (ii) regions with different numbers of stable steady states (see \cref{fig:2Dplot_default} for legend and details); 
	(iii) bifurcation diagrams of tumour steady state numbers, $m^*$, as $\sigma$ is varied and $\delta$ is fixed at its default value (see \cref{fig:bifurcation_diagram_default} for legend and details). }
	\label{fig:2Dplot_eta}
\end{figure*}

\smalltitle{Varying \texorpdfstring{$\sigma$}{\textsigma} and \texorpdfstring{$\delta$}{\textdelta}}{sec:vary_sigma_delta} We first study the bifurcation structure when $\sigma$ and $\delta$ 
co-vary in a neighbourhood of the default parameter regime. This provides a reference for benchmarking model outcomes associated with variation of parameters linked to immunotherapy. We numerically track the number of physically realistic steady states, and their (local) stability, in order to partition the $(\sigma, \delta)$ plane (see Figure \ref{fig:2Dplot_default}); transitions from one region to another indicate qualitative changes in the system dynamics. Using analytical results from \cref{sec:ss}, we also compute curves, on which the existence and stability of steady state solutions change. In the $(\sigma,\delta)$ plane, these curves
are in good agreement with the numerically determined region boundaries. 
This confirms our analysis and increases our insight into the behaviour of our model. We identify seven distinct regions listed in \cref{tab:region_descr}. 
Regions \crefrange{enum:reg1}{enum:reg4} exhibit behaviours equivalent to those in Regions \crefrange{enum:regA}{enum:regE}, as related in \cref{tab:region_descr}, and described in both \cref{sec:bifurcation_structure} and \cref{fig:bifurcation_diagram_schematic}. Behaviours in Regions \crefrange{enum:reg5}{enum:reg7} are described as follows:
\begin{enumerate}[label=(\arabic*), ref=\arabic*]
  \setcounter{enumi}{4}
  \item The system is monostable; for all positive initial conditions trajectories evolve to the large-tumour steady state (\textit{tumour escape}). The only other steady state is the tumour-free one, which is a saddle. \label{enum:reg5}
  \item The system exhibits bistability between the tumour-free steady state (\textit{tumour elimination}) and the larger of the large-tumour steady states (\textit{tumour escape}); depending on initial conditions trajectories evolve to one of these two stable steady states. 
  The smaller of the large-tumour steady states is a saddle. \label{enum:reg6}
  \item The system exhibits bistability between the inter\-mediate-sized-tumour steady state (\textit{tumour equilibrium}) and the larger of the large-tumour steady states (\textit{tumour escape}). 
  The smaller of the large-tumour steady states and the tumour-free steady state are both saddles. \label{enum:reg7}
\end{enumerate}

In \cref{fig:2Dplot_default} we see that varying $\delta$ typically has a similar effect to varying $\sigma$. The model predicts that therapies that increase $\sigma$ and/or decrease $\delta$ can drive the system to regions of parameter space where tumour elimination  or dormancy occur for all initial conditions (\Cref{enum:reg1} and a subregion of \Cref{enum:reg2} respectively).

We now investigate how varying the tumour-immune interactions parameters, $\mu$, $\rho$ and $\eta$, affects model behaviour in $(\sigma,\delta)$ parameter space. 

\smalltitle{Varying \texorpdfstring{$\rho$}{\textrho}}{}

As $\rho$, the tumour-stimulated supply rate of effector cells, decreases, we observe in panel (iii) of \cref{fig:2Dplot_rho} an increase in the region in which a locally stable large-tumour steady state exists; tumour cell numbers of this steady state decrease as $\sigma$ increases, and its confluence with the saddle large-tumour steady state occurs at a larger value of $\sigma$. The distinction among the nonzero-tumour steady states (see \cref{sec:asymp_ss}) is blurred for smaller values of $\rho$, and eventually one nonzero-tumour steady state is lost (e.g. at $\rho=0.01$, see panels (i) and (iii) of \cref{fig:2Dplot_rho}). This is supported by our asymptotic results (see \cref{sec:ss_phys}), which predict that the region given by \cref{eq:cond_ss2_real}, in which the intermediate-sized-tumour steady state is physically realistic, will shrink as $\rho$ is reduced. 

When $\rho$ is small, the presence of tumour cells does not stimulate effector cells and their proliferation sufficiently to eliminate or control the tumour, e.g. due to low tumour immunogenicity or immunosuppression. We see that, within the plotted $(\sigma, \delta)$ parameter region, decreasing $\rho$ shrinks \mycrefmultiregitem{enum:reg2}{enum:reg7tab} in which the intermediate-sized-tumour steady state is mono- or bistable, and dormancy may occur (see panels (ii) and (iii)  of \cref{fig:2Dplot_rho}). Since the fraction of the $(\sigma, \delta)$ parameter space with a locally stable large-tumour steady state (Regions \crefrange{enum:reg4}{enum:reg6tab}) increases, \Cref{enum:reg1} exhibiting tumour elimination contracts, while \Cref{enum:reg6tab} with bistability of escape and elimination expands.

Conversely, by increasing $\rho$, effector cells may increase their numbers at a greater rate in response to the tumour, and eradicate it more effectively. Therefore, as $\rho$ increases, there is a reduction in Regions \crefrange{enum:reg4}{enum:reg6tab} in which tumour escape may occur. \Cref{enum:reg1} with elimination thus expands, but only up to its maximum size below the line $\delta=\delta^\diamond(\sigma, \alpha, \kappa_2)$ (see \cref{sec:asymp_ss}). Above this line, \Cref{enum:reg3} with stable, large-amplitude, oscillatory solutions is considerably magnified.  While \mycrefmultiregitem{enum:reg2}{enum:reg7tab} in which tumour dormancy may occur also appear, these are small, and large-amplitude oscillations or escape are the dominant behaviour (see panel (ii) of \cref{fig:rho_1-5}).

\smalltitle{Varying \texorpdfstring{$\mu$}{\textmu}}{}
Increasing $\mu$, the rate of effector cell inactivation, has a similar effect on model behaviour as decreasing $\rho$, but with greater sensitivity (see \cref{app:bif}). This is expected since the term $\frac{\rho em}{\eta + m}$ in model equations \cref{eq:sys_full} may saturate, whereas the related term $-\mu em$ does not. 

\smalltitle{Varying \texorpdfstring{$\eta$}{\texteta}}{} 
Changes in parameter $\eta$, which determines the size of the tumour at which effector proliferation starts saturating, have a different effect on the model dynamics compared to changes in either $\rho$ or $\mu$. Varying $\eta$ mainly affects the zero-trace/Hopf bifurcation curve (solid grey), where the intermediate-sized-tumour steady state changes stability (see \cref{fig:2Dplot_eta}). Decreasing $\eta$ therefore increases the size of \mycrefmultiregitem{enum:reg2}{enum:reg7tab} that exhibit mono- and bistability of tumour dormancy, respectively, and reduces the size of \Cref{enum:reg3} with stable oscillatory solutions. We conclude that a therapy, which reduces $\eta$ (i.e. increases the threshold at which limitations in immune response start to occur)
, may push the system to a region where tumour dormancy is possible; due to the prospect of region bistability this may hold only for a subset of the phase space.

\smalltitle{Summary}{}
We observe that in $(\sigma, \delta)$ parameter space the line $\delta=\delta^{\diamond}(\sigma, \alpha, \kappa_2)$, which is only affected by changes in the tumour growth rate, $\alpha$, or the scaling parameter, $\kappa_2$, stratifies patients with parameters in $(\sigma, \delta)$ parameter space into those with strong immune responses (immunocompetent or vaccinated), whose tumours may be eliminated, and those with weaker responses (immunocompromised), whose tumours may at best be controlled. Assuming that the immune state of the patient, characterised by parameters $\sigma$ and $\delta$, is fixed, we note that changing parameters $\rho$ and $\mu$ separately or simultaneously is beneficial for immunocompetent patients. For immunocompromised patients, manipulating $\eta$ is more effective, as it is possible to achieve tumour dormancy. Ultimately, to achieve tumour elimination for immunocompromised patients, the basal function of the immune system must be improved, for example by targetting the parameters $\sigma$ and/or $\delta$.


\section{Discussion and conclusions}\label{sec:discussion}

We investigated a five-compartment mathematical model by \citet{Kuznetsov1994} to increase understanding of the impact that nonlinear relations between tumour and effector (immune) cells in the tumour microenvironment have on tumour growth dynamics. We performed a new %
asymptotic reduction of the model and characterised its behaviour. In particular, we systematically studied the model's sensitivity to changes in parameter values in order to understand how the efficacy of the immune system and its response to the tumour affect tumour growth. 
We showed further that the model is excitable, which would be a notable feature of tumour-immune interactions,  and revealed how this can give rise to complex dynamic behaviour not reported in analyses of alternative simplifications of \citeauthor{Kuznetsov1994}'s model.

In \citeauthor{Kuznetsov1994}'s original model a separation of timescales occurs due to differences in the rates of conjugate formation and dissociation without further effect and all remaining processes. 
Exploiting this to introduce a small parameter, we derived a two-dimensional matched asymptotic approximation that is more accurate in this distinguished limit than the original quasi-steady-state approximation (QSSA) proposed by \citet{Kuznetsov1994} (see \cref{app:qssa}). 
On the short timescale, the numbers of total tumour and total effector cells do not vary at leading order, while conjugates relax to a steady state. This allows for simplification of the full model on the long timescale, where conjugates react instantenously to changes in total effector and total tumour cell numbers, and so vary parametrically with respect to these variables at leading order. The accuracy of the asymptotically reduced model was compared numerically with the full model on both timescales and in multiple parameter regimes, and shown to preserve its excitable and bifurcating dynamics, unlike the QSSA model. This result highlights that special care is needed when applying the quasi-steady-state assumption. We note that parameter estimates for the rate of conjugate dynamics are needed in order to justify our asymptotic reduction; this may in future motivate the exploration of alternative distinguished limits of \citeauthor{Kuznetsov1994}'s original model.

On the long-timescale, in a neighbourhood of the default parameter regime, the tumour-immune dynamics of our reduced model are excitable due to the nullcline structure and the large ratio of timescales. Excitability manifests through large phase-plane excursions; these are characterised by alternation between short periods of large and rapid changes in total tumour cell numbers, and long periods of slow variation in total effector cell numbers, during which tumour cell numbers remain close to zero or the carrying capacity. The switch from slow to fast dynamics is driven by exponential growth of the tumour when effector cell numbers are much smaller than tumour cell numbers. This model feature illustrates how the underlying saturating growth, as modelled by logistic growth dynamics in this context, may contribute to excitability. It also provides a possible explanation for why 
tumours often return after periods of remission in clinic. Long-term tumour recurrence is exhibited in the model through long time periods with low tumour cell numbers, followed by rapid (exponential) tumour regrowth. We note that the length of the tumour remission period under small parameter changes varies considerably, which suggests our excitable model is sensitive to parameters at small cell numbers. In future work, we will therefore investigate how stochastic effects impact the excitable model dynamics, focusing on regimes when cell numbers are small. 

Like \citeauthor{Kuznetsov1994}'s QSSA model, our reduced model exhibits the three Es of immunoediting  -- elimination, equilibrium, and escape \citep{Dunn2004}. We investigated how the number and nature of steady state and limit cycle solutions change as we vary $\sigma$, the basal rate at which effector cells are introduced to the tumour microenvironment, and $\delta$, their death rate, in order to identify parameter regions in which each stage of immunoediting arises via a (locally) stable steady state. 
For large $\sigma$ and small $\delta$ (immunocompetent or vaccinated patient), tumour elimination is predicted. For small $\sigma$ and large $\delta$ (immunocompromised patient)%
, the tumours escape immune surveillance (and grow to carrying capacity). For intermediate values of $\sigma$ and $\delta$, the tumour responses are predicted to be diverse; bistability of the tumour elimination/equilibrium and escape solutions can occur; monostability of a homoclinic orbit with a large amplitude, or bistability of the latter with the tumour equilibrium solution can also arise. We conclude that inter-subject variability in immune function, particularly in the supply and death rates of effector cells, or in the initial numbers of tumour and effector cells, can result in large qualitative differences in how a particular tumour progresses. Experiments have indicated similar qualitative dichotomies in tumour progression curves between subjects that respond to cancer therapy, and those that do not, examples being preclinical mouse studies testing the efficacy of immune checkpoint inhibitors of the PD-1/PD-L1 axis in combination with a Bruton's tyrosine kinase inhibitor \citep{Sagiv-Barfi2015} or with radiotherapy \citep{Dovedi2017}. Variability in parameters and variability in initial cell numbers represent two different hypotheses via which our model may explain inter-subject heterogeneity of tumour responses observed experimentally. 

Our analysis has shown that the intrinsic excitability of the model generates a complex, non-intuitive, bifurcation structure that includes emergence of a homoclinic orbit. 
Model solutions traversing the vicinity of this orbit undergo long-term oscillations that demonstrate long-term tumour reccurence. Short-term oscillations, as also observed in leukemias \citep{Mehta1980,Rodriguez1976,Gatti1973}, can also be observed in the model's excitable parameter regime and away from it; here the system may evolve via damped oscillations to a nonzero-tumour steady state or to a limit cycle with smaller periods and amplitudes than for the large homoclinic orbit.  
While no limit cycle solutions, only damped oscillatory steady state solutions, exist in the QSSA model proposed by \citet{Kuznetsov1994}, long- and short-term oscillatory behaviour of the type presented in this paper was observed in a model developed by \citet{Kirschner1998} to describe interactions of cancer and immune cells in the presence of interleukin-2, a cytokine inducing effector cell proliferation. 
We note further that in parameter regimes where the homoclinic orbit and the equilibrium steady state solution are bistable, our reduced model exhibits sneaking through, whereby very small tumours may escape (in the vicinity of the orbit) while larger ones may be controlled, and, counter-intuitively, immunostimulation can be damaging. In \citeauthor{Kuznetsov1994}'s QSSA model these phenomena were shown to depend on the rate of effector cell inactivation, $\mu$, while in our model they are attributed to the model's excitable properties. 
Overall, having demonstrated that our reduced model is consistent with a range of clinically observed phenomena, we postulate that excitable dynamics may underpin tumour-immune interactions.

By varying parameters that regulate the dynamics of the effector cells, we used our model to investigate how tumour cell numbers may be controlled via immunotherapy in a heterogeneous population of patients. As previously suggested, immunotherapies that improve competence of a patient's immune system (i.e. increase baseline effector supply rate or decrease effector death rate), such as adoptive cell therapy or vaccination, could drive tumours from tumour escape to elimination or control at a small-tumour steady state. Other immunotherapies may directly target and manipulate the immune response to the tumour, such as immune checkpoint therapies. Treatments that target  $\rho$, the maximum rate of tumour-stimulated supply of effector cells, and $\mu$, the rate of tumour-induced effector cell inactivation, are mainly beneficial to immunocompetent patients that lie in a bistable region of tumour escape and elimination. For immunocompromised patients, the most promising therapy was found to be one that impacts $\eta$, the tumour cell density at which increased effector cell proliferation and infiltration due to the tumour become limited. Reducing this parameter may drive the system to a monostable parameter region exhibiting tumour equilibrium, or to a bistable region with only a subset of smaller tumours under immune control. We note, however, that tumour control in the latter region would likely be difficult to achieve in practice (e.g. via surgery, in which the effector and tumour cell numbers are manipulated) due to the small size of the basin of attraction of the dormant steady state.  For these patients, the system may be moved to a monostable tumour equilibrium regime via a therapy that perturbs $\mu$ or $\rho$ in addition to $\eta$. The model also suggests it may be more effective to give the majority of immunocompromised patients a combination therapy that initially increases the patient's immunocompetence; this way tumour elimination/control may be achieved or become more easily attainable with subsequent immunotherapies that directly modulate the immune response to the tumour.

In practice, immunotherapy will likely impact multiple parameters, as a result of interconnectedness of pathways in the tumour microenvironment. Also, combination (immuno)therapies may be synergistic or antagonistic \citep{Melero2015, Rojas2015, Lai2017}. 
These observations nonetheless do not diminish the insight our results provide. However, they raise questions about the practical feasibility and complexity management of tuning model parameters and (combination) therapies with high enough accuracy to drive the patient and their tumour into a desired parameter region. 
In our study we used parameter values estimated in \citet{Kuznetsov1994} by fitting their QSSA model to murine data \citep{Siu1986}. The full model would in future have to be sufficiently tuned against patient tumour volume data, in order to identify realistic parameter regimes and increase the model's predictive power. We therefore aim to fit our reduced model to human data as well as to murine data from \citet{Siu1986}.

There are several ways in which the original model could be modified to better describe tumour-immune interactions. We could generalise the logistic tumour growth term to include competition between all cells in the tumour region for resources, growth factors and space. The model assumption of a constant supply of activated cytotoxic T cells (CTLs), $s>0$, 
is only valid when the region has been previously exposed to tumour antigens, or if the organism has been treated with a cancer vaccine or adoptive cell therapy as in \citet{Kirschner1998}. In a model without treatment effects, we could omit the baseline effector influx term \citep{Kirschner1998, Kronik2008, Itik2010, Letellier2013} to take into account that activation of CTLs is predominantly tumour-dependent, and not instantenous. 
The current model also does not capture immunosuppression as dominant at large tumour sizes, since the number of effector cells does not decrease with a growing tumour population in model simulations; this could also be an aspect for further exploration. 

The simplifying assumptions in the original model by \citet{Kuznetsov1994} allowed us to asymptotically reduce it to a two-dimensional tumour-effector model, and to gain a thorough understanding of its behaviour through a systematic dynamical systems analysis. Despite its limitations, the model captures a variety of clinically observed phenomena through dynamics that are subject to excitability and a rich bifurcation structure. The model has thus demonstrated the complexity of tumour-immune interactions, and the heterogeneity in their final outcomes. Preferred strategies for controlling tumour size may differ between patients depending on the strength of their immune systems, which is a result that supports personalised approaches to cancer therapy. While the model, and the hypotheses generated from it, must be validated with experimental/clinical data, we emphasise that understanding the intricate behaviours of such models is one of the pivotal steps towards overcoming difficulties associated with their validation, and, in the longer term, using validated and evidence-based models to generate patient-specific predictions.


\section*{Acknowledgements}
\addcontentsline{toc}{section}{Acknowledgements}
This publication is based on work supported by the EPSRC Centre For Doctoral Training in Industrially Focused Mathematical Modelling (EP/L015803/1) in collaboration with AstraZeneca.

	\bibliography{library}

	\begin{appendices}
		\crefalias{section}{appsec}
		\numberwithin{equation}{section}
		\counterwithin{table}{section}
		\counterwithin{figure}{section}

\section{Nullclines}\label{app:nullclines}
Here we present equations for nullclines of the reduced, long-timescale system \eqref{eq:sys_slow2}. On the $w_1$- and $w_2$-nullclines we have that $\dot{w_1}=0$ and $\dot{w_2}=0$ respectively. We find that $w_1$-nullclines are a subset of solutions to the quadratic in $w_2$ given by
\begingroup
\allowdisplaybreaks
\begin{subequations}
\begin{align}
	\bar{a}(w_1) w_2^2 + \bar{b}(w_1) w_2 + \bar{c}(w_1) = 0,
	\label{eq:w1_nullcline}
\end{align}
where $\bar{a}$, $\bar{b}$ and $\bar{c}$ are polynomials in $w_1$ given by
\begin{align}
	\begin{aligned}
		&\bar{a}(w_1):= \bar{a}_1 w_1 + \bar{a}_0,\\
		&\bar{b}(w_1):= \bar{b}_2 w_1^2 + \bar{b}_1 w_1 + \bar{b}_0,\\
		&\bar{c}(w_1):=\bar{c}_3 w_1^3 + \bar{c}_2 w_1^2 + \bar{c}_1 w_1 + \bar{c}_0,
	\end{aligned}
\end{align}
with coefficients dependent on model parameters as follows
\begin{align}
	\begin{aligned}
	\bar{a}_0 & = \sigma \kappa _1 \left(\left(\eta  \kappa _1-1\right) \left(\delta  \kappa _1-\mu \right)+\rho\kappa _1  \right), \\
	\bar{a}_1 & = - \mu  \left(\left(\eta  \kappa _1-1\right) \left(\delta  \kappa _1-\mu \right)+\rho\kappa_1  \right),\\
	\begin{split}
		\bar{b}_0 &=  \sigma  \left(\left(\eta  \kappa _1+1\right) \left(\left(\eta  \kappa _1-1\right) \left(\delta  \kappa _1-\mu \right)+ \rho \kappa _1  \right)
		\right.\\ & \phantom{=}\left.
		+ \sigma\kappa _1 \kappa _2   \left(\eta  \kappa _1-1\right)\right),
	\end{split}
	\\
	\begin{split}
		\bar{b}_1 & = (\rho - \eta \mu - \delta) ((\eta \kappa_1 - 1)(\delta \kappa_1 - \mu) + \rho\kappa_1 ) 
		\\&\phantom{=}
		+ \sigma \kappa_2 (\mu - \rho\kappa_1 + \delta\kappa_1(1-2\eta\kappa_1)),
	\end{split}
	\\
	\bar{b}_2 & = \mu \kappa _1^2 \kappa _2^2 \left(\rho-\eta  \mu + \delta  \left(2 \eta  \kappa _1-1\right) \right),\end{aligned}\notag\\
	\begin{aligned}
	\bar{c}_0 &= \eta \sigma  ((\eta \kappa_1 - 1)(\delta \kappa_1 - \mu + \sigma \kappa_1 \kappa_2) + \rho \kappa_1),\\ 
	\begin{split}
		\bar{c}_1 &= -\eta \left(\delta \left( \left(\eta  \kappa _1-1\right)\left(\delta \kappa _1 - \mu \right) + \rho  \kappa _1\right)
		\right.\\&\phantom{=}\left.
		+ \sigma\kappa_2\left( \eta \kappa_1 (\delta \kappa_1 + \mu) - 2\mu - \rho \kappa_1 + \sigma \kappa_1\kappa_2\right) \right),
	\end{split}
	\\
	\bar{c}_2 &=  \eta \kappa _2 \left(\delta  \kappa _1 (\delta +\eta  \mu -\rho )+\sigma \kappa _2   \left(\delta  \kappa _1+\mu \right)-2 \delta  \mu \right),\\
	\bar{c}_3 &= - \delta  \eta \mu \kappa _2^2 .
	\end{aligned}
\end{align}
\end{subequations}
\endgroup
Analogous conditions hold for the $w_2$-nullclines, which satisfy
\begin{subequations}
\begin{align}
	w_2 = 0, \qquad \text{and} \qquad \tilde{a} w_2^2 + \tilde{b}(w_1) w_2 + \tilde{c}(w_1) = 0,
\end{align}
where $\tilde{a}$, $\tilde{b}$ and $\tilde{c}$ are given by
\begin{align}
	\begin{aligned}
	&\tilde{a} = \tilde{a}_0,\\
	&\tilde{b}(w_1) = \tilde{b}_1 w_1 + \tilde{b}_0,\\
	&\tilde{c}(w_1) = \tilde{c}_2 w_1^2 + \tilde{c}_1 w_1 + \tilde{c}_0,
	\end{aligned}
\end{align}
and
\begin{align}
	\begin{aligned}
		\tilde{a}_0 &=\alpha  \beta  \kappa _1^2,\\
		\tilde{b}_0&=\alpha  \left(\kappa _1 \left(\kappa _1-\beta \right)-\alpha  \beta  \left( \kappa _1 + \beta\right)\right),\\
		\tilde{b}_1&=2 \alpha  \beta  \kappa _1 \kappa _2,
		\\
		\tilde{c}_0&=\alpha  \left(\kappa _1+\alpha  \left(\beta +\kappa _1\right)\right),\\
		\tilde{c}_1&=\kappa _2	\left( \alpha \beta(\alpha - 1)-\kappa _1(\alpha + 1)\right),\\
		\tilde{c}_2&=-\alpha  \beta  \kappa _2^2.
	\end{aligned}
\end{align}
\end{subequations}

\section{Asymptotic approximation of the non\-zero-tumour steady states and conditions for their physicality}\label{app:ss}
Here we detail the derivation of asymptotic approximations of the nonzero-tumour steady states presented in \cref{sec:asymp_ss}, and discuss where they are physically realistic (real and nonnegative).

The nonzero-tumour steady states are defined as $(e^*, m^*)=(\frac{\alpha}{\kappa_2}(1-\beta m), m)$, where $m$ is a root of the cubic \eqref{eq:ss_cubic}. We assume $0<\xi\ll 1$, $\mu=\xi\hat{\mu}$ and $\beta=\xi\hat{\beta}$. Under this rescaling, \eqref{eq:ss_cubic} becomes
\begin{subequations}
\begin{align}
	\begin{split}
		p(m)&=\xi^2 \hat{a}_3 m^3 + \xi \hat{a}_{2} m^2 + \hat{a}_{1} m + a_0 
		\\&\phantom{=}
		+ \xi^2 \hat{a}_{21} m^2 + \xi \hat{a}_{11} m =0, 
	\end{split}
	\label{eq:ss_cubic_xi}
\end{align}
where 
\begin{align}
	\begin{aligned}
		&a_0=\eta (\tfrac{\sigma\kappa_2}{\alpha} - \delta),\\
		&\hat{a}_{1}=\tfrac{\sigma\kappa_2}{\alpha} + \rho - \delta,\\
		&\hat{a}_{11} = \eta(\hat{\beta}\delta - \hat{\mu}),
	\end{aligned}
	\qquad
	\!
	\begin{aligned}
		&\hat{a}_{2} =
		-\hat{\mu} + \hat{\beta}(\delta - \rho) ,\\
		&\hat{a}_{21} = 	\eta\hat{\mu}\hat{\beta},\\
		&\hat{a}_3 = 
		\hat{\mu}\hat{\beta}.
	\end{aligned}\label{eq:coeffs_asymp}
\end{align}
\end{subequations}
We seek approximate solutions to \eqref{eq:ss_cubic_xi} of the form $m=\xi^{-\gamma}\hat{m}$ with $\gamma$ a constant and $\hat{m}=\mathcal{O}(1)$, giving
\begin{align}
	\begin{split}
		p(\xi^{\gamma}\hat{m})&=\xi^{2-3\gamma} \hat{a}_3 \hat{m}^3 + \xi^{1-2\gamma} \hat{a}_2 \hat{m}^2 + \xi^{-\gamma} \hat{a}_1 \hat{m} 
		\\&\phantom{=}
		+ \hat{a}_0+\xi^{2-2\gamma} \hat{a}_{21} \hat{m}^2 + \xi^{1-\gamma}\hat{a}_{11} \hat{m} =0.
	\end{split}
	\label{eq:cubic_with_xi}
\end{align}
We find pairwise dominant balances among terms in the cubic to simplify it.

First, balancing the third and fourth terms gives $\gamma=0$. With $\gamma=0$, we obtain a leading-order solution $\hat{m}_2=\mathcal{O}(1)$ to \cref{eq:cubic_with_xi} of the form
\begin{subequations}
	\begin{align}
		\hat{m}_2 := - \frac{a_0}{\hat{a}_1} = 
		\frac{\eta( \alpha\delta - \sigma\kappa_2)}{\sigma \kappa_2  - \alpha\delta + \alpha\rho},
	\end{align}
	which gives
	\begin{align}
		m^*_2 \sim m_2 \sim \hat{m}_2\quad \text{and}\quad e^*_2\sim\frac{\alpha}{\kappa_2}(1-\xi \hat{\beta}\hat{m}_2)
		\sim\frac{\alpha}{\kappa_2}=:\hat{e}_2.
	\end{align}
\end{subequations}
These approximate solutions correspond to an inter\-me\-diate-sized-tumour steady state given by $(e_2^*, m_2^*)\sim(\hat{e}_2, \hat{m}_2)$.
Noting $\rho>0$, it is straightforward to show that this steady state is physically realistic if and only if $\frac{\sigma\kappa_2}{\alpha}\leq\delta<\rho + \frac{\sigma\kappa_2}{\alpha}$ (or equivalently $\frac{a_0}{\eta} \leq 0 < \hat{a}_1$).

We recover the other roots of the cubic by balancing the first, second and third terms of \eqref{eq:cubic_with_xi}, by setting $\gamma=1$. This gives leading-order solutions $\hat{m}_{3,4}=\mathcal{O}(1)$ to \cref{eq:cubic_with_xi} of the form
\begin{subequations}
\begin{align}
	\hat{m}_{3,4} := \frac{-\hat{a}_2 \pm \sqrt{\hat{a}_2^2 - 4\hat{a}_3 \hat{a}_1}}{2\hat{a}_3}, 
\end{align}
so that
\begin{align}
	m^*_{3,4}\sim m_{3,4}\sim \frac{1}{\xi}\hat{m}_{3,4} \quad \text{and} \quad e^*_{3,4}\sim\frac{\alpha}{\kappa_2}(1-\hat{\beta}\hat{m}_{3,4})=:\hat{e}_{3,4}.
\end{align}%
\label{eq:ss34_m_app}%
\end{subequations}
These approximate solutions define two large-tumour steady states given by $(e^*_{3, 4}, m^*_{3, 4})\sim(\hat{e}_{3,4}, \hat{m}_{3, 4}/\xi)$. 
We note that $\hat{e}_{3, 4}$ and $\hat{m}_{3, 4}$ are both nonnegative if $0\leq\hat{m}_{3,4}\leq\frac{1}{\hat{\beta}}$. Given $\hat{a}_3>0$, we identify regions of $(\hat{a}_1, \hat{a}_2)$ coefficient space with different numbers of real and nonnegative roots $\hat{m}_{3,4}$ (see \cref{tab:roots}). This allows us to identify regions in which there are zero, one or two physically realistic steady states with large tumours. 
In summary, provided $\hat{m}_{3,4} \leq \frac{1}{\hat{\beta}}$, there is a single large-tumour steady state solution when $\hat{a}_1<0$, two such solutions when $0<\hat{a}_1<\hat{a}_2^2/4\hat{a}_3=:\hat{a}^\#$ and $\hat{a}_2<0$, and none otherwise. 

\begin{table}[t!]
	\centering
	\footnotesize
	\begin{tabular}{c|cccc}
		\toprule
		\diagbox {$\hat{a}_2$}{$\hat{a}_1$} & $\hat{a}_1<0$ & $0<\hat{a}_1<\hat{a}_1^\#$ & $\hat{a}_1^\# < \hat{a}_1$ \\
		\midrule
		$\hat{a}_2 < 0$ & $\hat{m}_4 < 0 < \hat{m}_3$ & $\hat{m}_{3, 4} > 0$ & $\hat{m}_{3, 4}$ complex \\
		$\hat{a}_2 > 0$ & $\hat{m}_4 < 0 < \hat{m}_3$ & $\hat{m}_{3, 4} < 0$ & $\hat{m}_{3, 4}$ complex \\
		\bottomrule
	\end{tabular}
	\caption{Table showing where in $(\hat{a}_1, \hat{a}_2)$ coefficient space asymptotic approximations \cref{eq:ss34_m_app} of the large roots $m^*_{3, 4}$ of the cubic \cref{eq:ss_cubic} are real and nonnegative. Note that $\hat{a}_1^\# := \hat{a}_2^2/4\hat{a}_3$.}
	\label{tab:roots}
\end{table}

\section[Comparison with the QSSA model by Kuznetsov et al. (1994)]{Comparison with the QSSA model by \citet{Kuznetsov1994}}\label{app:qssa}

Here we discuss how the behaviour of our asymptotic long-timescale approximation \cref{eq:sys_slow2} (or equivalently \cref{eq:sys_slow_xy}) of the original model \cref{eq:sys_dim} by \citet{Kuznetsov1994} differs from the QSSA model that was proposed by \citet{Kuznetsov1994}, and is given in its dimensionless form as 
\begin{subequations}
	\begin{align}
		&\dv{e}{\tau} = \sigma + \frac{\rho e m}{\eta + m} - \delta e - \mu e m,\\
		&\dv{m}{\tau} = \alpha m (1 - \beta m) - \kappa_2 e m,
	\end{align}%
	\label{eq:qssa}%
\end{subequations}
with $c=em$, initial conditions $e(0)=e_0$ and $m(0)=m_0$, and dimensionless parameter groupings defined in \cref{eq:param_nondim}.

\begin{figure*}[t!]
	\centering
	\subfloat[]{%
		\includegraphics[width=0.33\columnwidth, valign=t]{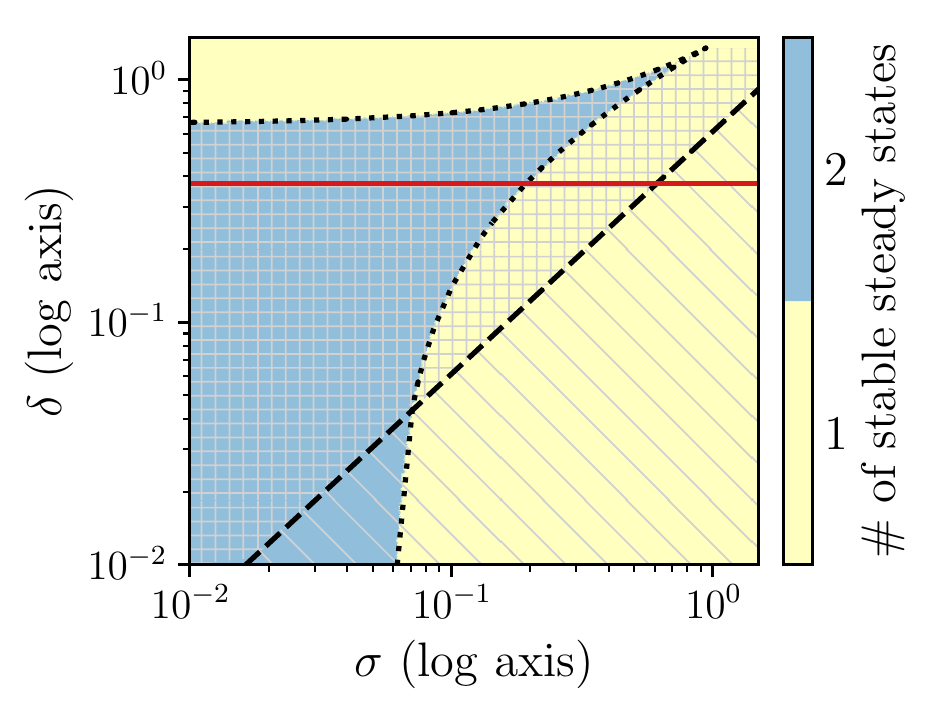}%
	}
	\hspace*{0.01\columnwidth}
	\includegraphics[width=0.25\columnwidth, valign=t]{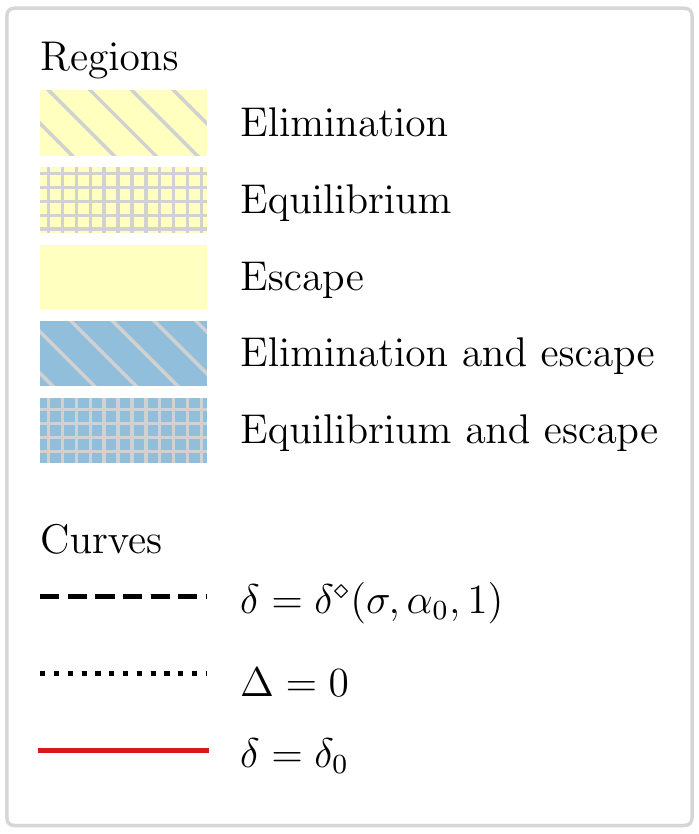}
	\hspace*{0.06\columnwidth}%
	\subfloat[]{%
		\includegraphics[width=0.33\columnwidth, valign=t]{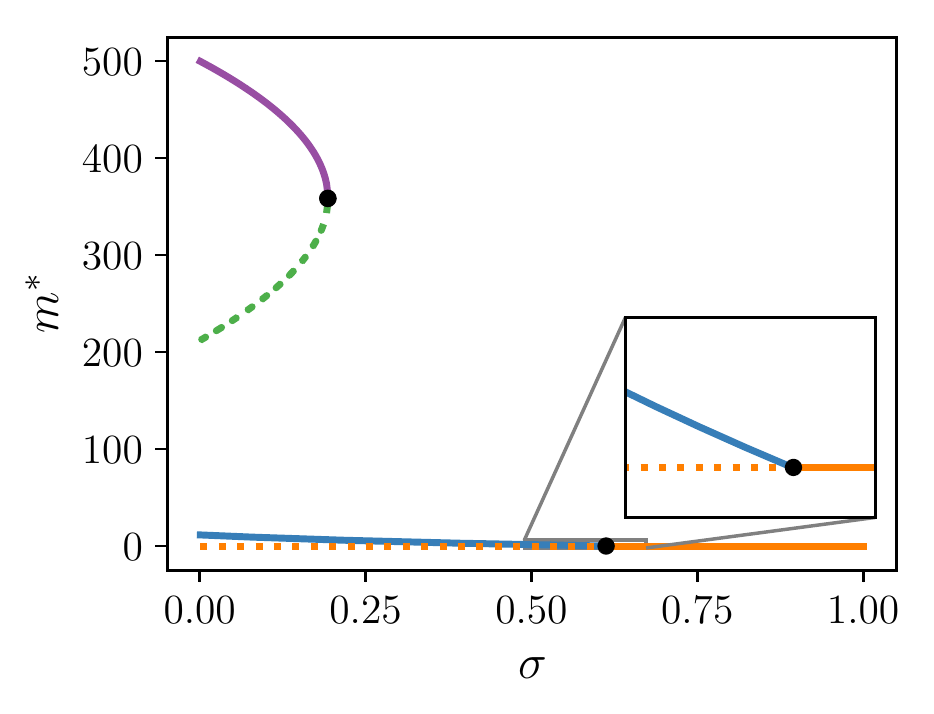}
	}
	\caption{Bifurcation structure of the QSSA model \eqref{eq:qssa} differs from that of the long-timescale model \cref{eq:sys_slow_xy} as $\sigma$ and $\delta$ are varied in the vicinity of the default parameter regime (see \cref{tab:param_nondim}); the key difference is that the intermediate-sized-tumour steady state is stable when physically realistic. Plot (a) illustrates this by showing how $(\sigma, \delta)$ parameter space is split into distinct regions depending on existence and stability of steady states in the QSSA model. Plot (b) is a bifurcation diagram of steady state tumour cell numbers in the QSSA model as $\sigma$ is varied, and $\delta$ is fixed at its default value, $\delta_0=0.374$. Plots for the long-timescale model \cref{eq:sys_slow2} that are comparable to (a) and (b) are \cref{fig:2Dplot_default,fig:bifurcation_diagram_default} respectively; see these figures for details on how (a) and (b) are generated and legend.}
	\label{fig:2Dplot_qssa}
\end{figure*}

Steady states \cref{eq:ss}, their asymptotic approximations \cref{eq:ss2,eq:ss34}, and the condition for local stability of the tumour-free steady state \cref{eq:stab_cond_zero} are equivalent in model \cref{eq:sys_slow_xy} and the QSSA model \cref{eq:qssa}. We derive a condition, analogous to \eqref{eq:trace_int1}, for stability of the intermediate-sized-tumour steady state in the QSSA model.
The Jacobian of the QSSA model, at leading order in $\xi$, is given by
$\mathcal{J}(e^*,m^*) = \mathcal{A}(e^*, m^*)$, 
with $\mathcal{A}$ defined in \eqref{eq:jac_intermediate}. For $(e^*, m^*)=(e_2^*, m_2^*)$ as in \cref{eq:ss2}, we obtain that
\begin{align}
	\det(\mathcal{A}(e^*_2, m^*_2))>0 \quad \text{as in \cref{eq:det2}},
\end{align}
and
\begin{align}
	\tr(\mathcal{A}(e^*_2, m^*_2)) = -\frac{\sigma\kappa_2}{\alpha}<0,
\end{align}
for physically realistic values of $(e_2^*, m_2^*)$ (see \cref{sec:lin_stab}). If the intermediate-sized-tumour steady state is physically realistic, then it is stable (see \cref{fig:2Dplot_qssa}). This is different from model \cref{eq:sys_slow_xy} where the trace condition \cref{eq:trace_int1} is not trivially satisfied.

\begin{figure*}[b!]
	\subfloat[]{%
		%
		\includegraphics[width=\columnwidth]{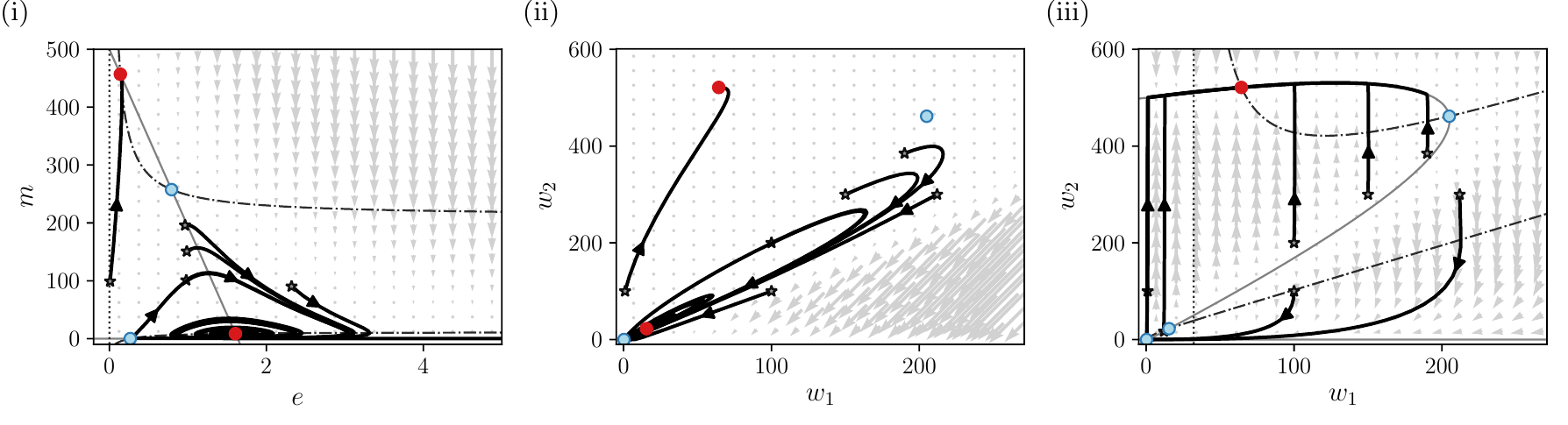}%
	}\\
	\
	\subfloat[]{%
		\includegraphics[width=\columnwidth]{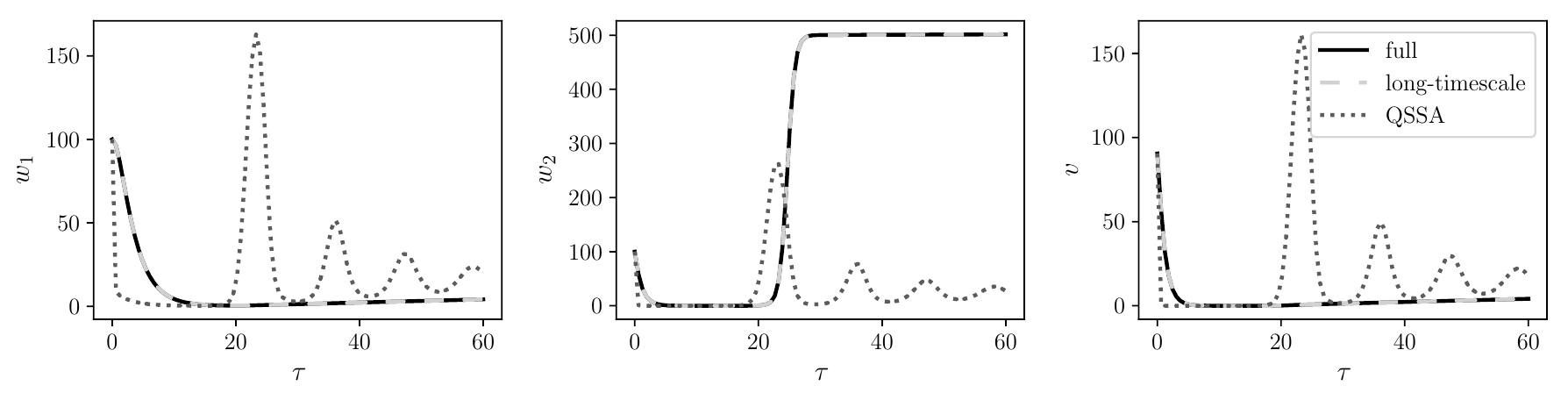}%
	}
	\caption{Series of plots (a) show phase portraits of (i) the QSSA model \cref{eq:qssa} in $(e, m)$ phase space, (ii) the same model in $(w_1, w_2)$ phase space, and (iii) the asymptotic long-timescale model \cref{eq:sys_slow2} in $(w_1, w_2)$ phase space. The QSSA model (see (i) and (ii)) does not exhibit excitable dynamics that are observed in model \cref{eq:sys_slow2} (see (iii) and also \cref{fig:phase_planes_regions,fig:phase_portraits_vary}); trajectories do not follow nullclines, and no limit cycles are observed. See \cref{fig:phase_planes_regions} for phase portrait legend. In all three plots, equivalent initial conditions, appropriately transformed, are used for computing solution trajectories via methods described in \cref{sec:num_sim}. Trajectories in plot (a)(ii) are generated by numerically solving equations \cref{eq:qssa} first; the obtained solutions for $e$ and $m$ are then transformed via $w_1=e + \kappa_1 e m$ and $w_2=m + \kappa_2 e m$ to compute numerical solutions for total effector and tumour cell numbers, $w_1$ and $w_2$. Series of plots (b) show different solutions for model variables (i) $w_1(\tau)$, (ii) $w_2(\tau)$ and (iii) $v(\tau)$ over time ($0\leq \tau\leq 60$) for the one set of initial conditions from (a), given as $w_1(0)=w_2(0)=100$, $v=v^\dagger(\boldsymbol{w}(0))=90.48$. Solid black, dashed light grey and dotted dark grey curves respectively correspond to solutions of the full model \cref{eq:sys_full}, the long-timescale model \cref{eq:sys_slow2} and the QSSA model \cref{eq:qssa}. The QSSA solution does not agree with the matching full and long-timescale solutions. For all plots in (a) and (b), parameter $\sigma$ is fixed at $\sigma=0.1$, and other parameters at their default values (see \cref{tab:param_nondim}).}\label{fig:qssa_sim}
\end{figure*}

In \cref{fig:2Dplot_qssa,fig:2Dplot_default} we observe that regions where tumour elimination may occur (yellow and blue, both with diagonal lines) correspond in both models. Due to differences in local stability of the intermediate-sized-tumour steady state mono- and bistable regions with dormancy (yellow and blue, both with cross-hatching) are larger in the QSSA model than in the long-timescale model \cref{eq:sys_slow_xy}; 
tumour equilibrium is therefore more plausible in the QSSA model. 
Simulations of the QSSA model also show simpler behaviour that is not excitable, and larger basins of attraction of the dormant steady state, than in model \cref{eq:sys_slow_xy} (see \cref{fig:qssa_sim}). The homoclinic orbit that arises in model \cref{eq:sys_slow_xy} (see \cref{fig:pp2,fig:pp3}) does not exist in the QSSA model as it was shown in \citet{Kuznetsov1994}, using the Dulac-Bendixson criterion, that there are no closed orbit solutions in the positive quadrant of $(e,m)$ phase plane.

\renewcommand{\thesubfigure}{\Alph{subfigure}}
\begin{figure*}[t!]
	\flushleft{\section[Schematics of phase portraits in Figure 3]{Schematics of phase portraits in \cref{fig:phase_portraits_vary}}\label{app:schematics_phase_portraits}%
	}
	\vspace*{-1em}
	\centering
	\subfloat[{$0<\sigma<\sigma^{\rm SH}$}]{\includegraphics[width=0.33\columnwidth]{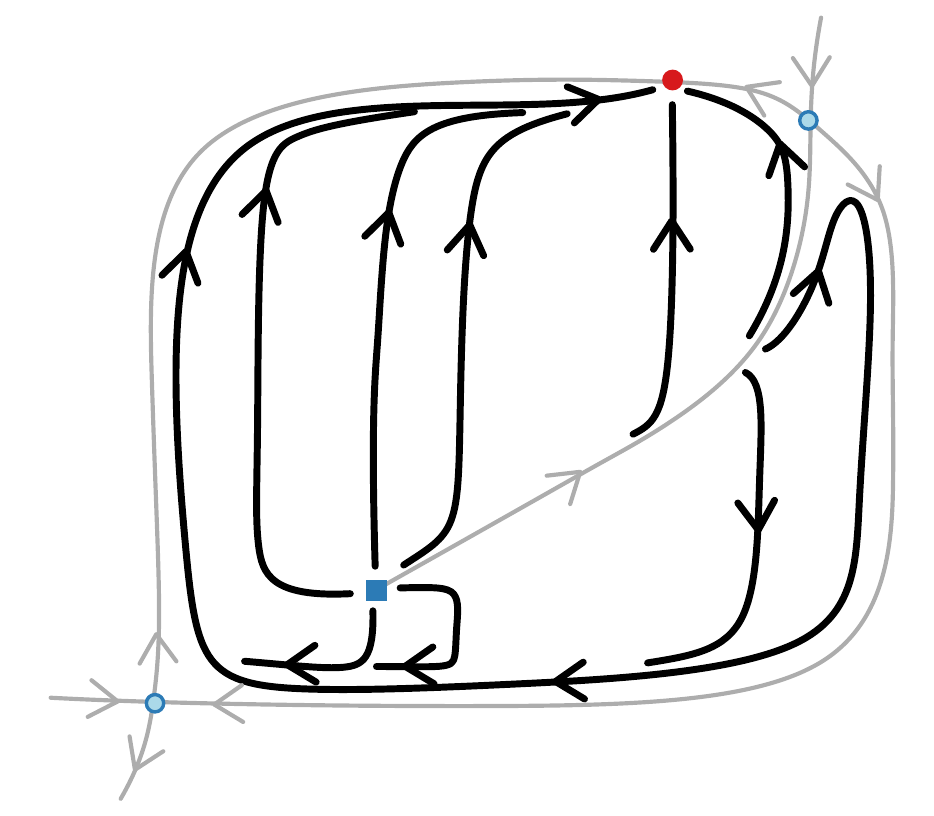}\label{fig:regA_schematic}}
	\subfloat[{$\sigma^{\rm SH} < \sigma < \sigma^{\rm Hopf}$}]{\includegraphics[width=0.33\columnwidth]{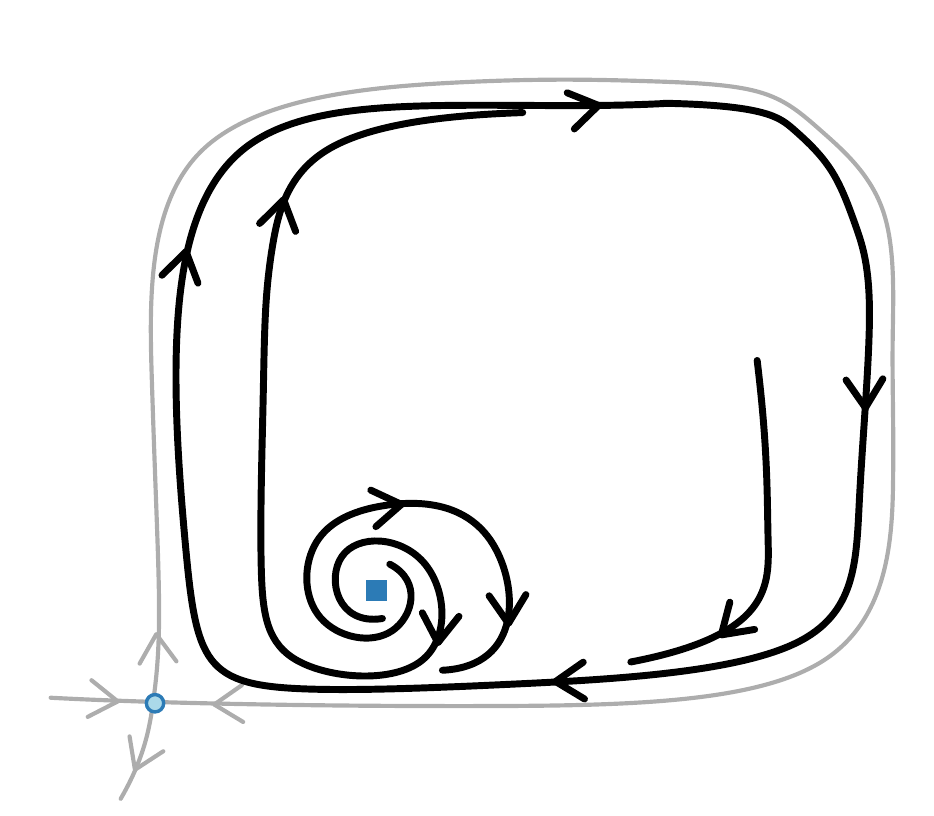}\label{fig:regB_schematic}}
	\subfloat[{$\sigma^{\rm Hopf}<\sigma<\sigma^{\rm SO}$}]{\includegraphics[width=0.33\columnwidth]{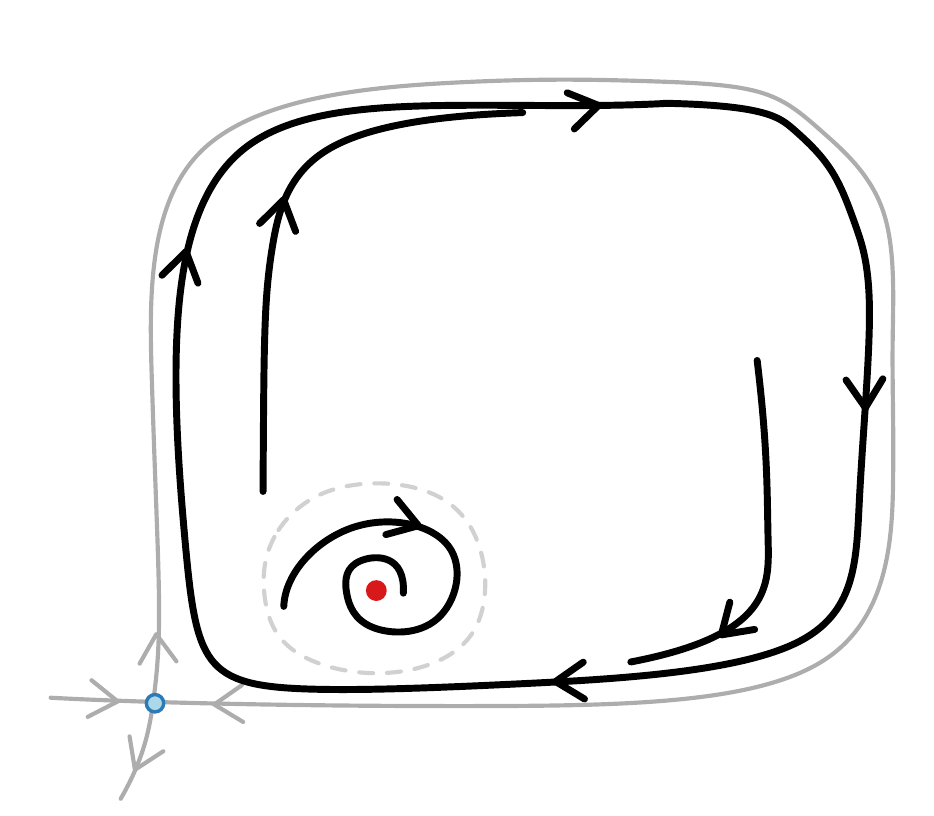}\label{fig:regC_schematic}}\\
	\subfloat[{$\sigma^{\rm SO}<\sigma<\sigma^{\rm T}$}]{\vspace*{-1em}\includegraphics[width=0.33\columnwidth]{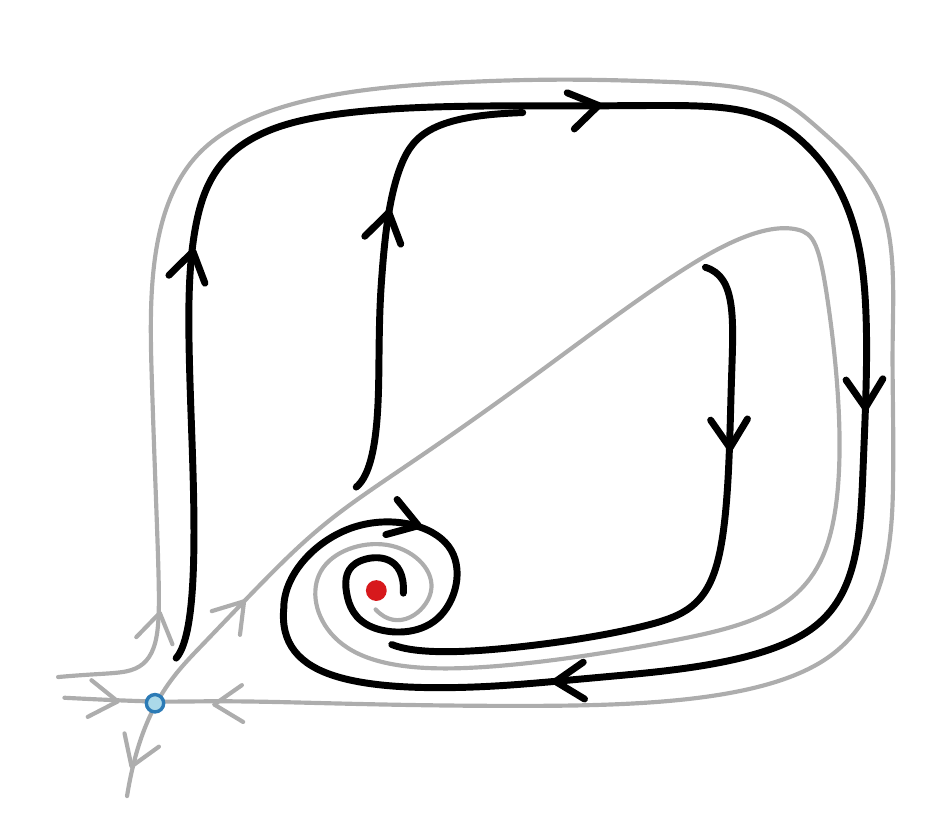}\label{fig:regD_schematic}}
	\subfloat[{$\sigma > \sigma^{\rm T}$}]{\includegraphics[width=0.33\columnwidth]{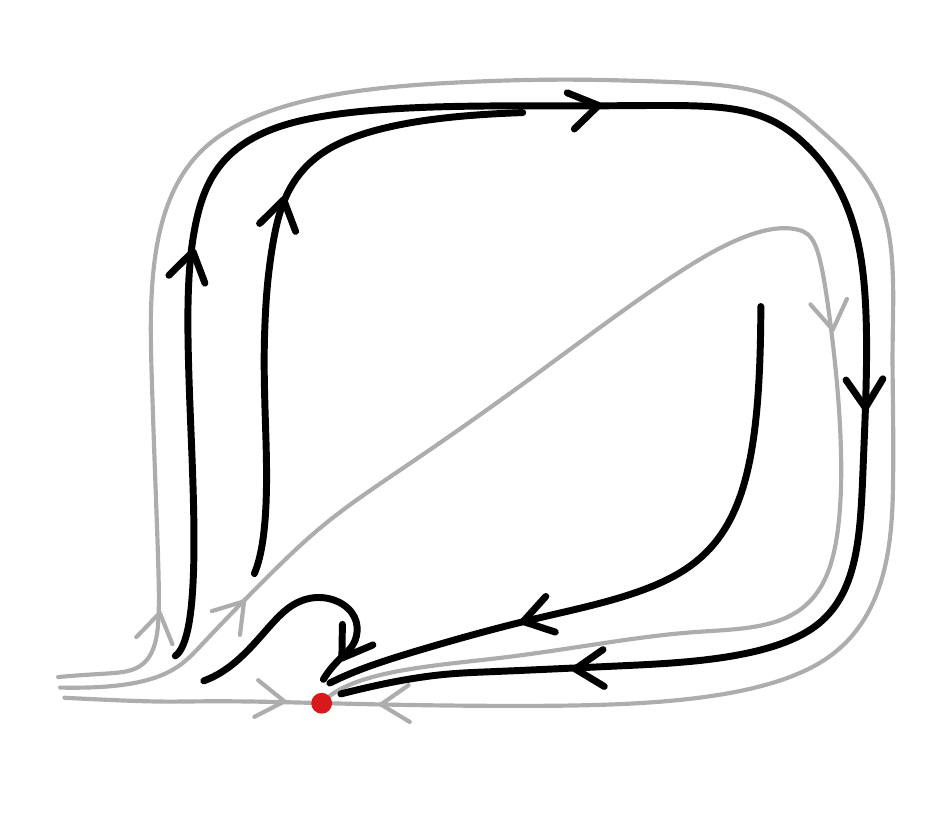}\label{fig:regE_schematic}}
	\caption{
		Schematic diagrams (not to scale) of phase portraits in \cref{fig:phase_planes_regions} as $\sigma$ varies. The vertical direction represents tumour cells, while the horizontal direction represents effector cells. Solid grey curves denote stable and unstable manifolds, solid black curves show example trajectories, and dashed grey circles represent unstable orbits. As in \cref{fig:phase_planes_regions}, stable, unstable and saddle nodes are respectively marked with red discs, blue squares and light blue discs with an outline of a darker shade of blue.
	}\label{fig:phase_portrait_schematics}
\end{figure*}%
\renewcommand{\thesubfigure}{\alph{subfigure}}%
%
\begin{figure*}[t!]
	\flushleft{%
	\section{Bifurcation structure as parameter \texorpdfstring{$\mu$}{\textmu} varies}\label{app:bif}%
	}
		\centering
		\subfloat[$\mu=\mu_0 /1.8=0.0017$]{%
			\includegraphics[width=\columnwidth]{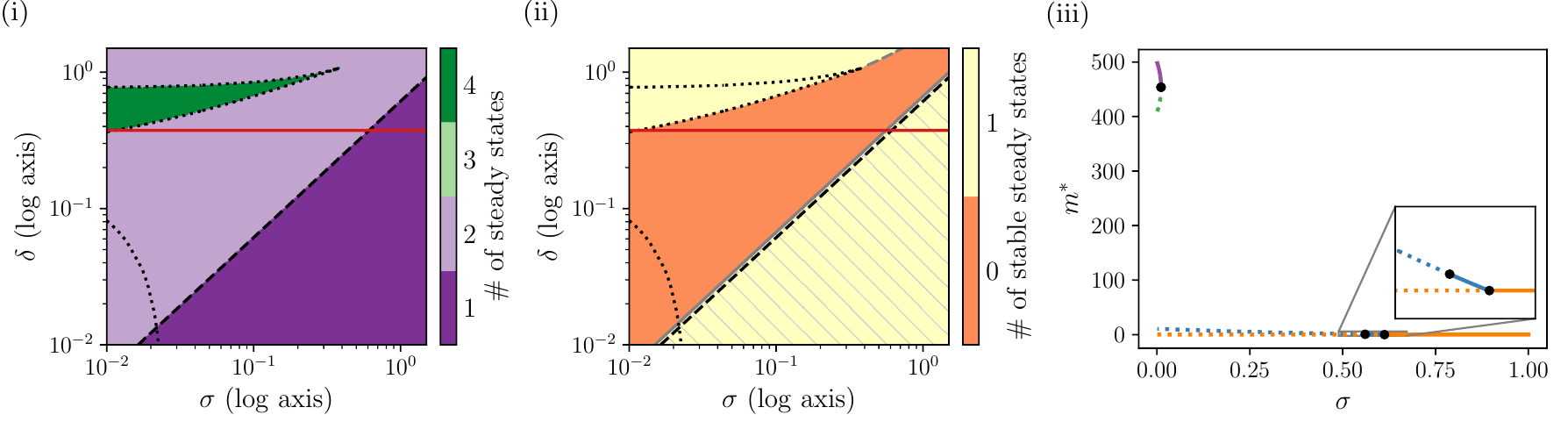}%
			}\\
		\subfloat[$\mu=2\mu_0=0.006$]{%
			\includegraphics[width=\columnwidth]{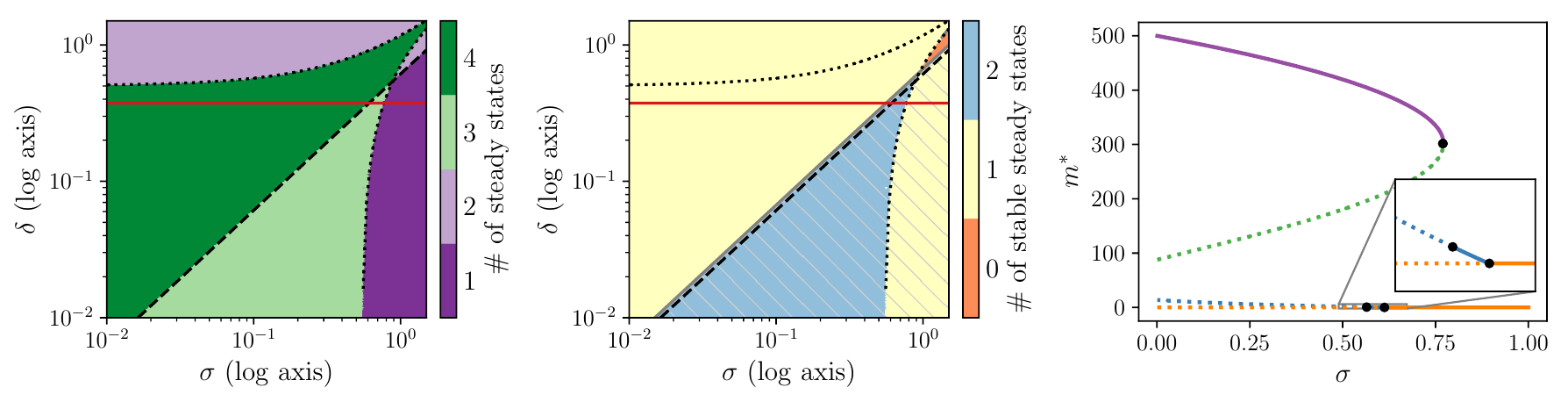}%
			}\\
		\subfloat[$\mu=32.2 \mu_0 =0.1$]{%
			\includegraphics[width=\columnwidth]{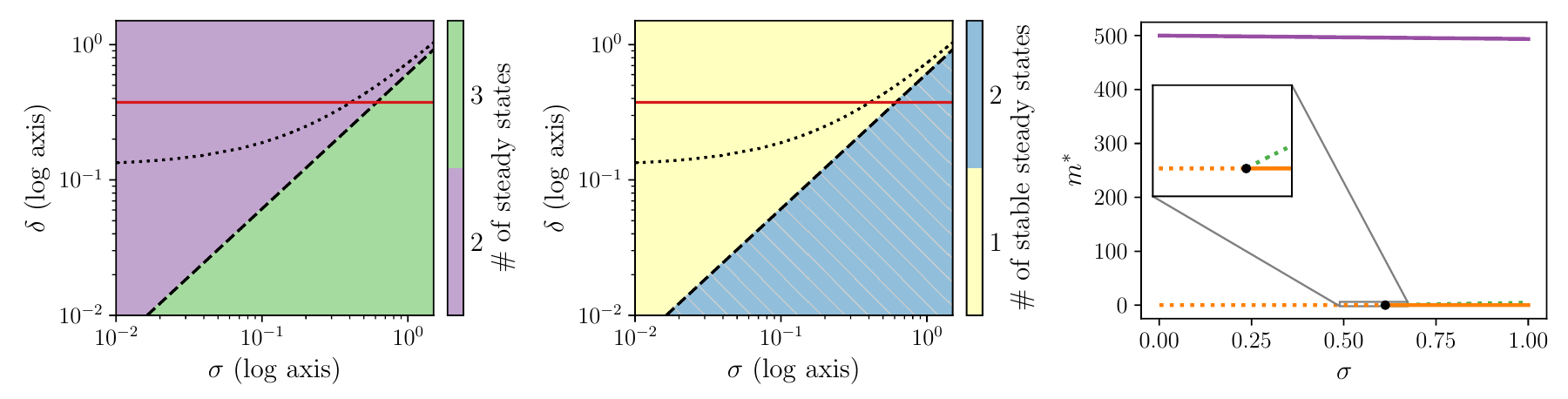}%
		}
		\caption{
		Series of plots (a)--(c) showing how the bifurcation structure of the long-timescale system \cref{eq:sys_slow2} changes as $\mu$ is varied from its default value $\mu_0=0.003$. 
		Effects of increasing $\mu$ on the system's bifurcation structure are similar to effects of decreasing $\rho$ (see \cref{fig:2Dplot_rho}). 
		For each value of $\mu$, we present three bifurcation diagrams: plots of $(\sigma, \delta)$ parameter space showing
		(i) regions with different numbers of physically realistic steady states (see \cref{fig:ss_borders} for legend and details), and (ii) regions with different numbers of stable steady states (see \cref{fig:2Dplot_default} for legend and details); 
		(iii) bifurcation diagrams of tumour steady state numbers, $m^*$, as $\sigma$ is varied and $\delta$ is fixed at its default value (see \cref{fig:bifurcation_diagram_default} for legend and details).\vspace{20em}
		}\label{fig:2Dplot_mu}
\end{figure*}
	\end{appendices}

\end{document}